\pdfoutput=1
\documentclass[PAPER, atlasdraft=false, cernpreprint, USenglish, texmf, orcidlogo]{atlasdoc}

\usepackage{atlaspackage}
\usepackage{atlasbiblatex}

\usepackage{atlasphysics}
\graphicspath{{logos/}{figures/}{figures/plots_reco_and_truth_paper/}{figures/systematics/}{figures/systematics_smooth/}}

\usepackage{ANA-STDM-2016-12-PAPER-defs}

\AtlasTitle{Measurements of isolated photon pair production in association with jets in $\boldsymbol{pp}$ collisions at $\boldsymbol{\sqrt{s}}$~=~8~TeV with the ATLAS detector}

\AtlasAbstract{
Measurements of integrated and differential production cross section of two isolated photons in events with and without associated hadronic jets based on the 2012 dataset of 20.3~fb$^{-1}$ integrated luminosity of proton--proton collisions at $\sqrt{s}=8$~TeV collected by the ATLAS experiment are presented. The measurements consider photons with pseudorapidities of $|\eta^{\gamma}|<1.37$ or $1.56<|\eta^{\gamma}|<2.37$ and transverse energy of $E^{\gamma}_{\mathrm{T}}>$~22~GeV. Jets are required to have $p^{\text{j}}_{\mathrm{T}}>$~25~GeV for $|\eta^{\mathrm{j}}|<2.4$ and $p^{\text{j}}_{\mathrm{T}}>$~50~GeV for $2.4<|\eta^{\mathrm{j}}|<4.4$.  Events are classified into categories based on the number of jets and integrated fiducial cross sections of diphoton production in association with inclusive and exclusive number of jets are presented. Fiducial cross sections, differential in sixteen diphoton, jet, dijet and photon--jet observables, are presented and compared with leading-order {\textsc{Sherpa~1.4.1}} predictions in all jet categories, fixed-order next-to-leading-order {\textsc{GoSam}} and {\textsc{BlackHat}} calculations in selected categories, and NLO-matched predictions from {\textsc{Sherpa~2.2.14}} in all categories.
Overall, the NLO predictions provide a reasonable description of the data, with significant discrepancies observed for increasing jet multiplicity, particularly for angular variables, attributable to the absence of higher-order terms in the calculations. The {\textsc{Sherpa~2.2.14}} NLO-matched predictions provide the best overall description of the data across all jet categories and observables.

}

\AtlasRefCode{STDM-2016-12}

\PreprintIdNumber{CERN-EP-2026-280}

\AtlasJournal{Phys. Rev. D}
\hypersetup{pdftitle={ATLAS document},pdfauthor={The ATLAS Collaboration}}

\usepackage{atlascontribute}

\usepackage{dcolumn}
\newcolumntype{.}{D{.}{.}{-1}}

\usepackage{siunitx} %
\usepackage{morefloats}
\usepackage{stackengine} %

\begin{document}

\maketitle

\section{Introduction}
\label{sec:intro}

The measurement of the production of two isolated prompt photons\footnote{Prompt photons are those not originating from the decay of hadrons.}
in association with jets, produced by Standard Model (SM) processes in proton--proton ($pp$) collisions,
is a useful tool to test perturbative QCD (pQCD) predictions. Cross sections for these processes are in fact difficult to compute using pQCD in particular regions of the phase space, such as back-to-back configurations of the two photons, or the diphoton system recoiling against a single jet, for which resummation techniques are necessary~\cite{Balazs:2006cc}, and good theoretical knowledge of the photon pair production in association with jets must be established. In particular, fixed-order calculations for this process are available at next-to-leading order (NLO) for multiple jets~\cite{th_gosam,Bern:2014vza}, and this order of the perturbative series might inaccurately predict the overall cross-section normalization and fail to capture the full dynamics of the process, especially in the presence of more than one jet and in particular regions of the phase space. Monte Carlo (MC) generators providing NLO matrix elements matched to parton showers (NLO+PS)~\cite{sherpa} can better describe the overall cross-section normalization and the process kinematics simultaneously. This paper aims therefore to provide comparisons of the $\gamma\gamma+\text{jet(s)}$ final state object kinematic and angular properties to various theoretical predictions, to provide insight on and lead toward the improvement of MC simulations of these processes.

In this paper, the measurement of prompt diphoton production in association with jets in $pp$ collisions at $\sqrt{s}$ = 8 \TeV\ by the ATLAS experiment using an integrated luminosity of 20.3~\ifb\ is presented. Diphoton events are categorized according to the number of jets ($\Nj$) into exclusive and inclusive \Nj\ categories and fiducial cross sections are measured for each jet category.  Fiducial cross sections are also measured differentially in up to sixteen observables, depending on the number of jets present in an event.

Diphoton production at the LHC is dominated by three production processes, as illustrated in Figure~\ref{fig:ggjj_diagrams}: the Born process ($q\bar{q}\rightarrow\gamma\gamma$), which is LO in the strong coupling constant $\alpha_\text{S}$, the NLO bremsstrahlung process ($qg\rightarrow q\gamma\gamma$) and the box diagram ($gg\rightarrow\gamma\gamma$) that appears at NNLO.  Additional jets in the event can be produced by emitting a quark or a gluon either in the initial or final state. The full NNLO QCD corrections to inclusive diphoton production have been computed in Ref.~\cite{Catani:2018krb}; for diphoton production in association with a single jet, NNLO calculations have more recently become available~\cite{Chawdhry:2021hkp}, while NLO remains the highest fixed-order accuracy for multiple jets~\cite{th_gosam,Bern:2014vza}.

\begin{figure}[b]
\centering
\subfloat[]{\includegraphics[width= 0.32\textwidth]{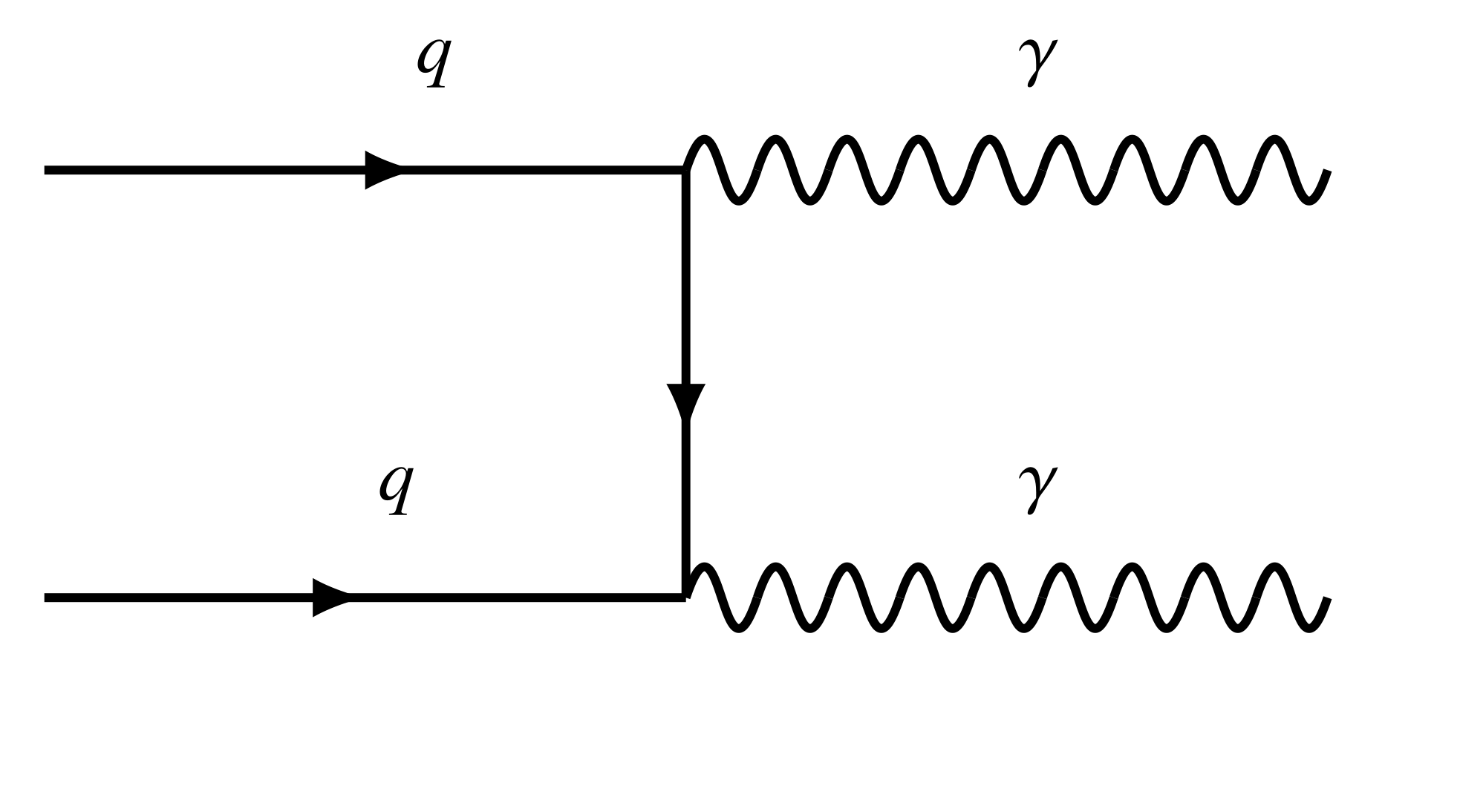}\label{fig:ggjj_born}}
\subfloat[]{\includegraphics[width= 0.32\textwidth]{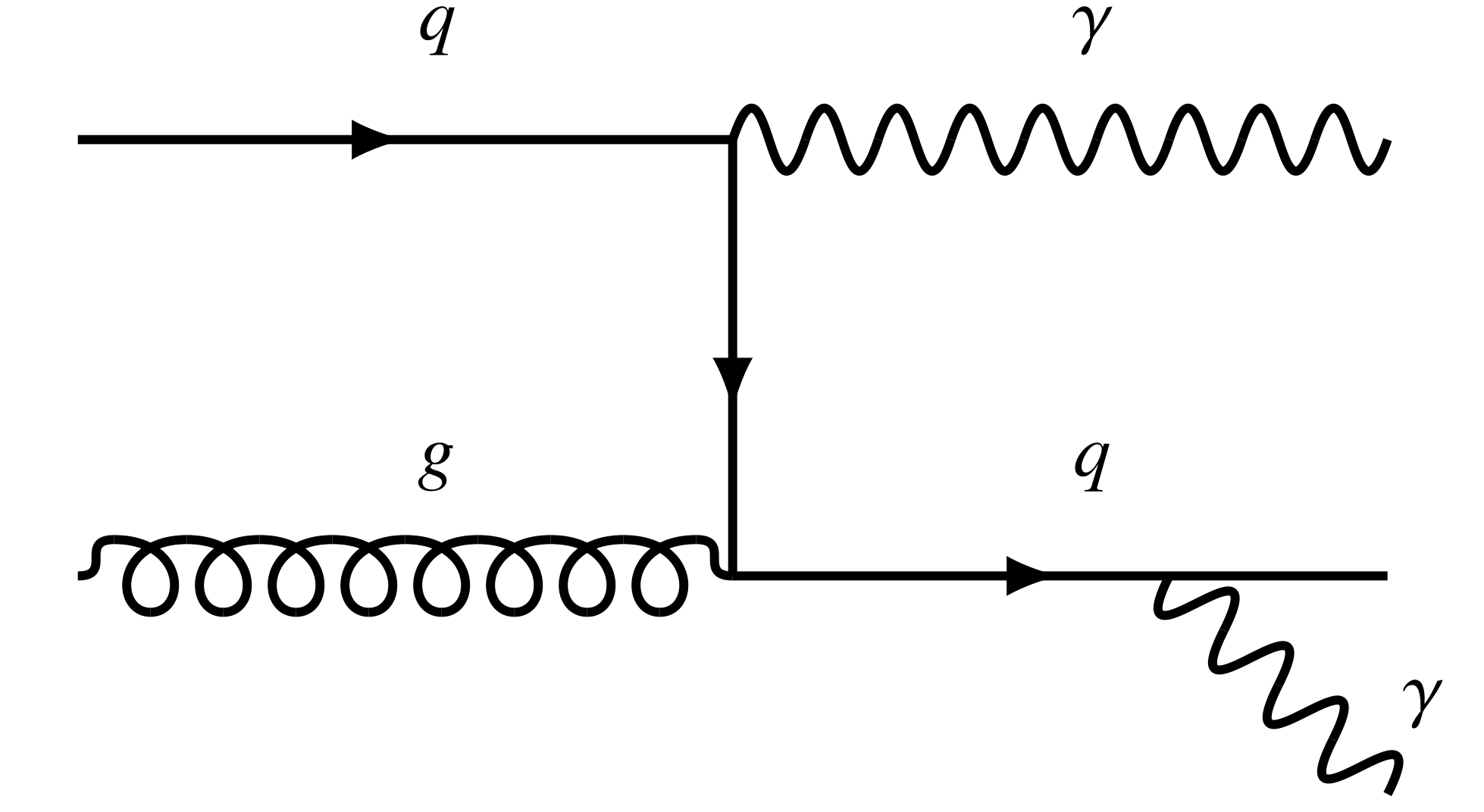}\label{fig:ggjj_brem}}
\subfloat[]{\includegraphics[width= 0.32\textwidth]{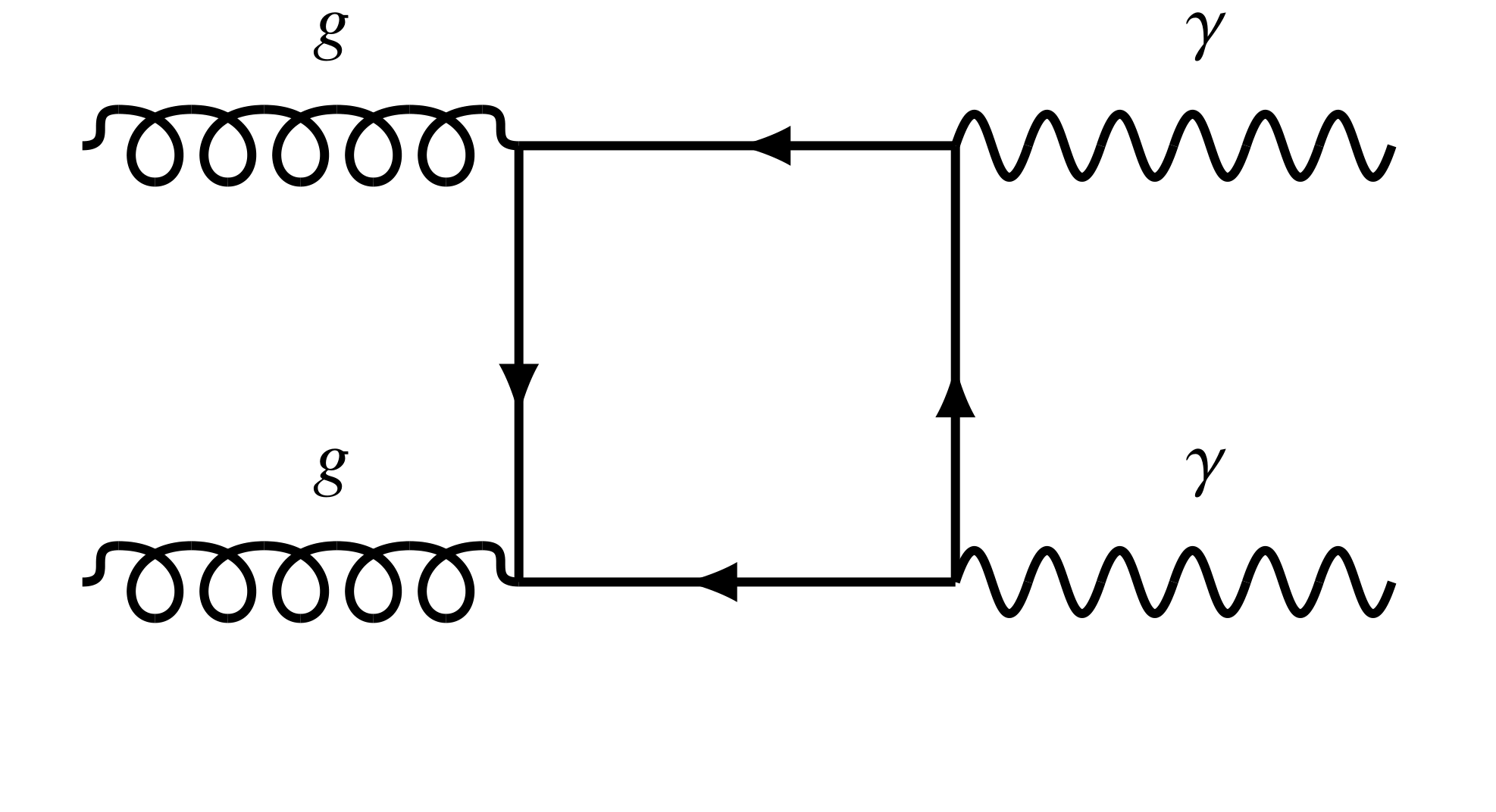}\label{fig:ggjj_box}}
\caption{Diphoton production at the LHC: (a) Born diagram (LO) (b) bremsstrahlung diagram (NLO) and (c) box diagram (NNLO).}
\label{fig:ggjj_diagrams}
\end{figure}

The SM production cross section of two isolated photons has been measured at the Tevatron by the D\O~and CDF experiments~\cite{Abazov:2010ah, th_gg_d0, th_gg_cdf} and at the LHC by the ATLAS and CMS experiments~\cite{STDM-2011-05, STDM-2012-05, STDM-2017-30, CMS-QCD-10-035, CMS-SMP-13-001}. More recently, ATLAS has also observed electroweak diphoton production in association with two jets~\cite{STDM-2022-16}. The measurement presented in this paper applies less restrictive kinematic selection criteria for the two photons compared with the published ATLAS measurement of inclusive diphoton cross section~\cite{STDM-2015-15}. This allows for a larger phase space, providing a larger data sample for the measurement that is split into jet categories.
The measured cross sections are compared with fixed-order NLO predictions from the \gosam\ and \blackhat\ frameworks, to LO predictions from {\textsc{Sherpa~1.4.1}} normalized to data, and to {\textsc{Sherpa~2.2.14}}, which provides NLO-matched matrix elements for diphoton production in association with up to three jets and represents the most advanced theoretical benchmark included in this study.

This paper is organized as follows. Section~\ref{sec:detector} gives a brief overview of the ATLAS detector. Section~\ref{sec:samples} introduces data and MC samples used in this analysis, while Section~\ref{sec:selection} describes the event and object selection. The extraction of the diphoton yield is described in Section~\ref{sec:background}. Section~\ref{sec:unfolding} describes the unfolding procedure, while the systematic uncertainties affecting the measured yields and cross sections are discussed in Section~\ref{sec:systematics}. In Section~\ref{sec:theory}, various theoretical predictions of the diphoton+jets process are discussed. The measured integrated and differential fiducial cross sections are discussed and compared with the corresponding theoretical predictions in Section~\ref{sec:results}. Section~\ref{sec:conclusion} summarizes the measurement.


\section{ATLAS detector}
\label{sec:detector}

The ATLAS detector~\cite{PERF-2007-01} is a multipurpose detector with a forward--backward symmetric
cylindrical geometry\footnote{ATLAS uses a right-handed coordinate system with its origin at the nominal interaction point in the center of the detector.
The positive $x$-axis is defined by the direction from the interaction point to the center of the LHC ring, with the positive $y$-axis
pointing upwards, while the beam direction defines the $z$-axis. Cylindrical coordinates (R, $\phi$) are used in the transverse plane, $\phi$
being the azimuthal angle around the $z$-axis. The pseudorapidity $\eta$ is defined in terms of the polar angle $\theta$ by $\eta = -\ln \tan(\theta/2)$
and the rapidity is defined as $y = (1/2) \ln((E+p_{z}) / (E-p_{z}))$ where $E$ is the energy and $p_{z}$ the longitudinal momentum of the
object of interest. The transverse momentum \pT\ and the transverse energy \ET\ are defined in the x-y plane.}.
At small radii, the inner detector (ID) is made up of fine-granularity pixel
and microstrip detectors. These silicon-based detectors cover the pseudorapidity range $|\eta|<2.5$.
A gas-filled straw-tube transition radiation tracker (TRT) complements the silicon tracker at
larger radii and also provides electron identification capabilities based on transition radiation.
Around the TRT, a thin superconducting solenoid produces a 2~T magnetic field.

The electromagnetic (EM) calorimeter is a lead/liquid-argon sampling
calorimeter with accordion geometry.  The calorimeter is divided into a barrel section covering
$|\eta|<1.475$ and two endcap sections covering $1.375<|\eta|<3.2$. For $|\eta|<2.5$ it is divided
into three layers in depth, which are finely segmented in $\eta$ and $\phi$. A thin presampler
layer, covering $|\eta|<1.8$, is used to correct for fluctuations in upstream energy losses.
Hadronic calorimetry in the region $|\eta|<1.7$ uses steel absorbers and scintillator tiles as an
active medium.  Liquid-argon calorimetry with copper absorbers is used in the hadronic endcap
calorimeters, which cover the region $1.5<|\eta|<3.2$.  A forward calorimeter using copper or
tungsten absorbers with liquid argon completes the calorimeter coverage up to $|\eta|=4.9$.

Surrounding the calorimeters is the muon spectrometer, which measures the deflection of muon tracks.
The deflection is provided by a toroidal magnetic field of air-core superconducting magnets.

Events are selected using a first-level trigger implemented in custom electronics,
which reduces the event rate to a design value of at most 75~kHz using a subset of detector information~\cite{PERF-2011-02}.
Software algorithms with access to the full detector information are then used in the high-level trigger to yield a
recorded event rate of about 400~Hz.


\section{Data and Monte Carlo samples}
\label{sec:samples}

The analyzed data set consists of $pp$ collision data at a center-of-mass energy $\sqrt{s}= 8~\TeV$, corresponding to an integrated luminosity
of 20.3~$\pm~0.4$~fb$^{-1}$, collected by the ATLAS detector at the LHC in 2012. The luminosity was determined following the methodology detailed in Ref.~\cite{DAPR-2013-01}.

Three sets of MC samples are used to model diphoton events in association with jets.
A {\textsc{Sherpa 1.4.1}} sample~\cite{sherpa}, generated with CT10 parton distribution functions (PDFs)~\cite{Lai:2010vv} is
used as the main MC sample, simulating the production
of two photons with up to three partons from the hard process in the final state at LO.
These processes with different parton multiplicities are combined with each other and with the parton shower using
the ME+PS@LO prescription~\cite{Catani:2001cc} to remove overlaps and effectively simulate all parton emissions.

The {\textsc{Pythia~8.163}} event generator~\cite{pythia8} is used to generate $q\bar{q}\rightarrow\gamma\gamma$ and
$gg\rightarrow\gamma\gamma$ diphoton events at LO.
Photon+jet events produced by $qg\rightarrow q\gamma$ and $\bar{q}g\rightarrow\bar{q}\gamma$ processes at LO are also generated.
Photons originating from hadronization of colored partons are considered at LO using $\gamma+\text{j}$ and dijet diagrams with initial
and final-state radiation. The event-generation uses CTEQ6L1~\cite{Pumplin:2002vw} PDFs at LO and its
parameters are set according to the ATLAS AU2 tune~\cite{ATL-PHYS-PUB-2012-003}.

Both the {\textsc{Sherpa~1.4.1}} and {\textsc{Pythia~8.163}} samples are used for event and object selection studies.
The {\textsc{Sherpa~1.4.1}} sample serves as the primary MC: it is used for background estimation, unfolding of detector effects, and non-perturbative corrections to the fixed-order predictions (see Section~\ref{sec:theory}).
The {\textsc{Pythia~8.163}} sample, which does not include additional jets from the hard matrix element beyond the two photons and therefore has limited applicability for multijet final states, is used exclusively as a cross-check for background estimation and systematic uncertainty evaluation; it is not included in the differential cross section comparisons presented in Section~\ref{sec:results}.

An additional {\textsc{Sherpa 2.2.14}} sample is used for theory comparison purposes, as described in Section~\ref{sec:theory}. It is not used in event and object selection studies, background subtraction, unfolding, or non-perturbative corrections, as this sample was not available at the time of the analysis.

A {\textsc{Powheg+Pythia 8.170}}~\cite{Nason:2004rx,Frixione:2007vw,pythia8} $Z\rightarrow e^{+}e^{-}$ sample at NLO with final state  radiation simulated by {\textsc{Photos}}~\cite{Golonka2006} and CT10 PDFs, is used to estimate the number of electrons misidentified as photons.

Dedicated {\textsc{Sherpa 1.4.1}} samples with two photons and up to two partons were generated with and without underlying event and hadronization mechanisms to compute non-perturbative corrections applied to the fixed-order predictions described in Section~\ref{sec:theory}.
These are referred to as non-perturbative corrections (NPC).

For all MC samples described in this section, generated events are passed through a full ATLAS detector simulation~\cite{SOFT-2010-01} based on {\textsc{Geant}}~4~\cite{AGOSTINELLI2003250}.
Pileup from additional $pp$ collisions in the same and neighboring bunch crossings is simulated by overlaying each MC event with a variable number of simulated
inelastic $pp$ collisions, corresponding to an average of approximately 21 interactions per bunch crossing.
The MC events are weighted to reproduce the observed distribution of the average number of interactions per bunch
crossing and the size of the luminous region along the beam axis.

NLO fixed-order predictions from \gosam\ and \blackhat, which are compared with the data in the 1-jet and 2-jet categories, are described in Section~\ref{sec:theory}.


\section{Event and object selection}
\label{sec:selection}

Events were collected in stable beam conditions and with all relevant ATLAS subsystems fully operational. The data were recorded using a diphoton trigger, which requires two electromagnetic clusters with transverse energies \ET\ above 20 \GeV, both fulfilling identification criteria based on the shape of the electromagnetic  shower in the EM calorimeter. The efficiency of the diphoton trigger is $(98.7 \pm 0.5)\%$ for signal events satisfying the analysis selection, which has been evaluated using methods of Refs.~\cite{PERF-2011-02,ATLAS-CONF-2012-048}. Events with at least one reconstructed primary vertex, which has at least three associated tracks with a transverse momentum above 400~\MeV, are selected to reject non-collision backgrounds. Almost twenty million events satisfy these criteria.

\subsection{Photon reconstruction and selection}
\label{subsec:photonsel}

Photon and electron candidates are reconstructed from clusters of energy deposited in the calorimeter, combined with tracking information from the inner detector. Clusters matched to a conversion vertex, reconstructed from either two tracks consistent with a vertex originating from a photon conversion, or one track that does not have any hits in the innermost pixel layer and has an electron-like response in the TRT, are reconstructed as converted photon candidates. Clusters without any matching track or conversion vertex are reconstructed as unconverted photon candidates, while clusters with a matching track are reconstructed as electrons~\cite{PERF-2013-04}. Depending on the detector region, 20--50\% of all photons convert into an electron--positron pair in the inner detector material~\cite{PERF-2007-01}.

The photon energy is calibrated using a simulation-based regression, the inputs of which from data are corrected to account for differences between data and MC using information obtained from \Zee, \Zmm\ and inclusive photon events. An overall energy scale calibration is derived from \Zee\ events~\cite{PERF-2013-05}.

The transverse energy of a photon candidate, $E^{\gamma}_{\text{T}}$, is computed using the pseudorapidity of the EM cluster measured in the second layer of the EM calorimeter, which provides the best photon direction measurement. Photon candidates are first required to satisfy $\ET>20~\GeV$ and have $|\etag|<2.37$, excluding the barrel--endcap transition region of $1.37<|\etag|<1.56$.  Photon candidates also have to satisfy \textit{loose} selection criteria for the shape of the shower in the EM calorimeter~\cite{PERF-2013-04}.  The remaining candidates are then ordered in \ET\ and the primary vertex of the hard interaction is determined using the photon pointing method~\cite{HIGG-2013-08}. The pseudorapidity and the transverse energy of the photons is recomputed relative to the primary vertex of the hard interaction, referred to as the diphoton vertex in the following.

Both photon candidates must then have $\ET>22$~\GeV\ and are required to be isolated and satisfy \textit{tight} identification criteria using nine shower-shape variables related to the energy leakage into the hadronic calorimeter and the lateral and longitudinal shower development in the EM calorimeter~\cite{STDM-2010-08,PERF-2013-04}. Non-isolated and non-\textit{tight} photons, satisfying all other selections, are used for the background subtraction, as described in Section~\ref{sec:background}. Photons are considered to be isolated if the calorimeter isolation transverse energy satisfies \ETiso\ $<4$~\GeV. The \ETiso\ is defined as the sum of transverse energies of the positive-energy topological clusters~\cite{PERF-2014-07,STDM-2010-08} within a cone of $\Delta R = 0.4$ around the photon candidate, excluding the center of the shower and correcting for leakage of the photon shower into the isolation cone. The contributions of the underlying event and pileup are subtracted based on the jet area/median method described in Ref.~\cite{Cacciari:2009dp}.
For each photon candidate a track isolation variable $\pTiso$ is defined as the scalar sum of the transverse momenta of tracks with $\pT>1$~\GeV\ within a cone of size $\Delta R= 0.2$ around the photon candidate. Tracks associated with a photon conversion and those not originating from the diphoton vertex are excluded from the sum. This cone size was optimized to correspond closely to the calorimeter isolation selection of $\ETiso < 4$~\GeV. The track isolation $\pTiso$ is not directly used as a selection, but to define control regions to measure backgrounds (see Section~\ref{subsec:jetbackground}).

The two photon candidates are required to be separated in the $\eta-\phi$ plane by $\dRgg > 0.4$ to avoid that they lie within their respective isolation cones, which would affect the isolation selection\footnote{Angular separation between particles or reconstructed objects are measured in the ($\eta-\phi$) plane using $\Delta R = \sqrt{(\Delta\phi)^{2}+(\Delta\eta)^{2}}$.}. The invariant mass of the two photon candidates \mgg\ is required to be $\mgg>60~\GeV$.

After applying these event and photon selection criteria, almost eight hundred thousand events with two \textit{tight} and isolated photon candidates are retained for further analysis.

\subsection{Jet reconstruction and selection}
\label{subsec:jetsel}

Jets are reconstructed using the anti-$k_t$~algorithm~\cite{antikt} with radius parameter of $R=0.4$. Jets are then calibrated at the electromagnetic energy scale. The energy deposited in the calorimeters by the EM and hadronic showers is measured and corrected relative to the response of the calorimeters to the true energy corresponding to a jet of stable particles entering the detector in MC simulation. Several corrections related to pileup, the jet origin, jet energy and direction, and an in situ correction accounting for differences between data and MC, are performed~\cite{PERF-2011-03}.

Calibrated jets are further required to have a rapidity of $|\yj|<4.4$. Jets within $\dRgj < 0.4$ of either photon are not considered, to avoid photon--jet double-counting. The transverse momentum of jets is required to be at least $\pTj>25~\GeV$ for $|\etaj|<2.4$ and $\pTj>50~\GeV$ for $|\etaj|>2.4$. These thresholds are applied as a function of pseudorapidity $|\etaj|$ rather than rapidity, since pileup suppression relies on charged-particle tracks available within the inner detector acceptance ($|\eta|<2.4$). The higher $\pT$ selection for jets with $|\etaj|>2.4$ is to suppress jets originating from pileup interactions, whereas for jets with $\pTj<50~\GeV\ $ and $|\etaj|<2.4$, a Jet Vertex Fraction (JVF) requirement is imposed, JVF~$>0.5$. The JVF is defined as the sum of the \pT\ of tracks associated with the jet that are produced at the diphoton vertex, divided by the sum of the \pT\ of the tracks associated with the jet from all collision vertices~\cite{PERF-2014-03}. The jet \pT\ and JVF selection were optimized to reach an at least 90\% purity of jets originating from a quark or a gluon from the hard process using \sherpa\ MC.  The residual contribution of pileup jets satisfying the selection is corrected for by the unfolding procedure.

Events are split into categories according to the number of jets satisfying the requirements listed above. Additionally, events in which any selected jet is within $\dRgj<0.6$ of either photon are removed, as such jets would contaminate the photon isolation cone.

For simulated events, the reconstruction-level selection follows the same criteria detailed above. In addition, reconstructed photons are required to be matched to photons originating in the hard process, ensuring that the MC efficiency is computed for signal photons only; in data, the contributions from fake photons and misidentified electrons are estimated and subtracted using the data-driven methods described in Section~\ref{sec:background}.

\subsection{Fiducial phase space definition}
\label{subsec:fid_vol}

The fiducial phase space in which the cross sections are reported is defined by the kinematic selection of the photons and jets in analogy to the selection described above. Events are required to have at least two photons from a hard interaction. The two highest-$\ET$ photons are required to have $\ETg>22$~\GeV\ and $|\etag|<1.37$ or $1.56<|\etag|<2.37$ and have to be separated by $\dRgg > 0.4$ in the $\eta-\phi$~plane.  The isolation criterion is implemented as a particle-level isolation criterion. The sum of transverse momenta of all generated particles in a cone of $\dR = 0.4$ around a photon, except muons, neutrinos and particles from interactions produced by pileup, is required to be below 9~\GeV.  This corresponds to the calorimeter isolation selection of $\ETiso < 4$~\GeV\ at reconstruction-level.

Jets at particle-level are built from all particles, including muons and neutrinos, with a lifetime greater than 10~ps using the anti-$k_t$ algorithm with radius parameter $R=0.4$. Jets are required to have the rapidity of $|\yj|<4.4$ and have to fulfill $\pTj>25~\GeV$ if $|\etaj|<2.4$ and $\pTj>50~\GeV$ otherwise. Jets are not considered if they are separated from any of the two photons by $\dRgj < 0.4$. Events, where a jet satisfying the $\pTj$ and $\yj$ requirements is within $\dRgj<0.6$ of either of the two selected photons, are removed.

\subsection{Measurement strategy}
\label{subsec:vars}

Fiducial cross sections are measured integrated per jet category and also differentially in up to sixteen observables. They are measured as a function of diphoton kinematic and angular properties: the diphoton invariant mass \mgg, the transverse momentum of the diphoton system \pTgg, the opening angle between the photons in the azimuthal plane \dphigg\ and \costhgg\ defined in the Collins--Soper frame~\cite{PhysRevD.16.2219}.  For events with jets, the cross section is also measured differentially in \pT\ of up to the first three highest-\pT\ jets in the event, and the separation of photons and jets in the ($\eta-\phi$) plane, $\Delta R$. Dijet kinematic variables, namely the dijet invariant mass \mjj, the transverse momentum of the dijet system \pTjj, the opening angle between the two highest-\pT\ jets in the azimuthal plane \dphijj\ and their difference in rapidity \dyjj\ are measured for events with at least two jets.  The azimuthal angle between the diphoton and dijet systems, $\Delta S_{\gamma\gamma,\text{jj}} \equiv |\Delta\phi_{\gamma\gamma,\text{jj}}| \in [0,\pi]$, is also measured. This observable can be sensitive to the contribution of multiple parton interactions. Table~\ref{tab:vars} lists all observables that are measured for each \Nj\ category.

\begin{table}
\caption{Observables measured in this paper. See text for definitions.}
\label{tab:vars}
\centering
\begin{tabular}{lcccc}
\toprule
Observable / \Nj                                                & 0 jets     & 1 jet      & 2 jets     & $\geq 3$ jets \\ \midrule
\mgg                                                            & \checkmark & \checkmark & \checkmark & \checkmark    \\
\pTgg                                                           & \checkmark & \checkmark & \checkmark & \checkmark    \\
\costhgg                                                        & \checkmark & \checkmark & \checkmark & \checkmark    \\
\dphigg                                                         & \checkmark & \checkmark & \checkmark & \checkmark    \\ \midrule
\mjj                                                            &            &            & \checkmark & \checkmark    \\
\pTjj                                                           &            &            & \checkmark & \checkmark    \\
\dphijj                                                         &            &            & \checkmark & \checkmark    \\
\dyjj                                                           &            &            & \checkmark & \checkmark    \\
\dRgonejone                                                     &            & \checkmark & \checkmark & \checkmark    \\
\dRgtwojone                                                     &            & \checkmark & \checkmark & \checkmark    \\
\dRgonejtwo                                                     &            &            & \checkmark & \checkmark    \\
\dRgtwojtwo                                                     &            &            & \checkmark & \checkmark    \\
\dSggjj                                                         &            &            & \checkmark & \checkmark    \\
\pTjone                                                         &            & \checkmark & \checkmark & \checkmark    \\
\pTjtwo                                                         &            &            & \checkmark & \checkmark    \\
\pTjthree                                                       &            &            &            & \checkmark    \\
\bottomrule
\end{tabular}
\end{table}


\section{Background decomposition and signal yield extraction}
\label{sec:background}

The sample of events satisfying the event and object selection still contains a sizable fraction of background. The main background consists of neutral mesons inside jets, where the mesons carry a high fraction of the jet momentum and subsequently decay into two colinear photons. These can be mistaken for a single prompt photon. The selected diphoton candidate spectrum, according to previous measurements~\cite{STDM-2011-05,STDM-2012-05,STDM-2015-15}, contains around 60\% prompt diphotons at low \mgg\ (below 100~\GeV) and more than 95\%\ at high \mgg\ (above 500~\GeV).

A smaller reducible background due to electrons is also present in the selected sample. It amounts to approximately 2--4\% of the selected sample, and is spread evenly in all measured distributions, except for the diphoton invariant mass distribution, where a localized peak is visible around the $Z$-boson mass ($\mgg \approx 90$~\GeV), since most of the electron background originates from $Z\rightarrow ee$ events. The peak in the \mgg\ spectrum helps to identify and quantify the electron background and its disappearance after the subtraction validates the approach.

The contribution of the Higgs boson decay $H\to\gamma\gamma$ is estimated to be negligible in all regions of the phase space, except for the diphoton invariant mass around 125~\GeV, where the contribution is predicted to be around 2\% of the diphoton yield using simulated events. Since photons from the Higgs boson are genuine prompt photons, this contribution is not subtracted.

\subsection{Fake-photon background}
\label{subsec:jetbackground}

A well-established 4$\times$4 matrix method, previously used in ATLAS~\cite{STDM-2011-05,STDM-2015-15}, is used to decompose the photon candidate spectrum into four components: $\gamma\gamma$, $\gamma \mathrm{j}$, $\mathrm{j}\gamma$ and $\mathrm{jj}$. The labels $\gamma\mathrm{j}$ and $\mathrm{j}\gamma$ distinguish the two configurations in which the subleading or leading photon candidate, respectively, is the fake. The matrix method uses background control regions populated with events with photon candidates that fail the isolation requirement, the \textit{tight} identification criteria, or both.

Yields of the four final states, $\gamma\gamma$, $\gamma$j, j$\gamma$, jj, are extracted from a sample of reconstructed photon pairs satisfying the full selection except the isolation criteria, referred to as the \T\T\ sample (\T\ for \textit{tight}) in the following. Events are further classified into four categories based on the photon pair isolation status: \I\I, \I\nI, \nI\I\ and \nI\nI, where \I\ indicates a candidate in the pair passing the isolation criterion, and \nI\ a candidate failing it. The event counts $N^{\T\T}_{\I\I},~N^{\T\T}_{\I\nI},~N^{\T\T}_{\nI\I},~N^{\T\T}_{\nI\nI}$ in these four categories are related to the yields $W^{\T\T}_{\gamgam},~W^{\T\T}_{\gamjet},~W^{\T\T}_{\jetgam},~W^{\T\T}_{\jetjet}$ by a matrix equation:
\begin{equation}
\renewcommand{\arraystretch}{1.3}
\left( \begin{array}{c} N^{\T\T}_{\I\I} \\ N^{\T\T}_{\I\nI} \\ N^{\T\T}_{\nI\I} \\ N^{\T\T}_{\nI\nI} \end{array} \right) =
\left(
\begin{array}{cccc}
\epsilon_1\epsilon_2 & \epsilon_1f_2^{\gamjet} & f_1^{\jetgam}\epsilon_2 & \xi f_1^{\jetjet}f_2^{\jetjet} \\
\epsilon_1(1-\epsilon_2) & \epsilon_1(1-f_2^{\gamjet}) & f_1^{\jetgam}(1-\epsilon_2) & f_1^{\jetjet}(1-\xi f_2^{\jetjet}) \\
(1-\epsilon_1)\epsilon_2 & (1-\epsilon_1)f_2^{\gamjet} & (1-f_1^{\jetgam})\epsilon_2 & (1-\xi f_1^{\jetjet})f_2^{\jetjet} \\
(1-\epsilon_1)(1-\epsilon_2) & (1-\epsilon_1)(1-f_2^{\gamjet}) & (1-f_1^{\jetgam})(1-\epsilon_2) & (1-f_1^{\jetjet}-f_2^{\jetjet}+\xi f_1^{\jetjet}f_2^{\jetjet})
\end{array}
\right)
\left( \begin{array}{c} W^{\T\T}_{\gamgam} \\ W^{\T\T}_{\gamjet} \\ W^{\T\T}_{\jetgam} \\ W^{\T\T}_{\jetjet} \end{array} \right)
\renewcommand{\arraystretch}{1}
\notag
\end{equation}
where the indices 1,2 refer to the leading/subleading photon candidate, and the matrix coefficients are:
$\epsilon_{1,2}$, the isolation efficiencies for prompt photons, assumed equal in diphoton and photon+jet events;
$f_1^{\jetgam},~f_2^{\gamjet}$, the photon--jet isolation fake rates for the leading/subleading candidate;
$f_{1,2}^{\jetjet}$, the dijet isolation fake rates.
Fake rates are defined separately for photon--jet and dijet events due to the different dominant jet origin (quark-initiated vs.\ gluon-initiated). In dijet events the isolation of the two fake photons is positively correlated because of the color connection between the outgoing partons; this is accounted for by the correction factor $\xi$.

The matrix method is applied in bins of each measured observable. The various coefficients are obtained as follows: $\epsilon$ is extracted purely from data; $f$ is measured with a tag-and-probe method using data with \sherpa\ MC input; $\xi$ is derived from MC simulation. Details are given below.

The isolation efficiencies $\epsilon$ are extracted from the \T\T\I\I\ signal region and control regions populated by candidates failing the isolation and \textit{tight} identification criteria. An assumption is made that isolation and \textit{tight} identification criteria are uncorrelated for fake photons.
A non-\textit{tight} photon ($\nT$)
is required to satisfy modified \textit{tight} criteria, where four out of the nine shower shape criteria defining the \textit{tight} selection are removed.
Track isolation $\pTiso$ is used to define an \emph{anti-isolated}
sample, which differs from the non-isolated \nI\ sample by having a negligible number of prompt photons and a high number of fakes. Candidates in the anti-isolated region are required to have track isolation $\pTiso>10$~\GeV, which removes prompt photons, and calorimeter isolation $\ETiso>3$~\GeV, removing fake photons that are misidentified electrons and leaving a pure jet sample.
The definition of the
anti-isolated
sample is optimized on MC and pseudo-data studies, and ensures no overlap with the signal region.
The computation of the isolation efficiencies is therefore purely data-driven.

The fake rates $f$ for $\gamma$j and j$\gamma$, the fake rates for dijet events, and the corresponding correlation correction factor $\xi$ are computed using a tag and probe method with input from data and \sherpa\ MC.

The yields extracted by the matrix method are affected by statistical fluctuations in the signal and control regions. The total statistical uncertainty in the measured yield is estimated using a bootstrap technique~\cite{bootstrap}.

\subsection{Electron background subtraction}
\label{subsec:electronbackground}

After the subtraction of the fake-photon background, a reducible background of electrons is still present. This component consists of final states with a prompt photon and at least one electron, such as $\gamma W^{\pm}\rightarrow \gamma e^{\pm}\nu$, or $\gamma Z\rightarrow \gamma e^{+}e^{-}$; and Drell--Yan electron pairs $Z/\gamma^{*}\rightarrow e^{+}e^{-}$ or $W^{+}W^{-}\rightarrow e^{+}e^{-}\nu\bar{\nu}$ where electrons are misidentified as photons.

The approach taken to subtract the electron background in this analysis is similar to that in Ref.~\cite{STDM-2011-05}. The electron to photon fake rate $f_{e\to\gamma}$ is measured in data as a single value, integrated over the photon pseudorapidity and transverse momentum, by counting $ee$ and $\gamma e$ events with invariant mass close to the $Z$-boson mass. In these control samples, electron candidates are required to satisfy the same kinematic requirements as photons and to satisfy the \textit{medium++} identification criterion~\cite{PERF-2016-01}; electrons within $\Delta R < 0.4$ of an identified photon are excluded.
The dependency of $f_{e\to\gamma}$ on the photons' kinematics is reflected in its effect on events with different number of jets, through the different spectrum of \pTgg. The value of $f_{e\to\gamma}$ is therefore computed both from the inclusive data set, which is dominated by events with $N_\mathrm{jet}=0$, and for a subset with $\pTgg>50~\GeV$, which is characteristic of events with $N_\mathrm{jet}\geq1$.
The electron background is then estimated by applying $f_{e\to\gamma}$ according to the jet multiplicity to the control samples containing one or two identified electrons.

The electron background depends strongly on \mgg, but it is much more evenly distributed for other observables. Therefore, a set of weights is derived based on narrow bins in \mgg, which are applied to the data before the fake-photon background subtraction is performed using the 4x4 matrix method. For this purpose, it is necessary to evaluate the electron contamination also for events not satisfying the isolation requirements. This approach is found to introduce a relative bias of below 1\% on the size of the electron background yield, which is already very small (2--4\%) and therefore it is neglected.

\subsection{Diphoton yield extraction}
\label{subsec:ggjjyield}

Results of the background subtraction are displayed in Figure~\ref{fig:ggjj_background}. The reconstructed \gamgam\ yield as a function
of \mgg\ features a peak around the $Z$-boson mass, which is due to $Z\to\ee$ events where both $e^\pm$ are wrongly
identified as photons. The \gamgam\ purity varies between the jet categories from 65\% for events with just two photons up to 74\% for
events with two photons and at least three jets. This can be explained by the fact that the $pp\rightarrow 5\text{j}$ cross section is smaller than 4j, 3j and so on.
Therefore, if there are already three jets in the event, the two photons are less likely to be misidentified jets than in the case of just two jets or one jet present in the event.

\begin{figure}
\centering
\subfloat[]{\includegraphics[width= 0.497\textwidth]{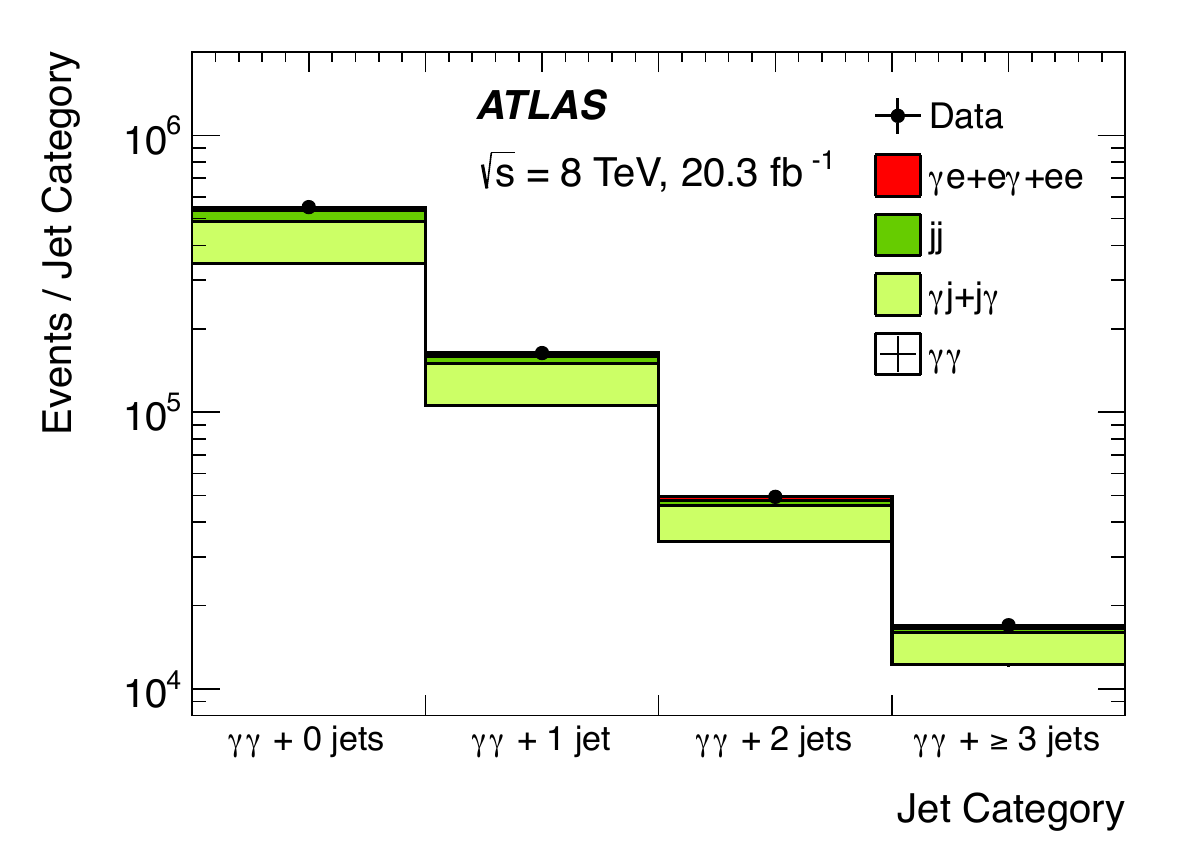}\label{fig:ggjj_back_njet}}
\subfloat[]{\includegraphics[width= 0.497\textwidth]{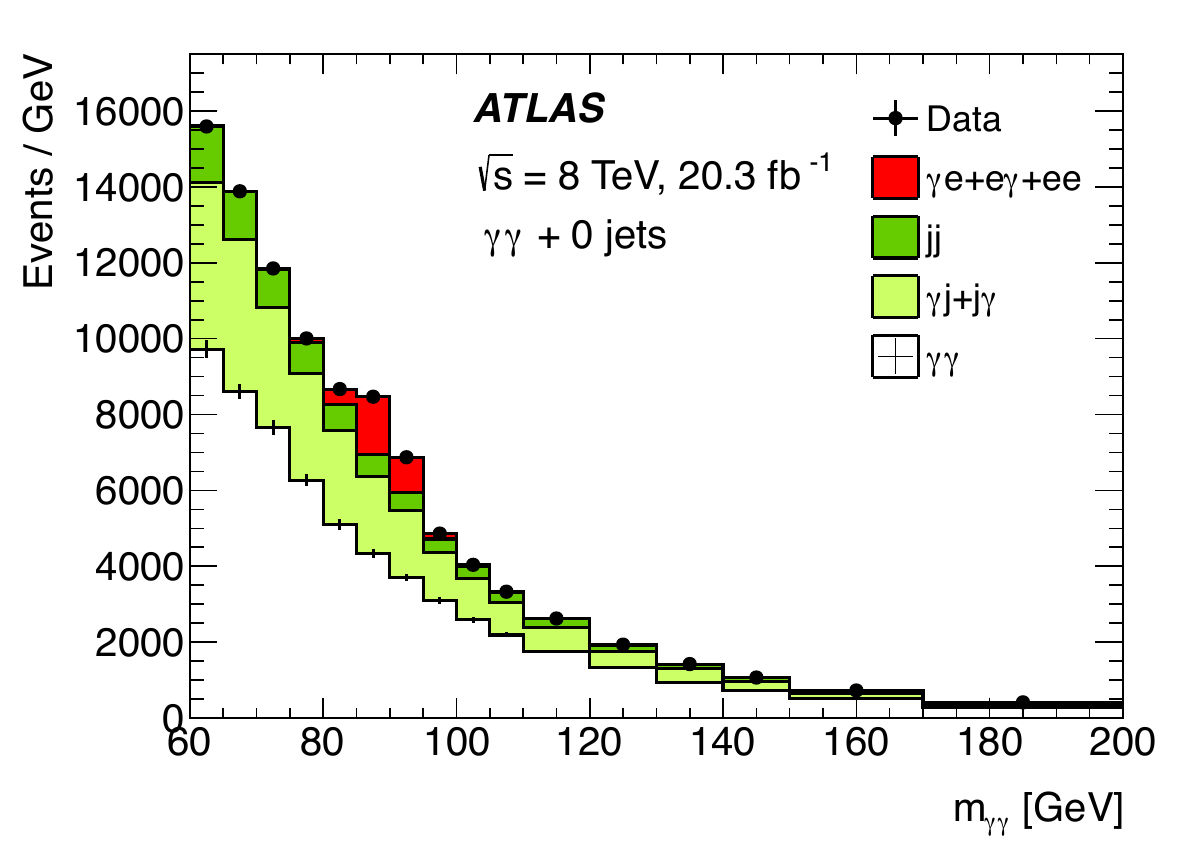}\label{fig:ggjj_back_mgg}} \\
\subfloat[]{\includegraphics[width= 0.497\textwidth]{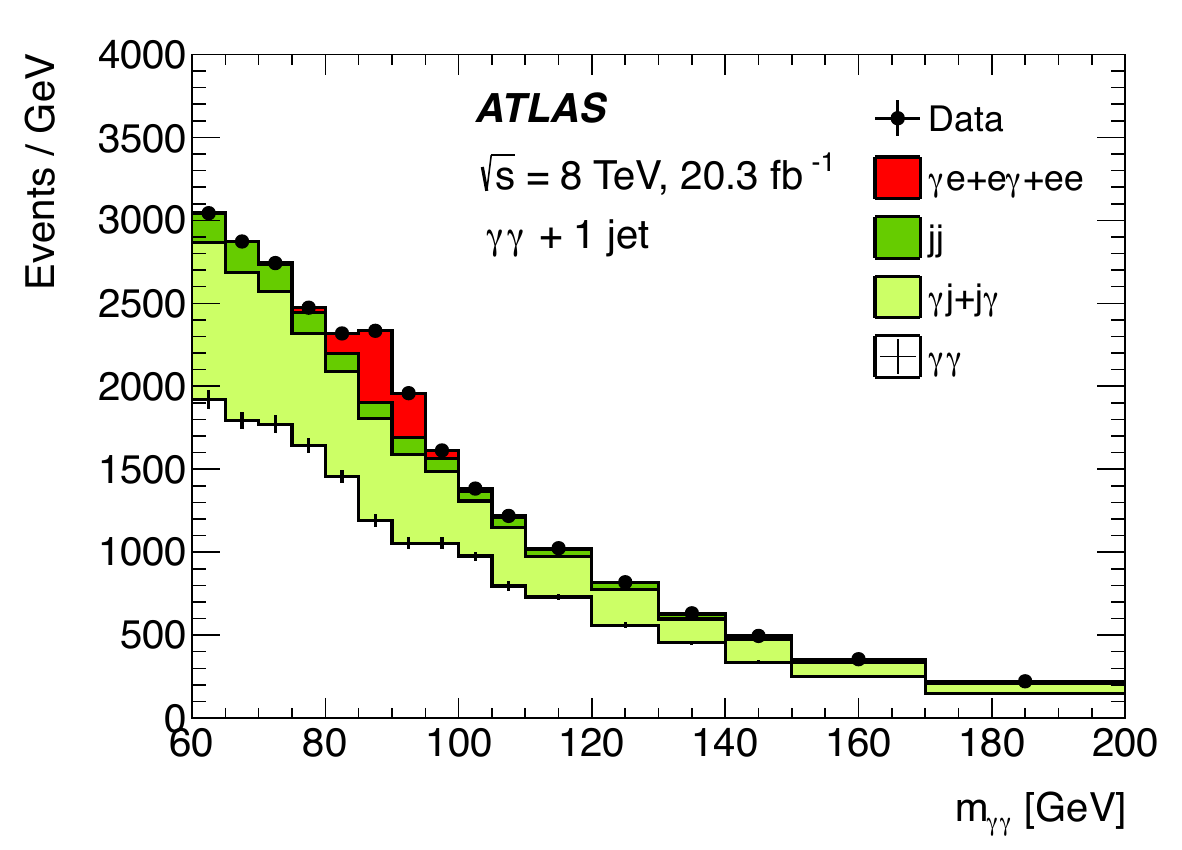}\label{fig:ggjj_back_mgg1}}
\subfloat[]{\includegraphics[width= 0.497\textwidth]{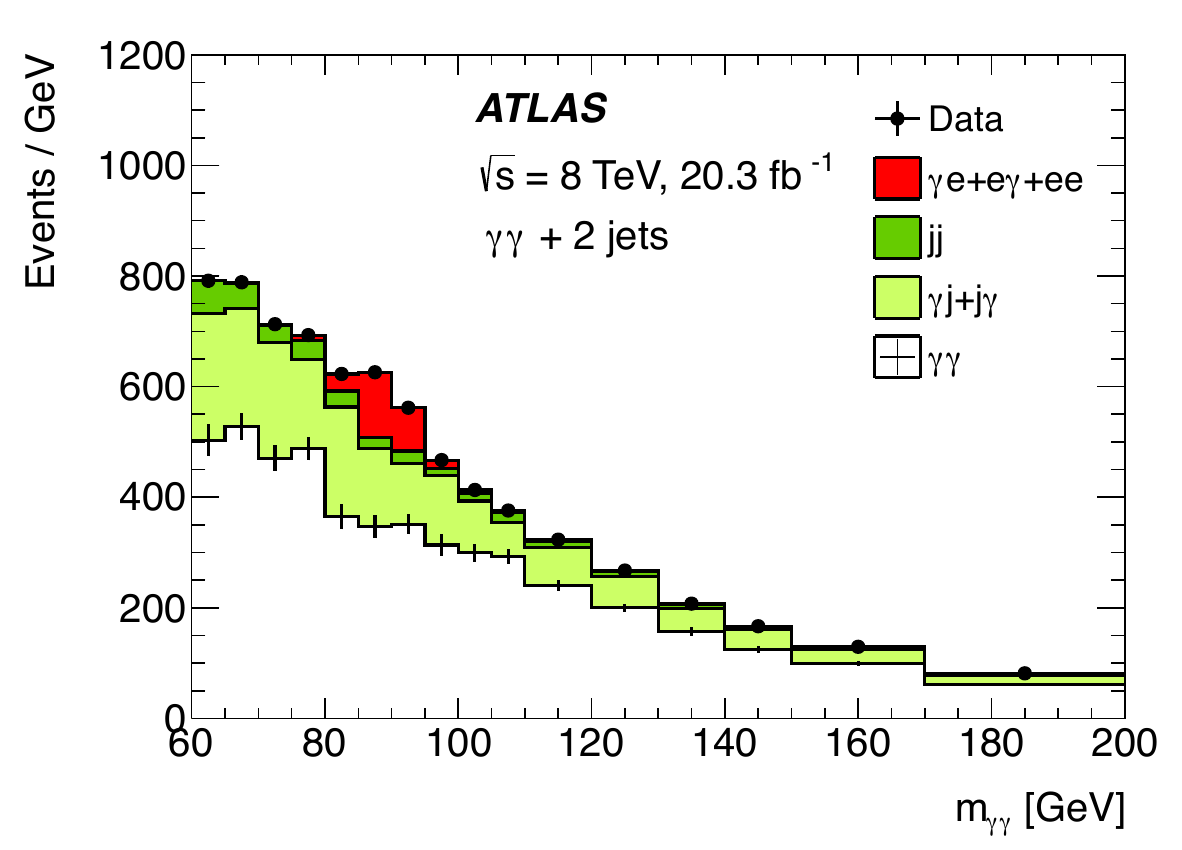}\label{fig:ggjj_back_mgg2}}
\caption{Background decomposition as a function of (a) the number of jets and as a function of the diphoton invariant mass for (b) $\Nj = 0$, (c) 1 and (d) 2, with emphasis on low-\mgg\ region. Uncertainties in data points are statistical only and too small to be visible.
Uncertainties in the diphoton distribution after background subtraction obtained from the 4x4 matrix method are statistical, using a bootstrap method.
Uncertainties for other components are not shown.}
\label{fig:ggjj_background}
\end{figure}

Figure~\ref{fig:ggjj_reco} shows the diphoton yields after background subtraction for selected observables in the $\Nj=1$ and $\Nj=2$ categories compared with Sherpa MC normalized to data, which are used
in the unfolding procedure described in the next section.

\begin{figure}
\centering
\subfloat[]{\includegraphics[width= 0.497\textwidth]{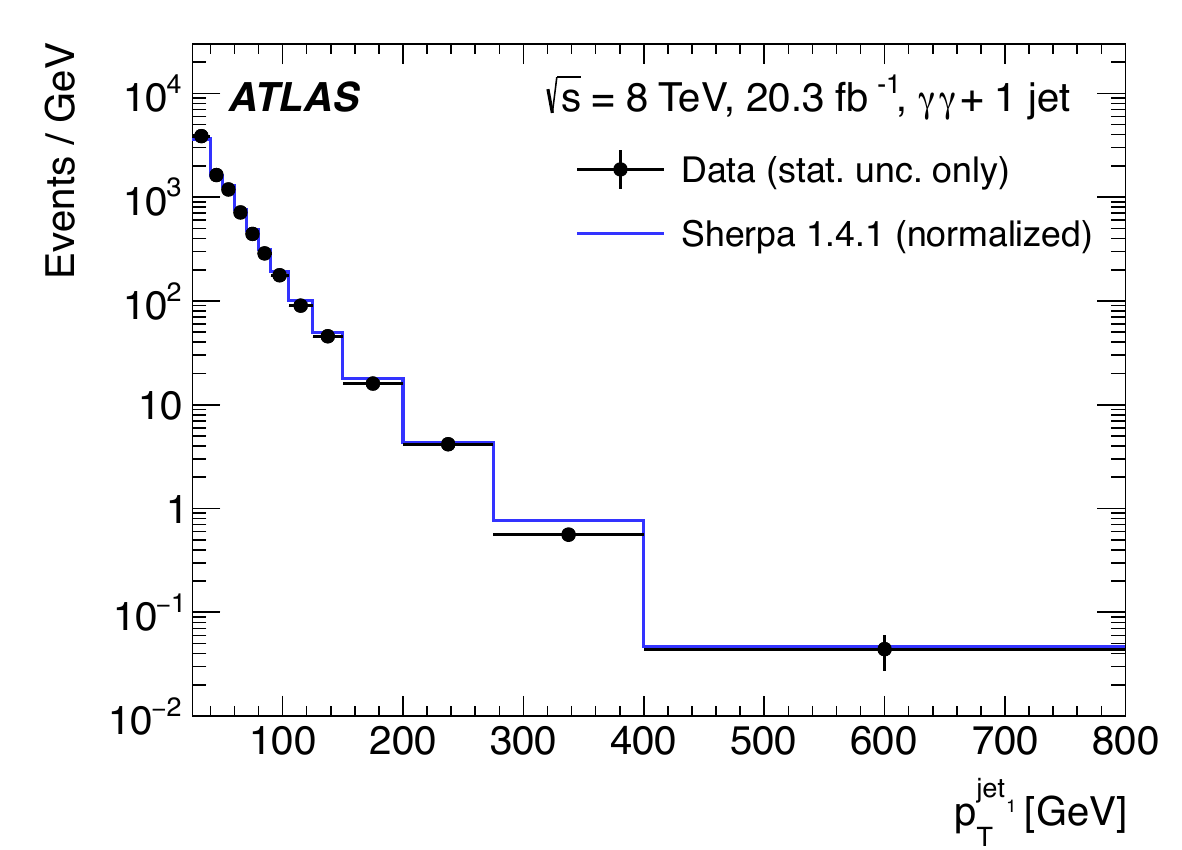}\label{fig:ggjj_reco2}}
\subfloat[]{\includegraphics[width= 0.497\textwidth]{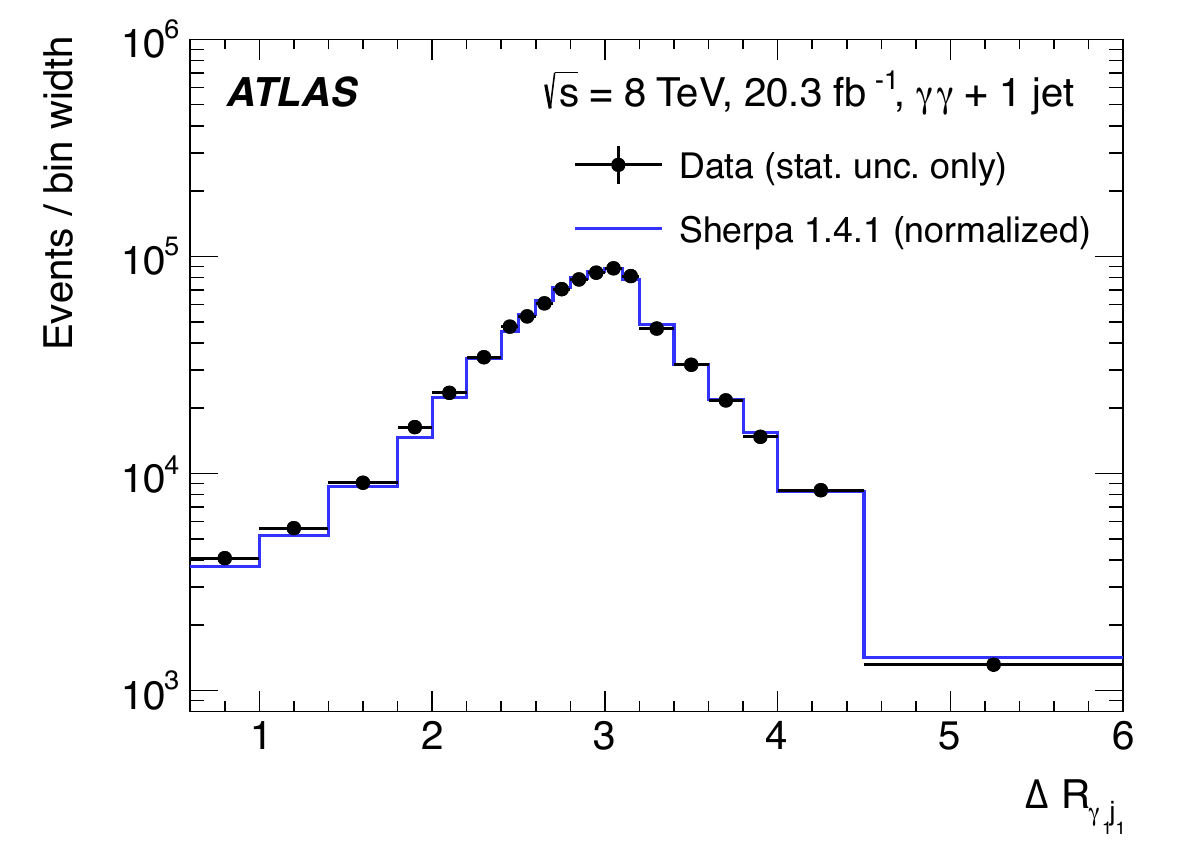}\label{fig:ggjj_reco3}} \\
\subfloat[]{\includegraphics[width= 0.497\textwidth]{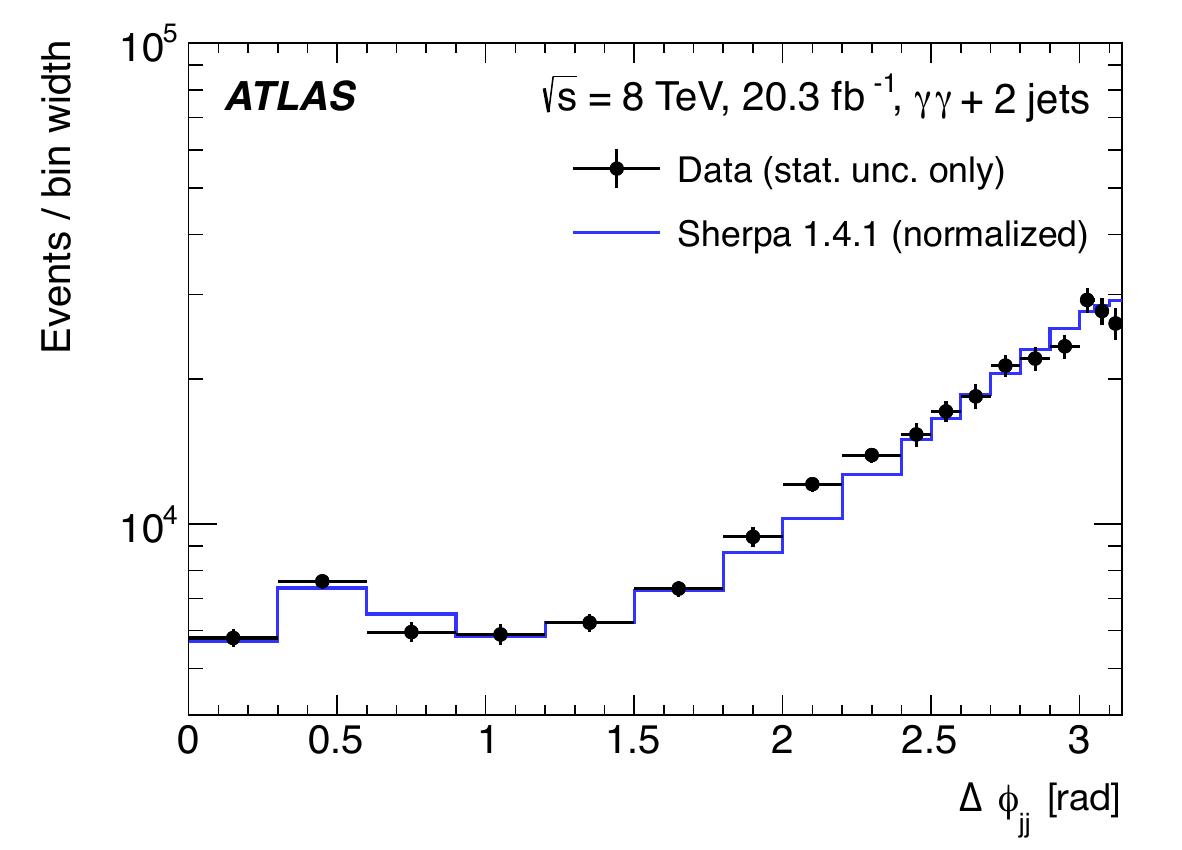}\label{fig:ggjj_reco1}}
\subfloat[]{\includegraphics[width= 0.497\textwidth]{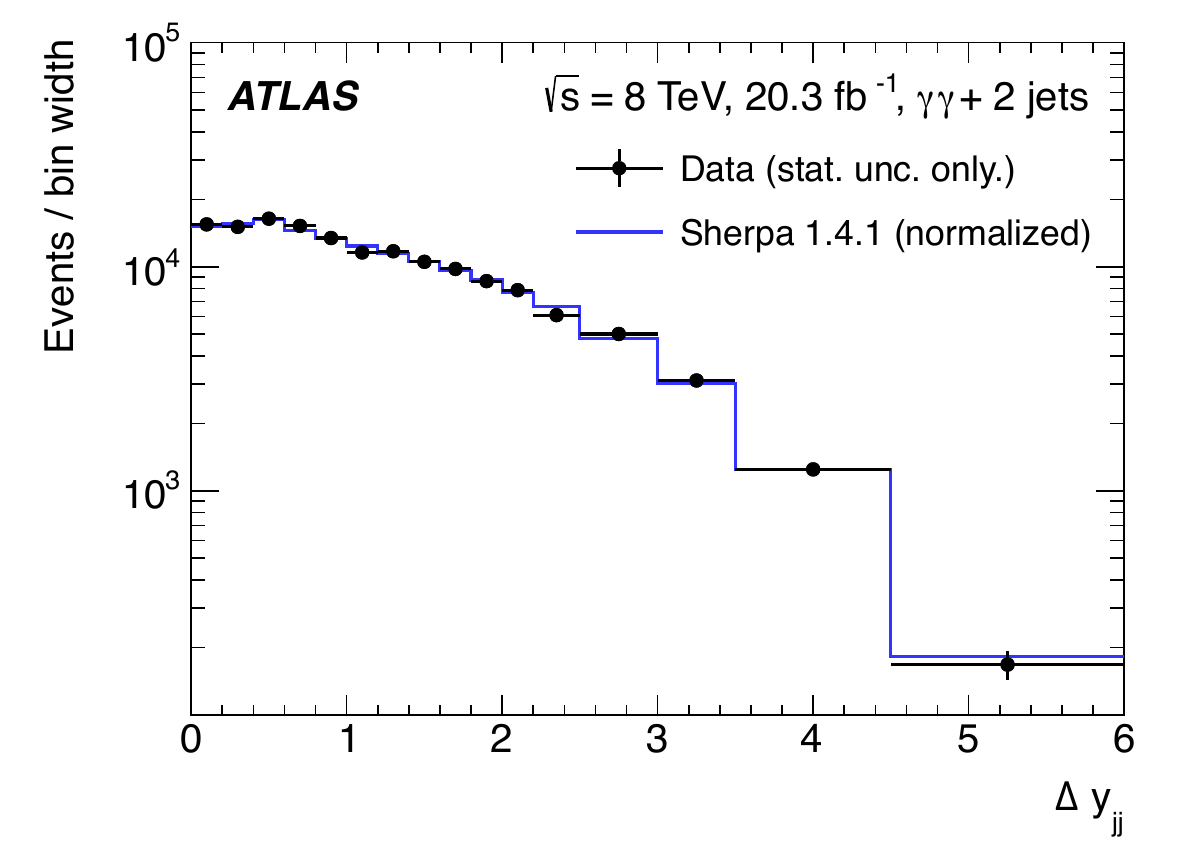}\label{fig:ggjj_reco4}}
\caption{Comparison of data (points) with statistical uncertainties to Sherpa 1.4.1 MC (line) at reconstruction level for (a) \pTjone\ and (b) \dRgonejone\ distribution for $\Nj=1$; and (c) \dphijj\ and (d) \dyjj\ distributions for $\Nj = 2$. Data points are plotted on bin centers. The level of agreement between data and simulation can be assessed by comparing the shapes.}
\label{fig:ggjj_reco}
\end{figure}


\section{Unfolding}
\label{sec:unfolding}

The measured yields of $\gamgam+\text{jet(s)}$ events are unfolded to particle level. Unfolding takes into account the trigger efficiency, signal reconstruction efficiency, selection criteria efficiencies, detector-induced distortions of the measured distributions and the presence of pileup jets. The unfolded spectra are then divided by the luminosity to derive differential cross sections.

Singular Value Decomposition (SVD) unfolding~\cite{HOCKER1996469} is used in this paper to account for detector effects, relating reconstruction and generation level information from the {\textsc{Sherpa}}~1.4.1 diphoton samples described in Section~\ref{sec:samples}.
Originally developed for single differential cross sections, the SVD unfolding is generalized to double differential cross sections. To avoid overestimated uncertainties, SVD unfolding uses a regularization to reduce sudden changes in the MC to unfolded data ratio in adjacent bins. With two-dimensional distributions, there are adjacent bins both in the measured observable and in the number of jets ($\pm$~1 jet). A regularization in the latter has been studied. Due to the small number of bins in this dimension (two to four) and the not sufficiently accurate description of the jet multiplicity in the MC samples used, it does not improve the accuracy of the measurement, but it can lead to a notable bias. Therefore, a regularization is applied only to the adjacent bins in the measured observable, not to adjacent bins in the number of jets.

As a deviation between MC spectrum and the true physical distribution can lead to a bias in the estimated yield, studies are performed to estimate the impact of MC to data differences. For these the MC shape is distorted by reweighting, and unfolding of the original MC shape is carried out with the distorted simulation. This is performed for different shapes of distortions, following MC-to-data shape deviations observed for representative observables in previous ATLAS diphoton measurements. In addition, the observed difference between MC and unfolded data is used as a distortion in a pseudo-data sample and the impact on the unfolded results is evaluated. In each bin, the largest observed bias is taken as a systematic uncertainty in both directions.

SVD unfolding has a free parameter describing the regularization strength $k$, an integer from 1 to the total number of bins, where smaller values lead to a stronger regularization.
A regularization that is too strong leads to large biases, while a regularization that is too weak leads to large statistical uncertainties.
To choose the regularization strength, a figure of merit is defined combining the statistical uncertainties and the expected biases across all bins using a sub-linear (cube-root) scaling to reduce the influence of outlier bins, with the bias term weighted by a factor of four to favor low-bias solutions.
The regularization strength is then chosen to minimize this figure of merit.
For observables where the figure of merit did not show a clear minimum, $k$ was chosen to be three quarters of the number of bins.


\section{Systematic uncertainties}
\label{sec:systematics}

\begin{table}
\centering
\caption{Summary of relative uncertainties in the integrated cross section for exclusive $\Nj$ categories, taken from the \dphigg\ observable.}
\label{tab:ggjj_systematics_njet}
\renewcommand{\arraystretch}{1.6}
\setstackgap{L}{0.8\normalbaselineskip}
\def\stackalignment{r}
\begin{tabular}{lccccccc}
\toprule
Uncertainty source [\%]         & $\Nj = 0$                 & $\Nj = 1$                 & $\Nj = 2$                 & $\Nj \geq 3$              \\
\midrule
Photon energy scale             & \Vectorstack{+0.5 -0.6}   & \Vectorstack{+0.7 -0.4}   & \Vectorstack{+0.7 -0.4}   & \Vectorstack{+1.1 -0.4}   \\
Photon energy resolution        & $\pm$0.03                 & \Vectorstack{+0.07 -0.04} & $\pm$0.07                 & $\pm$0.1                  \\
Photon identification           & $\pm$0.4                  & $\pm$0.3                  & $\pm$0.3                  & $\pm$0.3                  \\
Jet energy scale                & $\pm$3.7                  & $\pm$9.2                  & $\pm$9.3                  & \Vectorstack{+15.8 -14.5} \\
Jet energy resolution           & $\pm$0.5                  & $\pm$1.8                  & $\pm$0.7                  & $\pm$1.0                  \\
Jet vertex fraction             & $\pm$1.3                  & $\pm$3.1                  & $\pm$3.9                  & $\pm$4.4                  \\
Fake-photon background          & \Vectorstack{+8.5 -5.1}   & \Vectorstack{+5.7 -3.9}   & \Vectorstack{+5.1 -2.4}   & \Vectorstack{+4.7 -2.8}   \\
Electron background subtraction & $\pm$0.5                  & $\pm$2.0                  & $\pm$2.0                  & $\pm$2.0                  \\
Unfolding bias                  & $\pm$0.03                 & $\pm$0.07                 & $\pm$0.2                  & $\pm$0.3                  \\
MC sample size                  & $\pm$0.2                  & $\pm$0.5                  & $\pm$0.9                  & $\pm$1.2                  \\
Trigger efficiency              & $\pm$0.2                  & $\pm$0.2                  & $\pm$0.2                  & $\pm$0.2                  \\
Luminosity                      & $\pm$1.9                  & $\pm$1.9                  & $\pm$1.9                  & $\pm$1.9                  \\
\midrule
Total systematic uncertainty    & \Vectorstack{+10 -7}      & \Vectorstack{+12 -11}     & \Vectorstack{+11 -10}     & \Vectorstack{+17 -15}     \\
\midrule
Statistical uncertainty         & $\pm$0.7                  & $\pm$1.0                  & $\pm$1.4                  & $\pm$1.9                  \\
\bottomrule
\end{tabular}
\renewcommand{\arraystretch}{1}
\end{table}

Table~\ref{tab:ggjj_systematics_njet} lists the contributions of each systematic uncertainty component for the different $\Nj$ categories. The dominant systematic uncertainty in the cross-section measurement, in most regions of the phase space, comes from the definition of the non-\textit{tight} control region in the fake-photon background subtraction method.

Several systematic effects of the fake-photon background subtraction on the diphoton purity measurement are evaluated by varying the control regions described above.  A systematic uncertainty is assigned to the assumption that for fake photons the isolation properties are independent of the identification.  As explained previously, the nominal \nT\ selection corresponds to removing four criteria out of the nine that define the \textit{tight} region. The systematic uncertainty is computed by calculating the yields using different definitions of the \nT\ control region, removing not four but two, three and five of the nine criteria defining the \emph{tight} selection and taking the envelope of the largest difference relative to the nominal definition. This is the dominant systematic uncertainty in the yield.

Anti-isolated control regions are used to calculate the matrix method coefficients, as studies based on MC suggest that the fraction of genuine photons in the \T-anti-isolated selection is $\sim$1\%, and even much smaller in the \nT-anti-isolated selection. To assess the effect of this assumption, and of possible mismodeling by the MC of the isolation energy distribution at high value of \ETiso\ because of the poor MC modeling of photons issued from parton fragmentation, the fake-photon background subtraction is repeated changing the track isolation requirement of anti-isolated definition to $\pTiso>~5~\GeV$, and the result is taken as a systematic uncertainty in the yield, amounting to about 1\%.

\sherpa\ is used as default MC for the calculation of fake rates and correlations; when an alternative MC generator implementing different parton shower and hadronization models is used to compute those quantities, i.e.{} \pythia, negligible differences are observed in the measured yields; therefore, no systematic uncertainty is assigned to this specific choice.

Since a single value of the electron-to-photon fake rate is used, integrated over the photon pseudorapidity and transverse momentum, its dependence on the photon kinematics is accounted for by systematic variations. For events with $N_\mathrm{jet}\geq1$, the dominant uncertainty in the electron background subtraction arises from the difference between the $f_{e\to\gamma}$ values extracted from the inclusive sample and the high-\pTgg\ sample, to account for the non-negligible portion of such events that occur at lower \pTgg. This uncertainty is almost uniform around $\pm2\%$ for all kinematic distributions, except for \mgg, where it increases up to $\pm10\%$ around the $Z$-boson mass. A smaller contribution, which is the only one for events with $N_\mathrm{jet}=0$, is evaluated by using different values of $f_{e\to\gamma}$ for the barrel and endcap parts of the calorimeter, derived from simulation, and by varying $f_{e\to\gamma}$ to account for its dependence on the transverse momentum of the electron faking a photon.

\begin{figure}
\centering
\subfloat[]{\includegraphics[width= 0.497\textwidth]{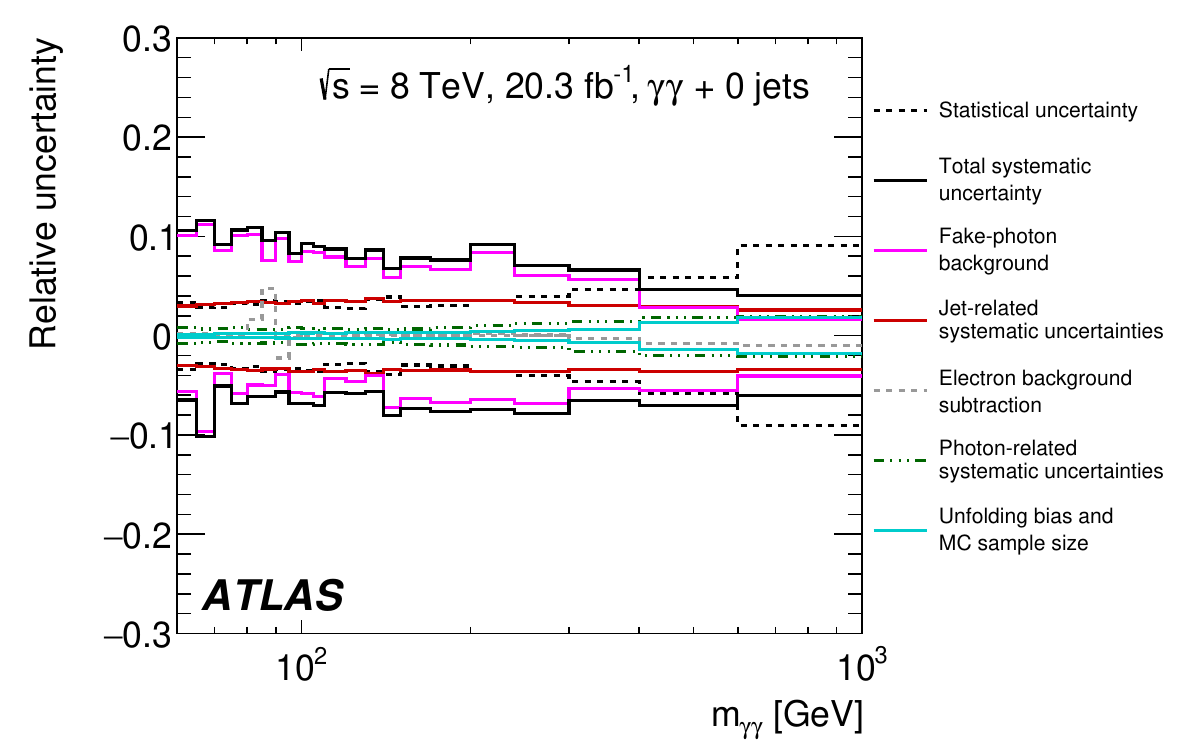}\label{fig:ggjj_syst0}}
\subfloat[]{\includegraphics[width= 0.497\textwidth]{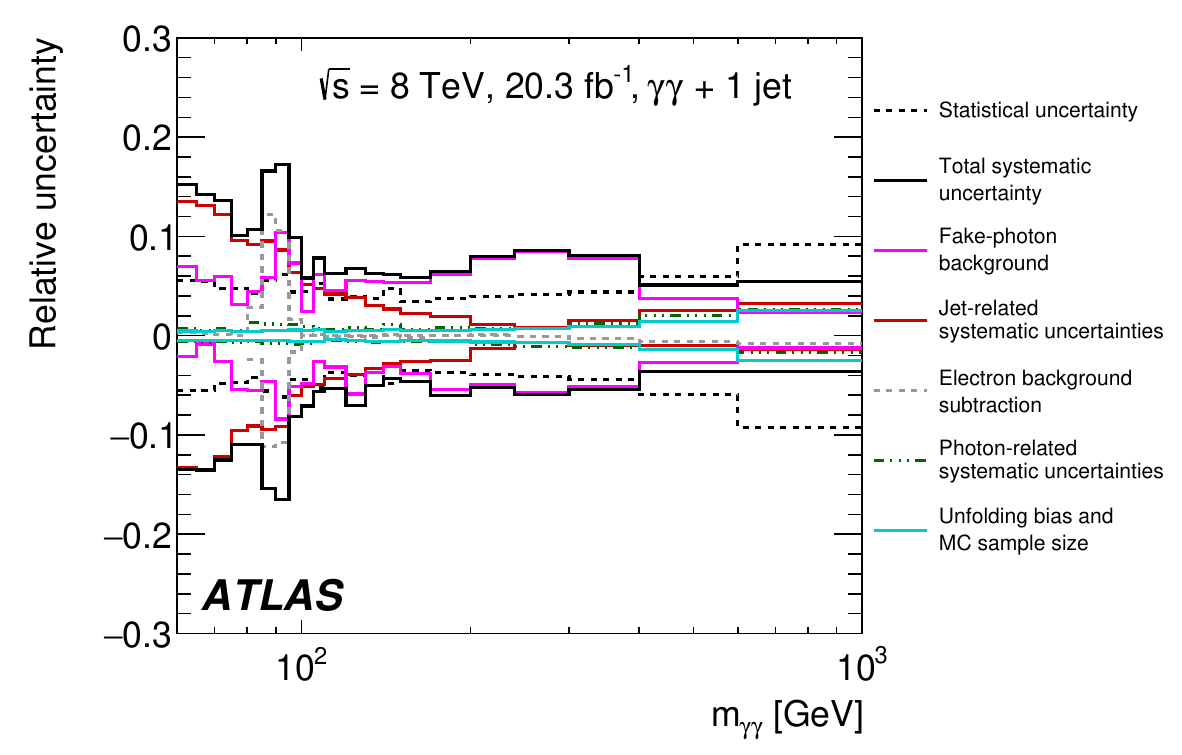}\label{fig:ggjj_syst1}} \\
\subfloat[]{\includegraphics[width= 0.497\textwidth]{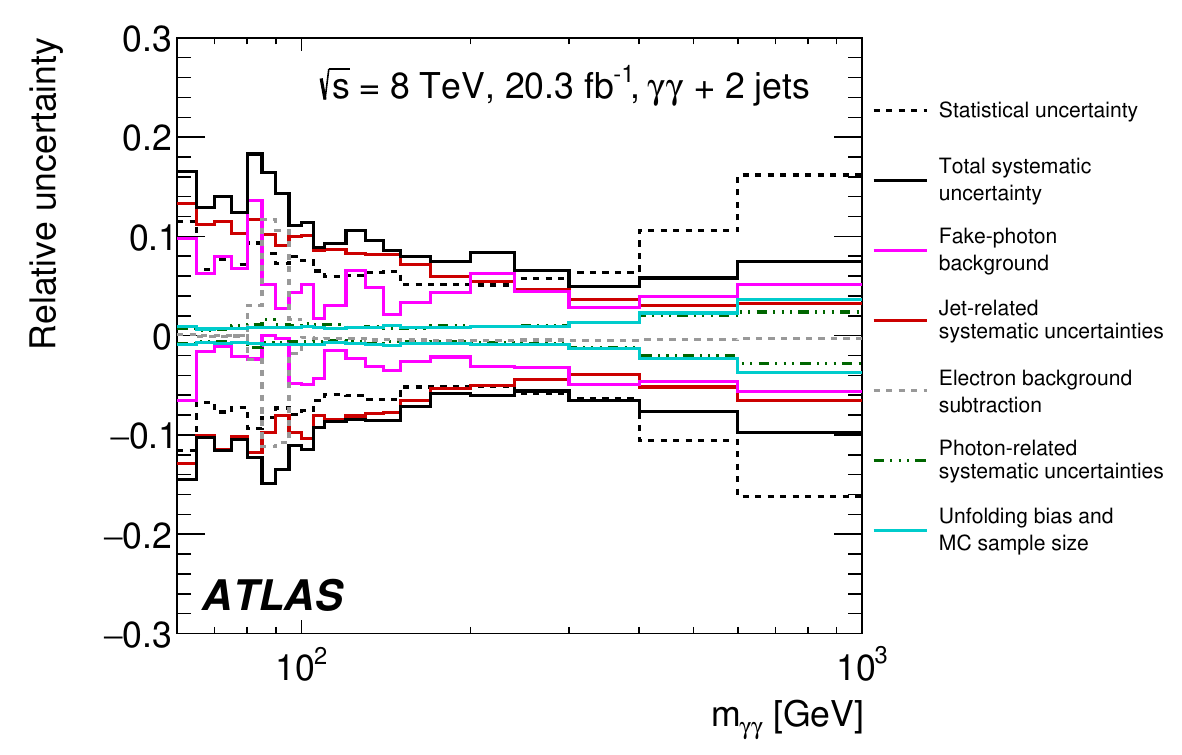}\label{fig:ggjj_syst2}}
\subfloat[]{\includegraphics[width= 0.497\textwidth]{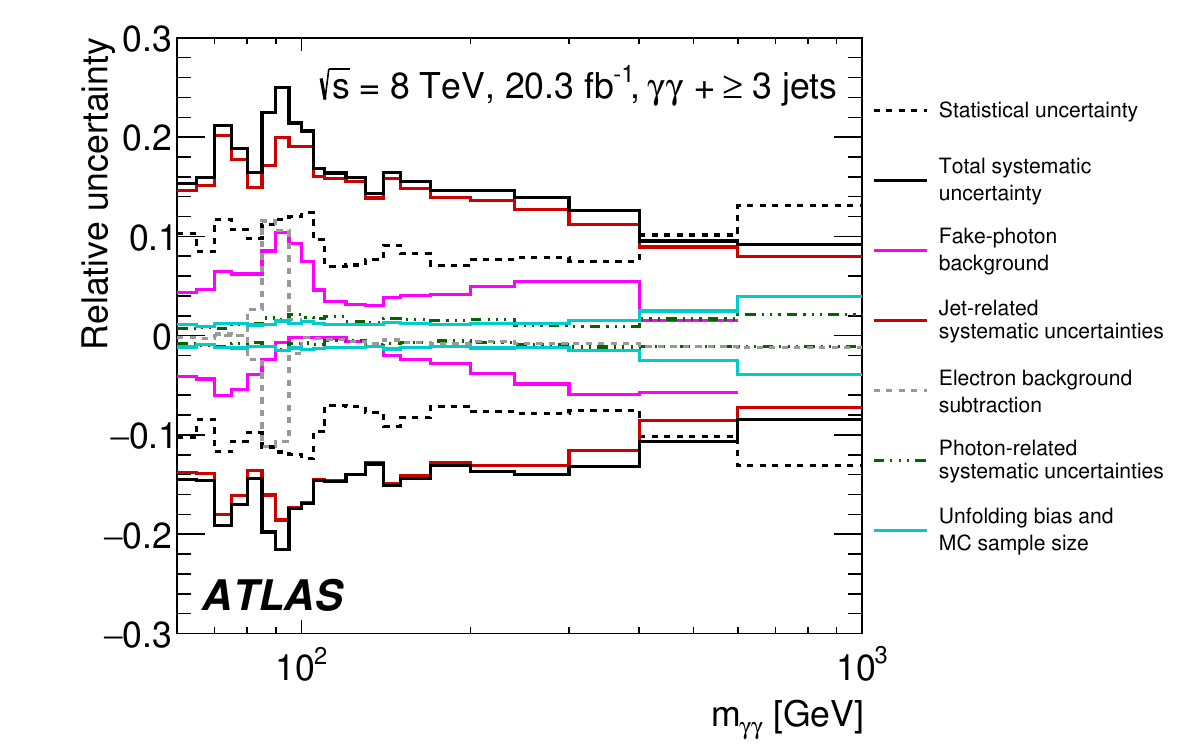}\label{fig:ggjj_syst3}}
\caption{Breakdown of the systematic uncertainties for the unfolded \mgg\ distribution for the exclusive jet categories (a) $\Nj=0$, (b) 1, (c) 2 and (d) the inclusive $\Nj\geq 3$~category. The total systematic uncertainty includes all systematic uncertainties except the luminosity uncertainty.}
\label{fig:ggjj_systematics}
\end{figure}

Figure~\ref{fig:ggjj_systematics} shows the breakdown of systematic uncertainties in the measured cross section as a function of the diphoton invariant mass in selected $\Nj$ categories. In most bins, the jet energy scale and the non-\textit{tight} control region definition are the dominant systematic uncertainties. In the 0-jet category, the electron background subtraction uncertainty, described in Section~\ref{subsec:electronbackground}, becomes the second largest systematic uncertainty at around $\mgg\approx 90~\GeV$.  Subdominant background-related uncertainties arise from the photon isolation shape and the photon identification scale factors. These uncertainties are propagated through the unfolding procedure.

Two systematic uncertainties are assigned to the unfolding procedure, associated with the statistical uncertainty of the MC sample and the unfolding bias.

Other experimental uncertainties are evaluated by modifying the detector response matrices according to the considered effect, which includes the photon energy scale, photon energy resolution, jet energy scale, jet energy resolution, photon \textit{tight} identification scale factors, JVF signal efficiency and the definition of pileup jets in the JVF selection. These effects are mostly subdominant, except for jet energy scale becoming more significant when $\Nj>2$ and JVF signal efficiency being the largest systematic uncertainty for low $\pTj$. The jet definition and pileup uncertainty becomes dominant for $\Delta\phi_{\gamma\gamma}\approx\pi$. The luminosity uncertainty, measured to be 1.9\%~\cite{DAPR-2013-01}, is also included in the final cross sections.  All uncertainty sources are treated as uncorrelated relative to other uncertainty sources, except for the correlation model recommendations on photon and jet energy scale and resolution according to methods in Refs.~\cite{PERF-2013-05,PERF-2013-04,PERF-2011-03,PERF-2012-01,PERF-2014-02,PERF-2014-03}.


\section{Theoretical predictions}
\label{sec:theory}

The measured fiducial cross sections, both integrated (per \Nj\ category) and differential in up to sixteen kinematic and angular variables, are compared with several theoretical predictions.

One prediction is based on the {\textsc{Sherpa~1.4.1}} LO event generator interfaced with a parton shower model, the same generator used for background subtraction and unfolding (see Section~\ref{sec:samples}). Particle-level events are selected in the fiducial phase space defined in Section~\ref{subsec:fid_vol}. Since this prediction is only LO, it is not expected to accurately predict the integrated fiducial cross section; accordingly, for differential comparisons, \sherpa\ 1.4.1 is shown normalized to data. The normalization factors are 1.5, 1.4, and 1.25 for $\Nj=0$,~1,~and~2, respectively; in the $\Nj\geq 3$ category \sherpa\ 1.4.1 is shown at absolute normalization, since its predicted cross section coincidentally matches the measurement. No theory uncertainties are quoted and MC statistical uncertainties are negligible.

NLO fixed-order pQCD predictions from the \gosam\ framework~\cite{th_gosam} are compared with data in the exclusive 1-jet and 2-jet categories. The predictions are computed using CT14 PDFs~\cite{Dulat:2015mca}. Virtual top-quark contributions are neglected and loop-induced processes are not included. The renormalization and factorization scales are set to $\mu_{r}=\mu_{f}=0.5\sqrt{\mgg^{2} + \sum{(\pTj)^{2}}}$ and varied by factors of 0.5 and 2 to evaluate scale uncertainties. Frixione isolation~\cite{Frixione1998369,Catani:2018krb} is applied with $R=0.4$, $\epsilon=0.2$, and $n=1$. The PDF uncertainty is negligible relative to the scale uncertainty. For the \dSggjj\ observable, which requires a $\gamma\gamma{+}2j$ final state, the \gosam\ prediction is computed at LO accuracy.

NLO fixed-order pQCD predictions from the \blackhat\ framework are obtained similarly to Refs.~\cite{Bern:2014vza,blackhat_atlas}. They use CT10 PDFs~\cite{Lai:2010vv} and the same central scale definition, $\mu_{r}=\mu_{f}=0.5\sqrt{\mgg^{2} + \sum{(\pTj)^{2}}}$, with factor-two up/down scale variations. The PDF uncertainty is also negligible relative to the scale uncertainty. All subprocesses are included except loop-induced ones, and a leading-color approximation is used for finite parts of the one-loop matrix elements, with an expected impact of about 1\% on the cross section. Frixione isolation is applied with $R=0.4$, $\epsilon=0.5$, and $n=1$. The \blackhat\ predictions were provided for all observables except \costhgg\ and \dSggjj.

Non-perturbative corrections (NPC) to the fixed-order predictions are computed from dedicated {\textsc{Sherpa 1.4.1}} samples, as described in Section~\ref{sec:samples}, with and without hadronization and underlying event using a Rivet routine~\cite{Buckley:2010ar}.
The total uncertainty in the fixed-order predictions is obtained as the sum in quadrature of statistical, scale, and NPC uncertainties.

An additional comparison is provided by {\textsc{Sherpa 2.2.14}}~\cite{Sherpa:2019gpd}. Matrix elements are calculated at NLO in QCD for diphoton$+0,1$ jet production and at LO for diphoton$+2,3$ jet production, merged with the MEPS@NLO prescription. Virtual QCD corrections are provided by \textsc{OpenLoops}; the \texttt{NNPDF3.0\_nnlo} PDF set~\cite{Ball:2014uwa} is used, and the loop-induced $gg\rightarrow\gamma\gamma$ contribution is included at LO. Given the higher perturbative accuracy, no normalization to data is applied for this prediction. Scale uncertainties are evaluated from a seven-point $\mu_R$/$\mu_F$ envelope with consistent parton-shower scale variations; PDF uncertainties are assessed following the PDF4LHC21 prescription~\cite{PDF4LHCWorkingGroup:2022cjn} and are at the 2--3\% level, small compared with the scale uncertainty.

Since the contribution of the Higgs boson decay into two photons is negligible relative to the total diphoton yield, it is not included in the theoretical comparisons.


\section{Results}
\label{sec:results}

\subsection{Integrated cross sections}

Background subtraction and unfolding are performed in exclusive jet categories. Due to the limited number of data events, the 3-jet category is inclusive. Results for inclusive 0, 1 and 2-jet categories and for integrated cross sections are derived from the exclusive jet category results, taking correlations between bins of different jet categories into account. Table~\ref{tab:ggjj_int_cs_justdata} shows the integrated cross sections after unfolding with their  statistical and systematic uncertainties, containing also the luminosity uncertainty, for each jet category.

\begin{table}
\centering
\caption{Integrated cross sections measured in data for each jet category, with statistical, and total systematic uncertainties.
All cross sections are measured in the fiducial region defined in Section~\ref{subsec:fid_vol}.}
\label{tab:ggjj_int_cs_justdata}
\renewcommand{\arraystretch}{1.6}
\setstackgap{L}{0.8\normalbaselineskip}
\def\stackalignment{r}
\begin{tabular}{l.cc}
\toprule
$\Nj$         & \multicolumn{1}{c}{$\sigma^{\text{fid}}$ [pb]} & \multicolumn{1}{c}{Stat. unc. [pb]} & \multicolumn{1}{c}{Syst. unc. [pb]} \\
\midrule
0-jet        & 28.5 & $\pm$0.21 & \Vectorstack{+2.8 -2.0} \\
1-jet        & 7.4  & $\pm$0.07 & $\pm$0.9 \\
2-jet        & 2.44  & $\pm$0.04 & \Vectorstack{+0.28 -0.24} \\
\midrule
$\geq$ 0-jet & 39.2 & $\pm$0.20 & \Vectorstack{+3.1 -1.9} \\
$\geq$ 1-jet & 10.7 & $\pm$0.07 & $\pm$1.2 \\
$\geq$ 2-jet & 3.3  & $\pm$0.03 & $\pm$0.4 \\
$\geq$ 3-jet & 0.82  & $\pm$0.02 & $\pm$0.14 \\
\bottomrule
\end{tabular}
\renewcommand{\arraystretch}{1}
\end{table}

As expected for a LO generator, \sherpa\ 1.4.1 greatly underestimates the measured integrated cross sections across all jet categories, with the notable exception of the $\geq 3$-jet inclusive category, where the predicted cross section coincidentally matches the data. Table~\ref{tab:ggjj_int_cross_sections} compares the measured cross sections with the NLO-level predictions.
\gosam\ NLO predictions are compatible with the data within uncertainties in both the 1-jet and 2-jet categories, with central values 11\% and 9\% below the data, respectively.
\blackhat\ is also compatible with the data within uncertainties in the 2-jet category, with a central value 3\% below data.
{\textsc{Sherpa~2.2.14}} predictions are compatible with the data within the scale uncertainties in all categories, with some underestimation in the 0-jet (9\%) and 2-jet (15\%) exclusive categories and good agreement in the 1-jet category.

\begin{table}
\centering
\caption{Integrated cross sections per jet category as measured in data, and the corresponding NLO theoretical predictions. The total uncertainty for data values includes the statistical, systematic and luminosity uncertainty. NLO fixed-order predictions are listed with the total statistical, scale and NPC uncertainty. For {\textsc{Sherpa 2.2.14}}, the scale uncertainty from the seven-point envelope is shown; PDF uncertainties are small in comparison. \gosam\ and \blackhat\ predictions are available only for the exclusive 1-jet and 2-jet categories. The small upper uncertainties for \gosam\ and \blackhat\ in the 2-jet category reflect the highly asymmetric scale uncertainty envelope in this exclusive jet multiplicity category.}
\label{tab:ggjj_int_cross_sections}
\renewcommand{\arraystretch}{1.6}
\setstackgap{L}{0.8\normalbaselineskip}
\def\stackalignment{r}
\begin{tabular}{l.cccccccc}
\toprule
$\Nj$ & \multicolumn{2}{c}{$\sigma^{\text{fid}}$ [pb]} & \multicolumn{2}{c}{$\sigma_{\text{Sherpa 2.2.14}}$ [pb]} & \multicolumn{2}{c}{$\sigma_{\gosam}$ [pb]} & \multicolumn{2}{c}{$\sigma_{\blackhat}$ [pb]} \\
\midrule
0-jet        & 28.5 & {\small\Vectorstack{+2.8 -2.0}} & 25.9 & {\small\Vectorstack{+5.0 -2.6}} &      &                                   &      &                                   \\
1-jet        & 7.4  & {\small $\pm$0.9}               &  7.7 & {\small\Vectorstack{+2.4 -1.4}} & 6.6  & {\small\Vectorstack{+0.4 -0.3}}   &      &                                   \\
2-jet        & 2.44  & {\small\Vectorstack{+0.28 -0.25}} &  2.1 & {\small\Vectorstack{+1.0 -0.5}} & 2.22 & {\small\Vectorstack{+0.01 -0.12}} & 2.36 & {\small\Vectorstack{+0.02 -0.15}} \\
\midrule
$\geq$ 0-jet & 39.2 & {\small\Vectorstack{+3.1 -1.9}} & 36.4 & {\small\Vectorstack{+8.6 -4.6}} &      &                                   &      &                                   \\
$\geq$ 1-jet & 10.7 & {\small $\pm$1.2}               & 10.5 & {\small\Vectorstack{+3.7 -2.1}} &      &                                   &      &                                   \\
$\geq$ 2-jet & 3.3  & {\small $\pm$0.4}               &  2.8 & {\small\Vectorstack{+1.4 -0.8}} &      &                                   &      &                                   \\
$\geq$ 3-jet & 0.82  & {\small $\pm$0.14}             &  0.71 & {\small\Vectorstack{+0.39 -0.22}} &      &                                   &      &                                   \\
\bottomrule
\end{tabular}
\renewcommand{\arraystretch}{1}
\end{table}

\subsection{Differential cross sections}

Differential cross sections for each jet multiplicity category are shown in Figures~\ref{fig:ggjj_0jet_excl_diphoton}--\ref{fig:ggjj_3jet_incl_ptj123}. {\textsc{Sherpa~1.4.1}} is normalized to the measured integrated cross section in each category (except in the $\geq 3$-jet category, where it is shown at absolute normalization since its predicted cross section matches the data); the normalization factor is quoted in the plot legends. {\textsc{Sherpa~2.2.14}}, \gosam, and \blackhat\ predictions are shown at absolute normalization. The \sherpa~2.2.14 scale and PDF uncertainty band is shown in both the main and the ratio panels.
The scale uncertainty bands for \gosam\ and \blackhat\ are also shown;
in bins where the NLO scale uncertainty exceeds the central NLO prediction,
only the central value is displayed as an open marker and the uncertainty band is not drawn.
Integrated NLO uncertainty values are listed in Table~\ref{tab:ggjj_int_cross_sections}.

\begin{figure}
\centering
\subfloat[]{\includegraphics[width= 0.49\textwidth]{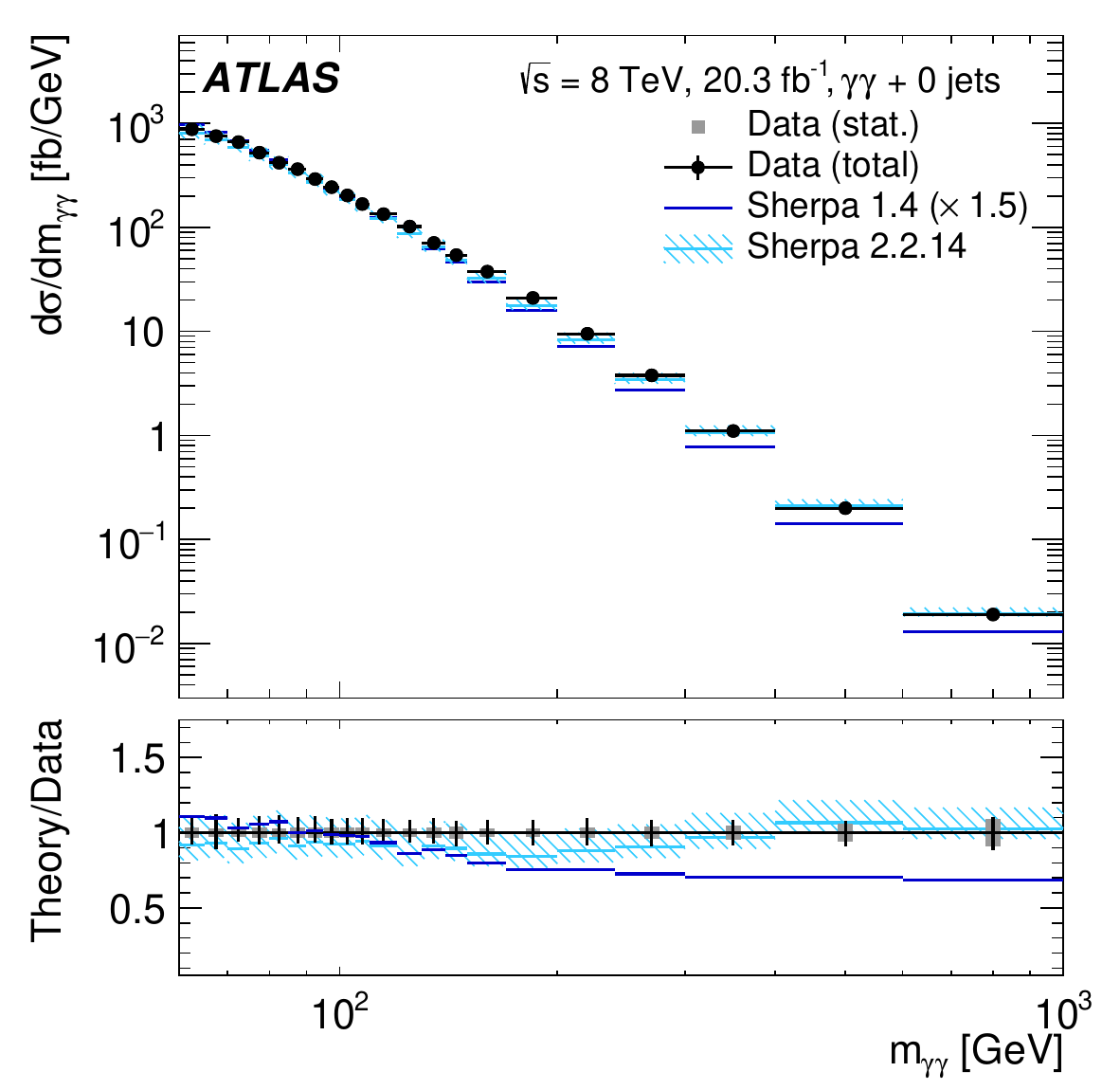}\label{fig:ggjj_0jet_excl_mgg}}
\subfloat[]{\includegraphics[width= 0.49\textwidth]{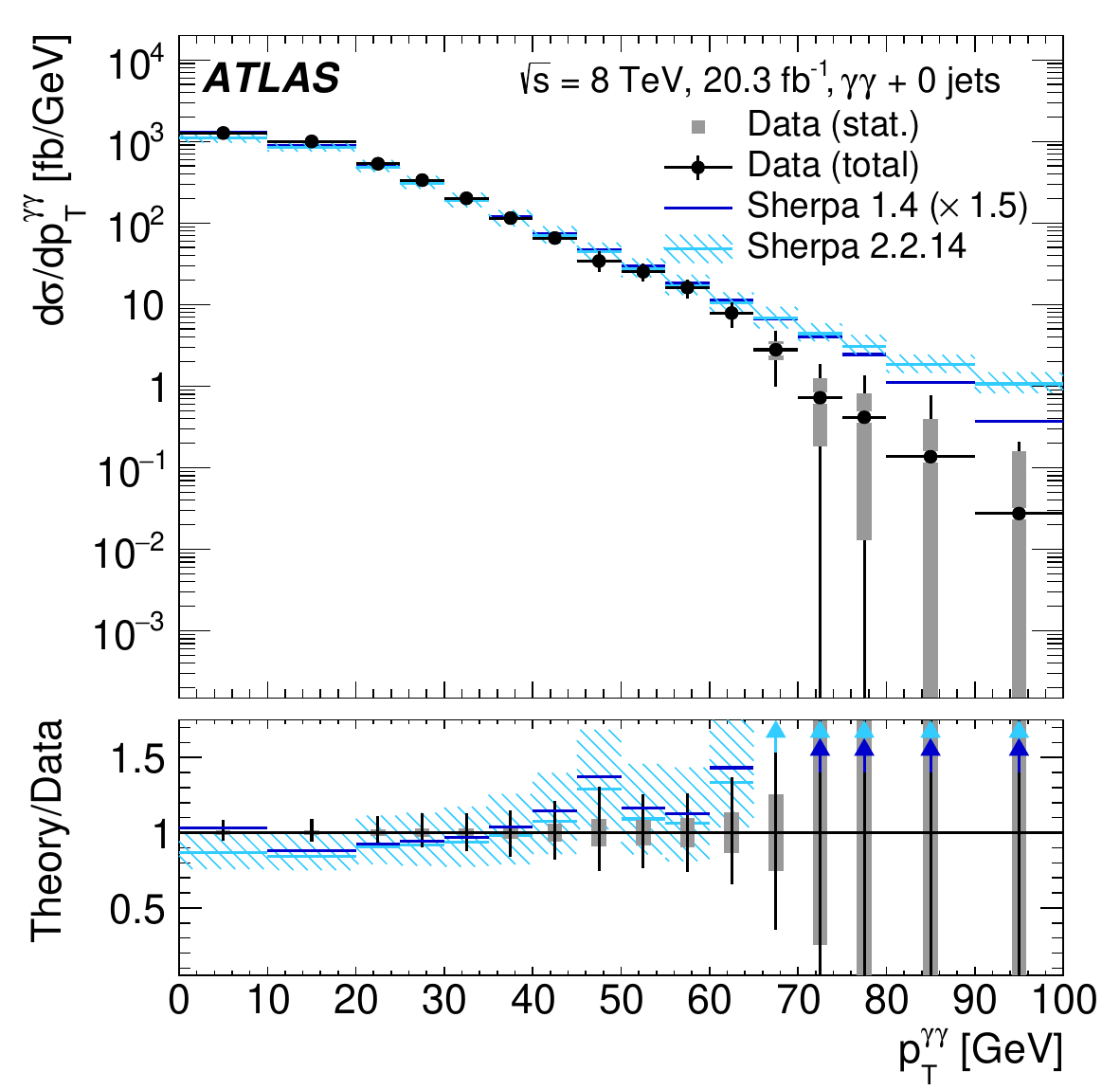}\label{fig:ggjj_0jet_excl_ptgg}} \\
\subfloat[]{\includegraphics[width= 0.49\textwidth]{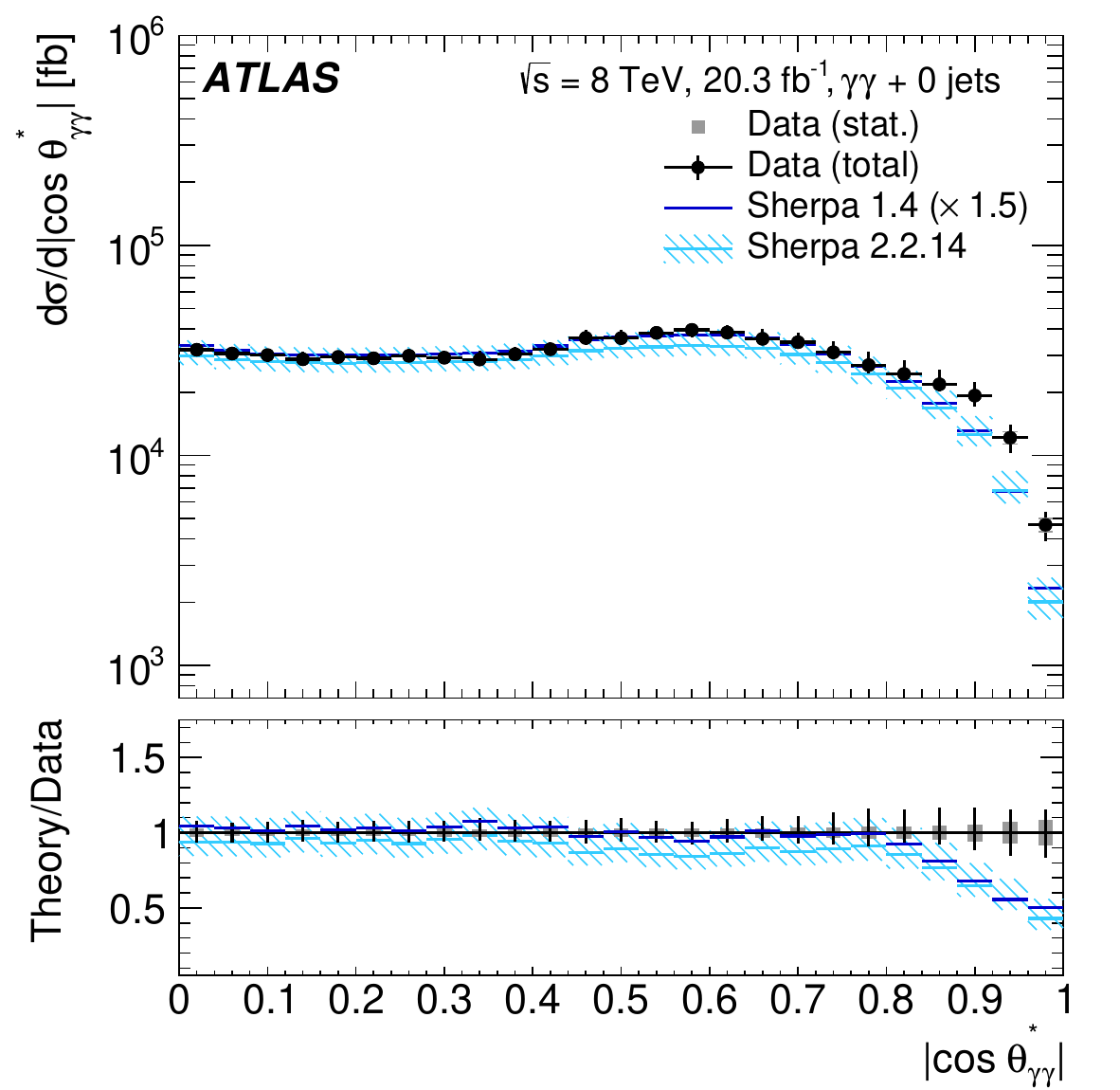}\label{fig:ggjj_0jet_excl_costhgg}}
\subfloat[]{\includegraphics[width= 0.49\textwidth]{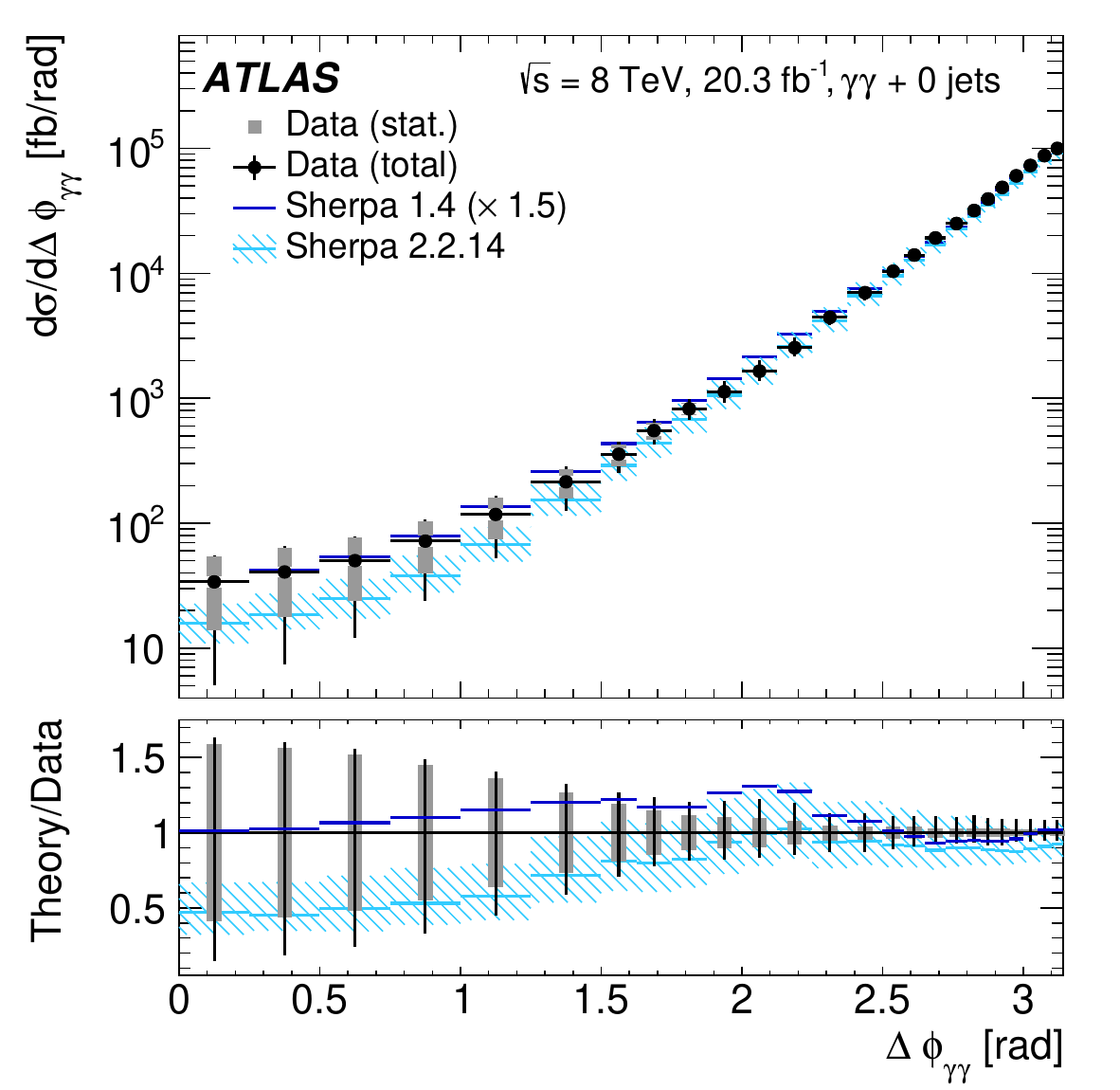}\label{fig:ggjj_0jet_excl_dphigg}}
\caption{
Differential cross sections in the $\Nj = 0$ category as a function of (a) \mgg, (b) \pTgg, (c) \costhgg\ and (d) \dphigg,
compared with theoretical predictions from \sherpa\ 2.2.14,
and to LO \sherpa\ 1.4.1 predictions normalized to data.
Ratios of theory to data are shown in the lower panel.
The normalization factor for \sherpa\ 1.4.1 is quoted in the plot legend.
}
\label{fig:ggjj_0jet_excl_diphoton}
\end{figure}

Figure~\ref{fig:ggjj_0jet_excl_diphoton} shows the differential cross sections as a function of the diphoton observables for events with zero jets satisfying the analysis selection, compared with \sherpa\ 1.4.1 and to the absolute {\textsc{Sherpa~2.2.14}} prediction.
In the exclusive 0-jet category, the high-\pTgg\ and large-\costhgg regions are kinematically suppressed, as events with large diphoton boost or strong back-to-back asymmetry tend to produce accompanying jets; this is reflected in the increased measurement uncertainties in those bins.
\textsc{Sherpa~2.2.14} describes the \mgg\ spectrum well across the full measured range, with ratios within 10\% of data at all values. The \pTgg\ distribution is well described at low and moderate values ($\pTgg \lesssim 50~\GeV$), but the prediction overshoots in the high-\pTgg\ region where the data uncertainties are large. For \costhgg, the description is good for $\costhgg < 0.75$, with a progressive undershoot at higher values. For \dphigg, good agreement is found for $\dphigg > 2~\text{rad}$, while \sherpa\ 2.2.14 significantly underestimates the data at small separations ($\dphigg \lesssim 1.5~\text{rad}$).
The normalized \sherpa\ 1.4.1 prediction describes the shape of low \mgg\ and low \pTgg\ regions well. The shape of the \costhgg\ distribution is well described up to 0.8, where it starts to deviate. The $\phi$-separation of the two photons is described well by \sherpa\ 1.4.1, except for the region around $\dphigg\approx2$.

\begin{figure}
\centering
\subfloat[]{\includegraphics[width= 0.49\textwidth]{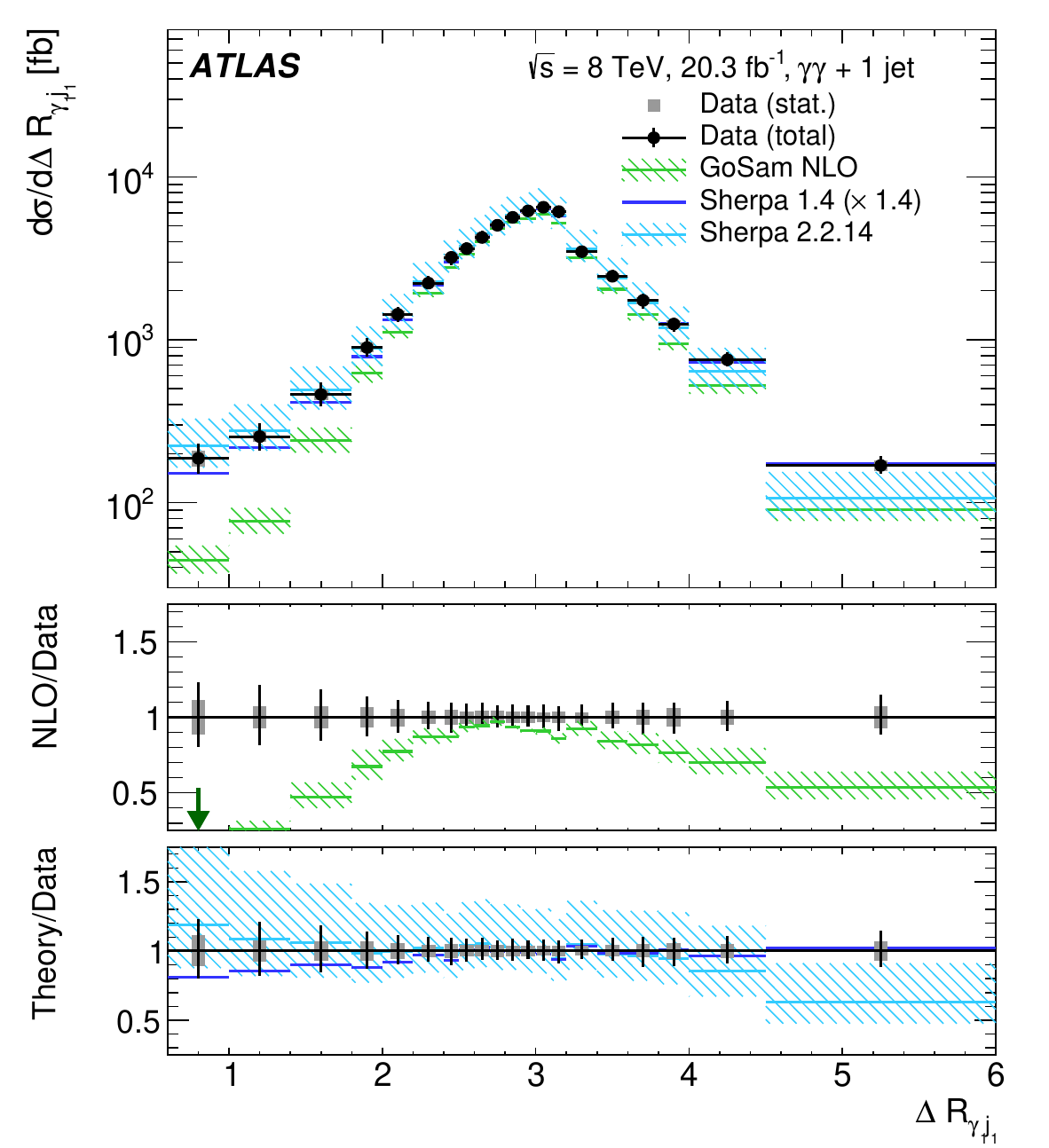}\label{fig:ggjj_1jet_excl_dRg1j1}}
\subfloat[]{\includegraphics[width= 0.49\textwidth]{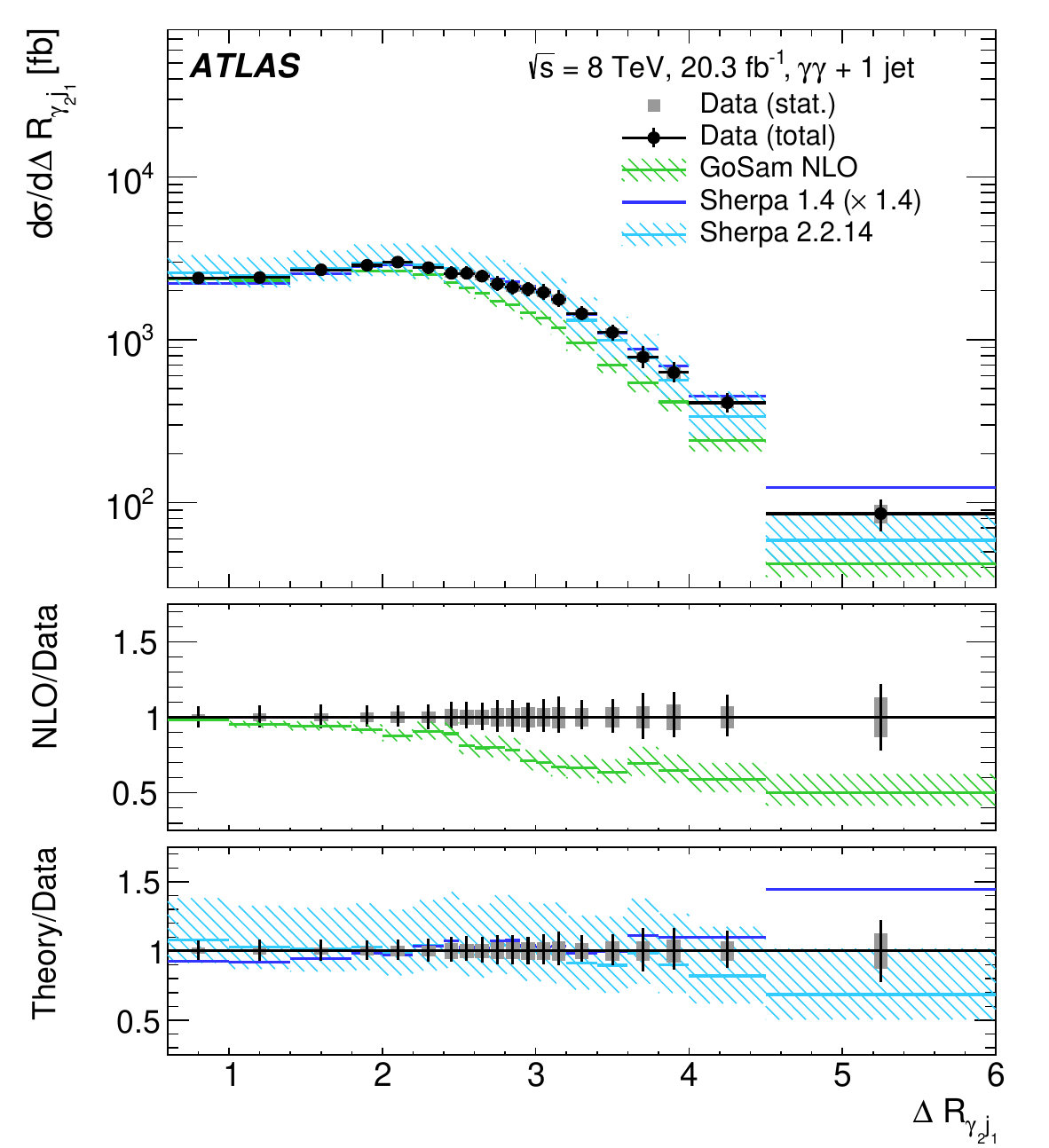}\label{fig:ggjj_1jet_excl_dRg2j1}} \\
\subfloat[]{\includegraphics[width= 0.49\textwidth]{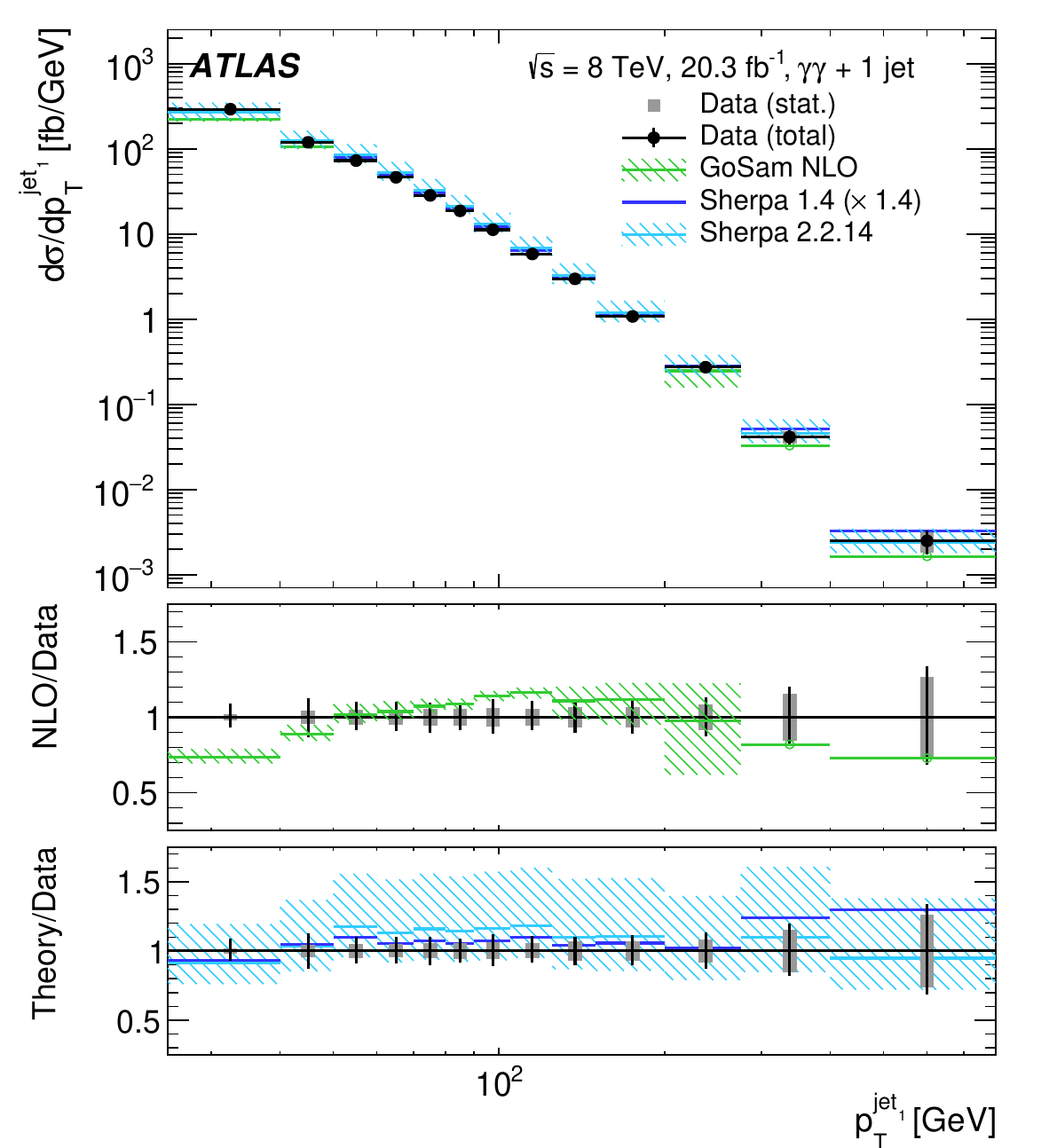}\label{fig:ggjj_1jet_excl_ptj1}}
\caption{Differential cross sections in the $\Nj = 1$ category as a function of (a) \dRgonejone, (b) \dRgtwojone\ and (c) \pTjone,
compared with \gosam\ NLO and \sherpa\ 2.2.14 predictions,
and to LO \sherpa\ 1.4.1 predictions normalized to data.
Ratios of theory to data are shown in the lower panel.
Arrows indicate data points outside the displayed ratio range.
Bins where the \gosam\ NLO scale uncertainty exceeds the central prediction are indicated by open markers without uncertainty bands.
The normalization factor for \sherpa\ 1.4.1 is quoted in the plot legend.
}
\label{fig:ggjj_1jet_excl_dR_ptj1}
\end{figure}

In Figure~\ref{fig:ggjj_1jet_excl_dR_ptj1}, measured cross sections for the exclusive $\Nj = 1$ category are shown compared with \gosam\ NLO and \sherpa\ 2.2.14 predictions for the photon--jet separation and leading jet \pT; shape comparisons to \sherpa\ 1.4.1 are also shown.
The cross sections are measured differentially in the photon--jet separation \dRgj, showing a poor description by \gosam\ for the leading photon at low and high-\dRgonejone, but a good description of the peak. For the subleading photon, the \gosam\ description is best for $\dRgtwojone<2.2$, after that, it starts to deviate by more than 10\% and up to 50\% at high-$\dRgtwojone$. The leading jet transverse momentum is described by \gosam\ within uncertainties for $\pTjone>40$~\GeV.
\textsc{Sherpa~2.2.14} provides a good description of both photon--jet separation observables over the bulk of the distribution. For \dRgonejone, the prediction matches the measured peak position and falls within approximately 15\% of data for $1 < \dRgonejone < 4.5$. A similar level of agreement is found for \dRgtwojone\ up to $\dRgtwojone \approx 4$. At the largest separation values, both observables are slightly underestimated, consistent with the behavior seen in the \gosam\ NLO prediction. The leading jet \pT\ is described well across the full measured range.
\sherpa\ 1.4.1 performs well in describing the shape of \dRgonejone\ and \dRgtwojone, except for the last bin of \dRgtwojone. The shape of the \pTjone\ observable is very well described by \sherpa\ 1.4.1.

\begin{figure}
\centering
\subfloat[]{\includegraphics[width= 0.47\textwidth]{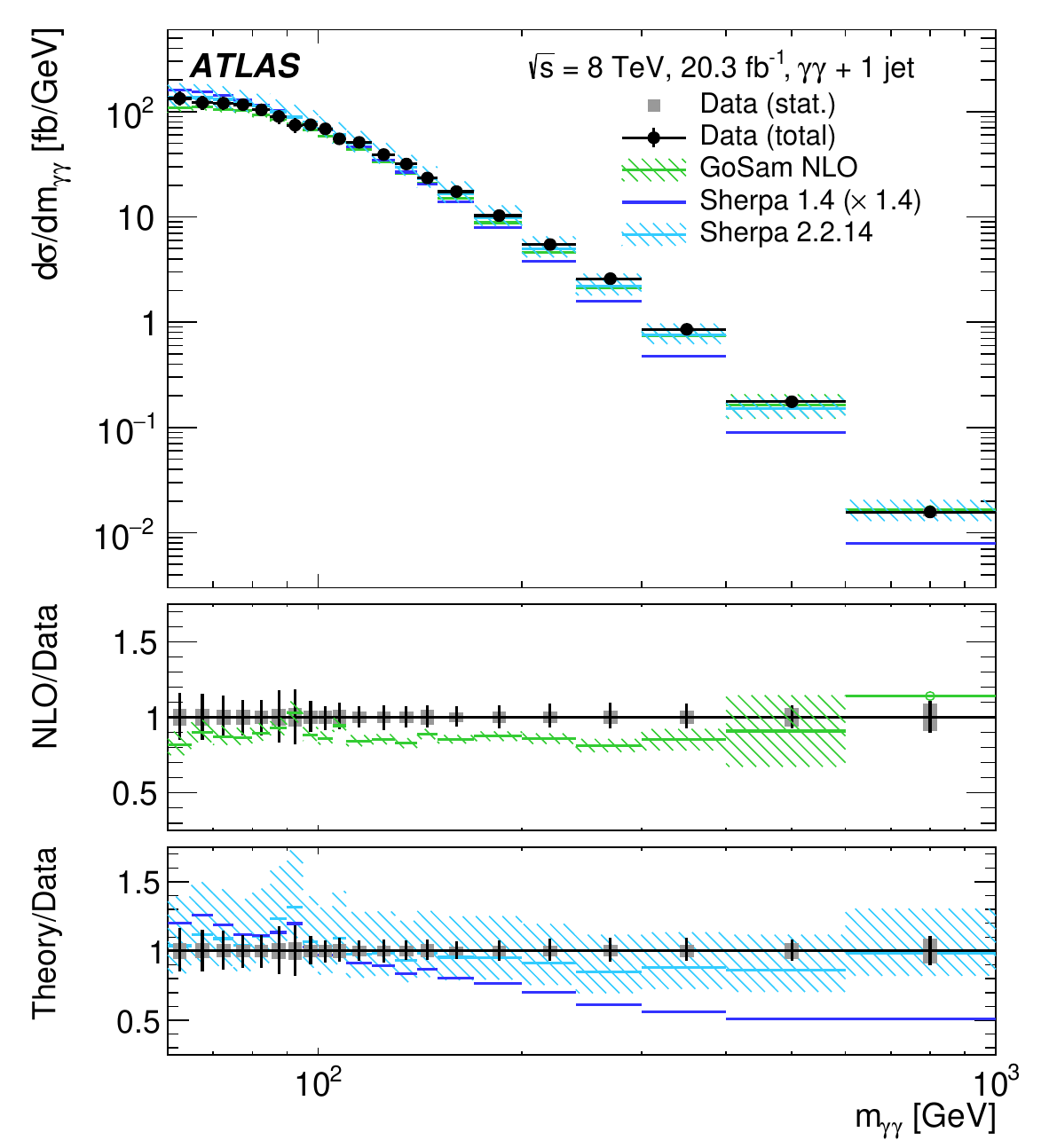}\label{fig:ggjj_1jet_excl_mgg}}
\subfloat[]{\includegraphics[width= 0.47\textwidth]{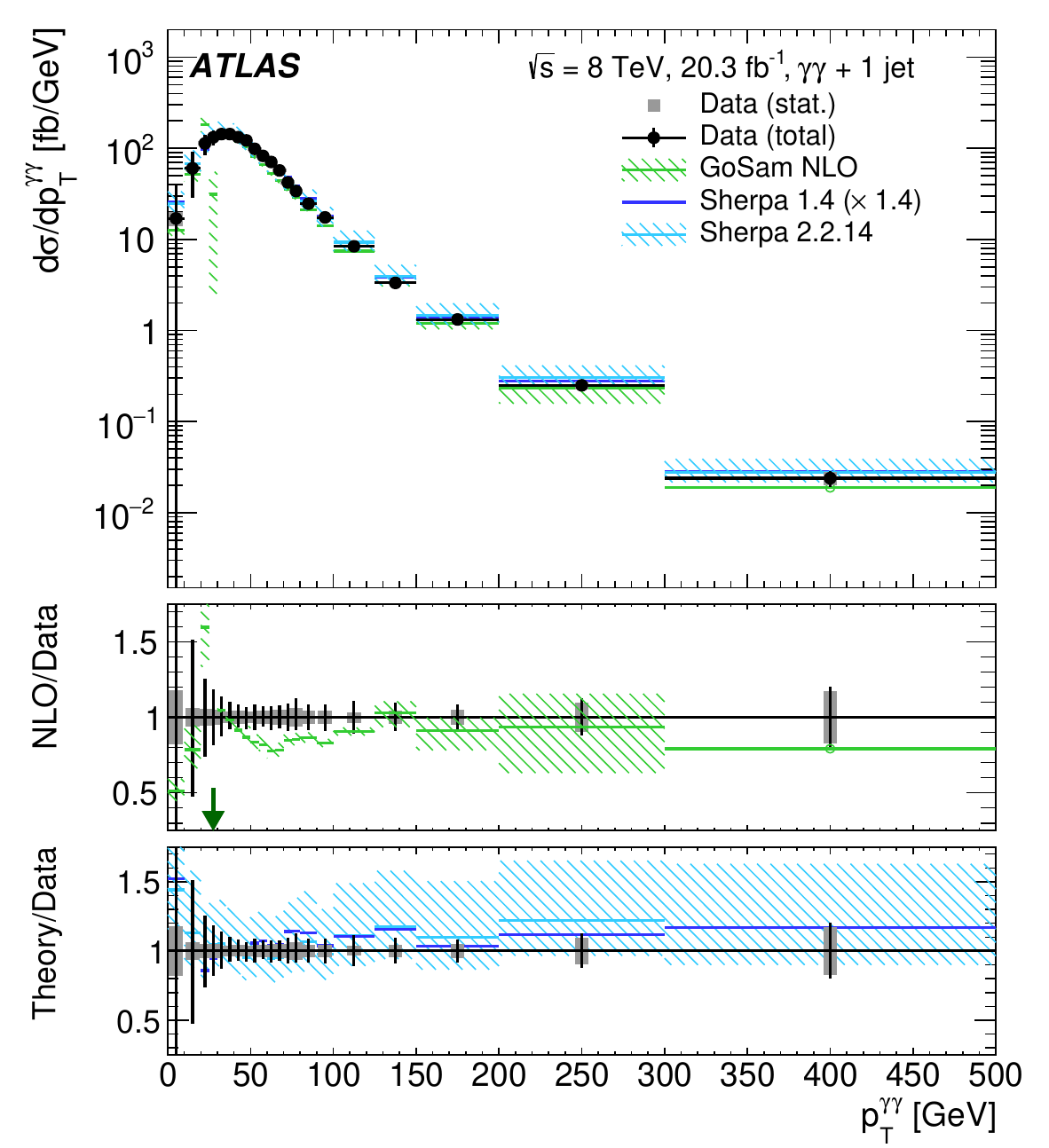}\label{fig:ggjj_1jet_excl_ptgg}} \\
\subfloat[]{\includegraphics[width= 0.47\textwidth]{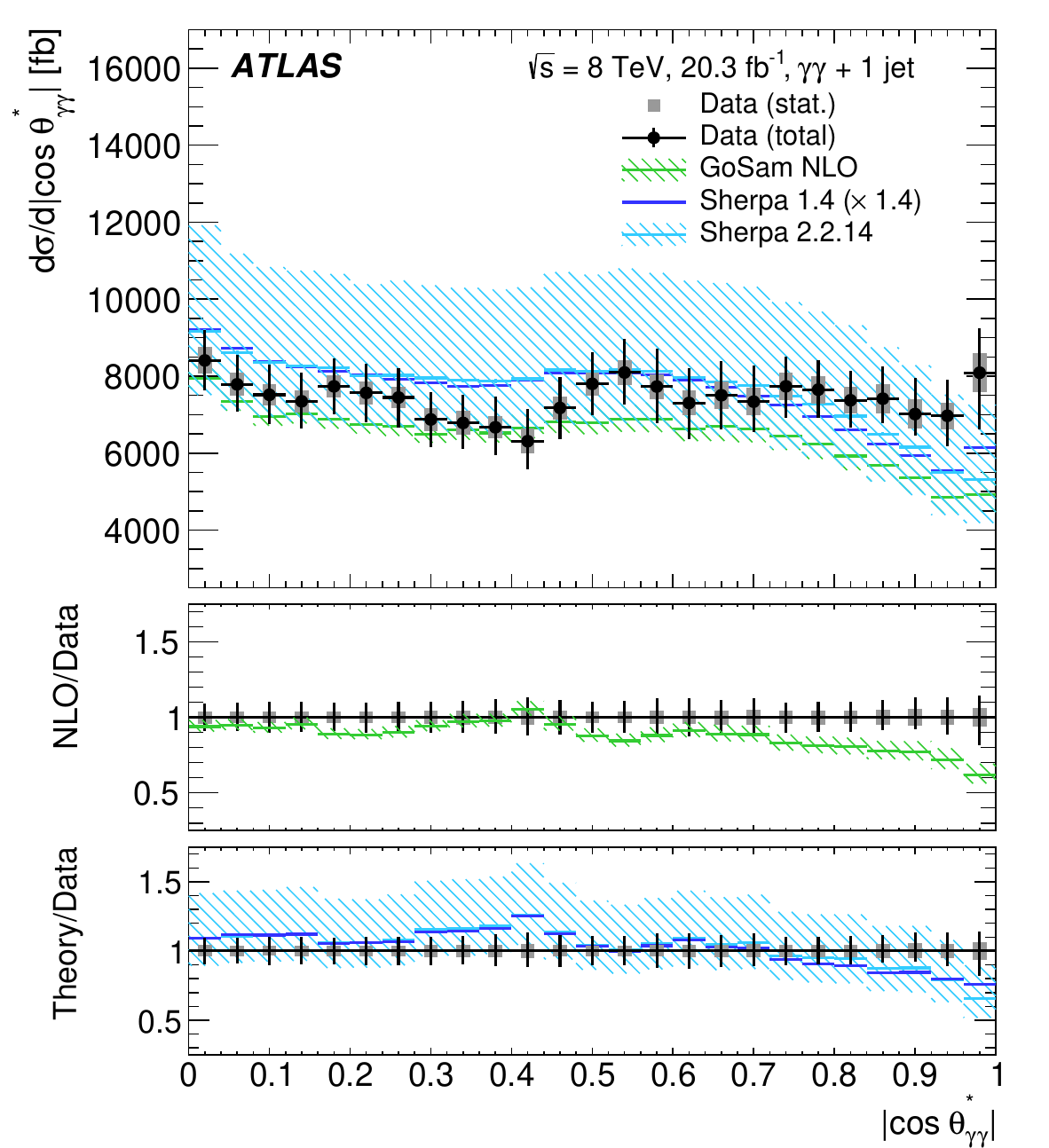}\label{fig:ggjj_1jet_excl_costhgg}}
\subfloat[]{\includegraphics[width= 0.47\textwidth]{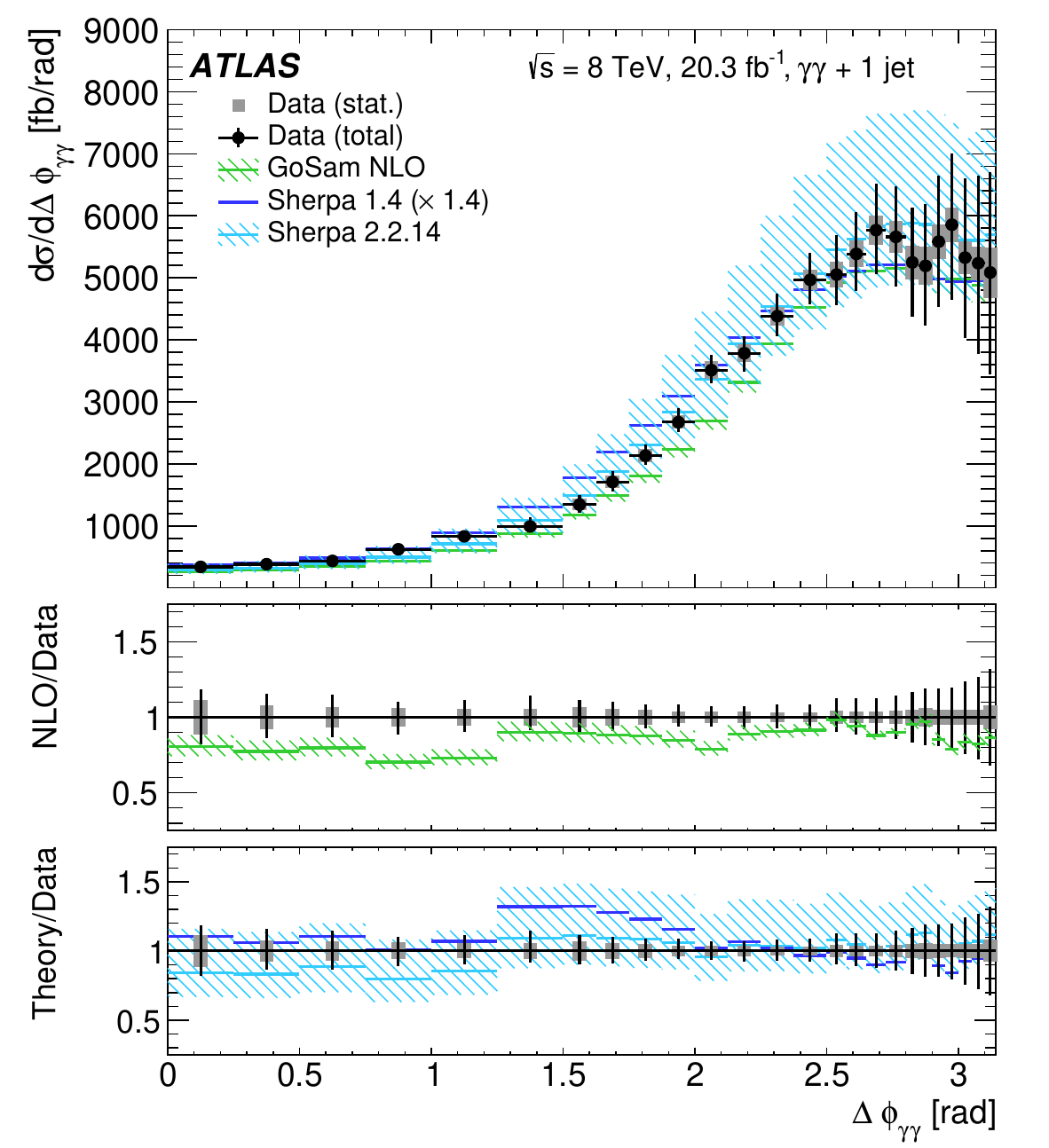}\label{fig:ggjj_1jet_excl_dphigg}}
\caption{
Differential cross sections as a function of (a) \mgg, (b) \pTgg, (c) \costhgg\ and (d) \dphigg\ for the $\Nj = 1$ category,
compared with \gosam\ NLO and \sherpa\ 2.2.14 predictions,
and to LO \sherpa\ 1.4.1 predictions normalized to data.
Ratios of theory to data are shown in the lower panel.
Arrows indicate data points outside the displayed ratio range.
Bins where the \gosam\ NLO scale uncertainty exceeds the central prediction are indicated by open markers without uncertainty bands.
The normalization factor for \sherpa\ 1.4.1 is quoted in the plot legend.
}
\label{fig:ggjj_1jet_excl_diphoton}
\end{figure}

The diphoton observables for the $\Nj = 1$ category are shown in Figure~\ref{fig:ggjj_1jet_excl_diphoton}. The \gosam\ \pTgg\ distribution shows a feature at around 25~\GeV, which is not observed in data. This can be explained as follows. The diphoton system needs to recoil against the one jet in the event in the Born contributions, with the $\pTj$ starting at 25~\GeV. This is no longer true for real contributions, where \pTj\ can start from 0~\GeV.  Adding up the contributions, the \pTgg\ distribution displays a dip at the \pTj\ threshold. A similar feature is observed in Figure~5 of Ref.~\cite{Greiner:2015jha}.
{\textsc{Sherpa~2.2.14}} provides a good description of the diphoton observables in the $\Nj=1$ category, except at high \costhgg\ and small \dphigg. The \mgg\ distribution is well described across the full range, at a level comparable to the \gosam\ NLO prediction. The \pTgg\ distribution is well reproduced without the low-\pTgg\ threshold feature present in the fixed-order calculation. The \dphigg\ distribution is well described above about 1.25~rad, including the intermediate-separation region where \gosam\ shows deviations from data, while at smaller separations the prediction lies 15--20\% below the data, within its scale uncertainty. For \costhgg, the agreement is good for $\costhgg < 0.7$, with a moderate undershoot at higher values that is common to all predictions.
\sherpa\ 1.4.1 fails to describe the shape of \mgg, \dphigg\ and $\costhgg>0.8$.

\begin{figure}
\centering
\subfloat[]{\includegraphics[width= 0.49\textwidth]{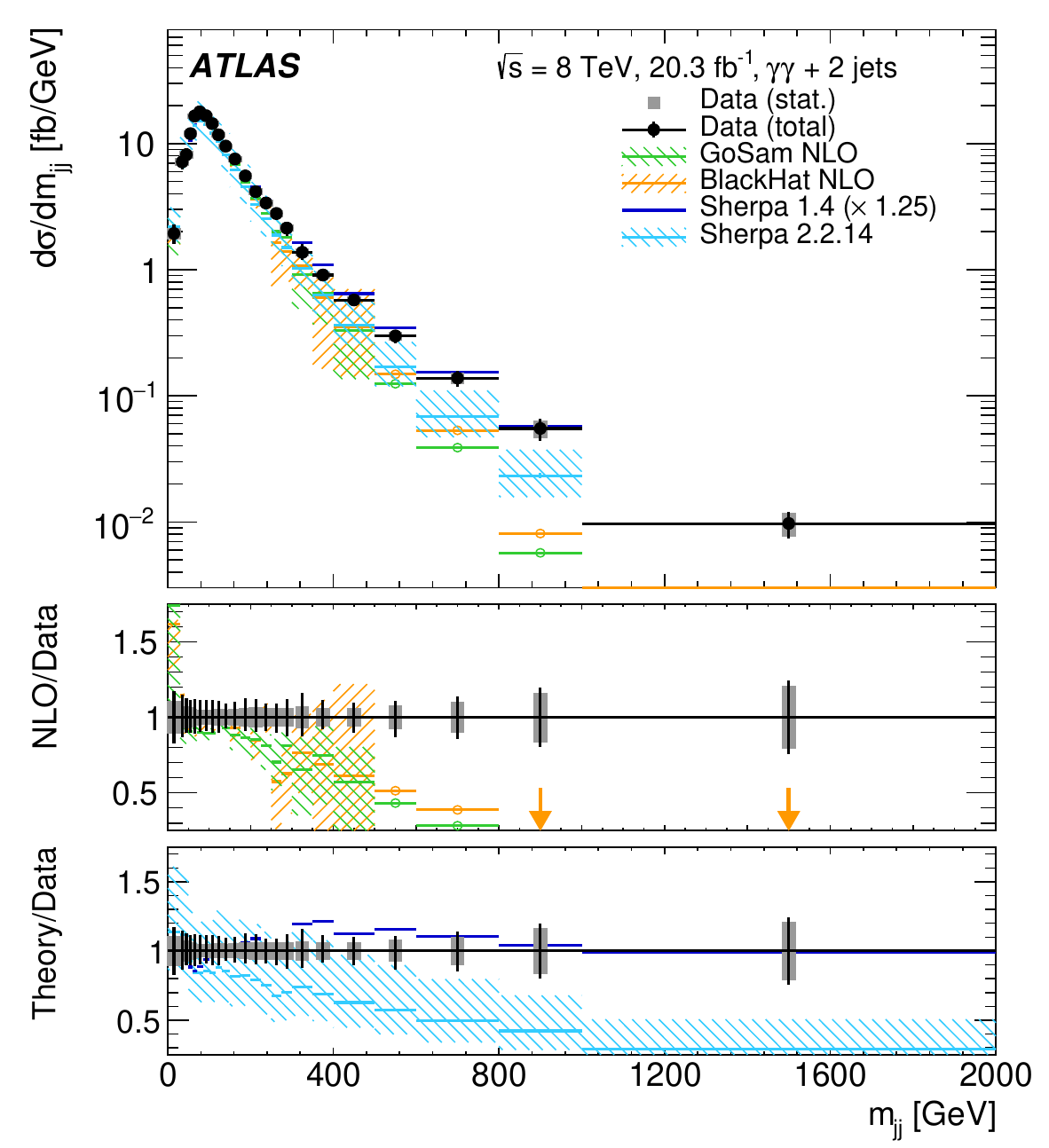}\label{fig:ggjj_2jet_excl_mjj}}
\subfloat[]{\includegraphics[width= 0.49\textwidth]{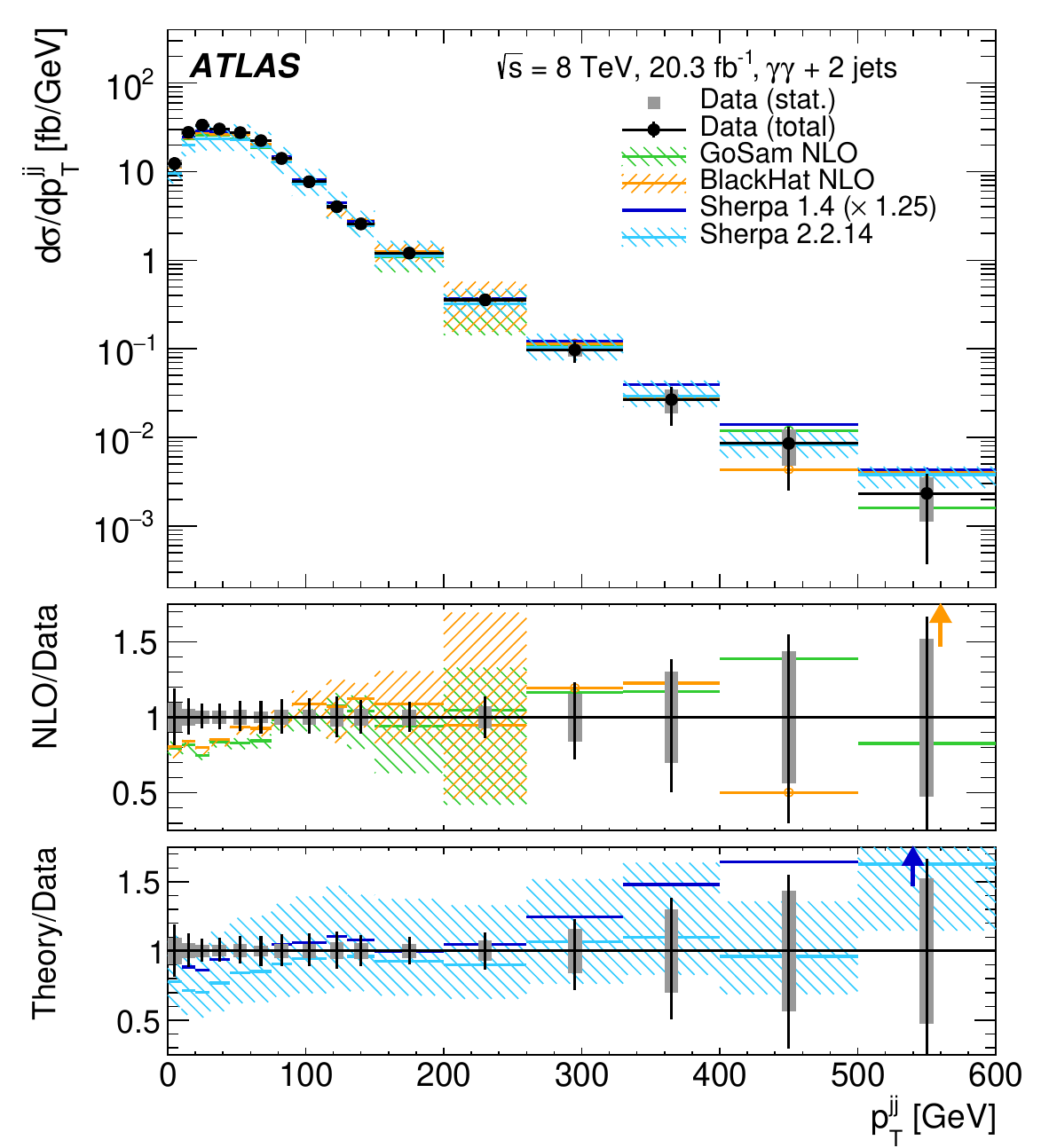}\label{fig:ggjj_2jet_excl_ptjj}} \\
\subfloat[]{\includegraphics[width= 0.49\textwidth]{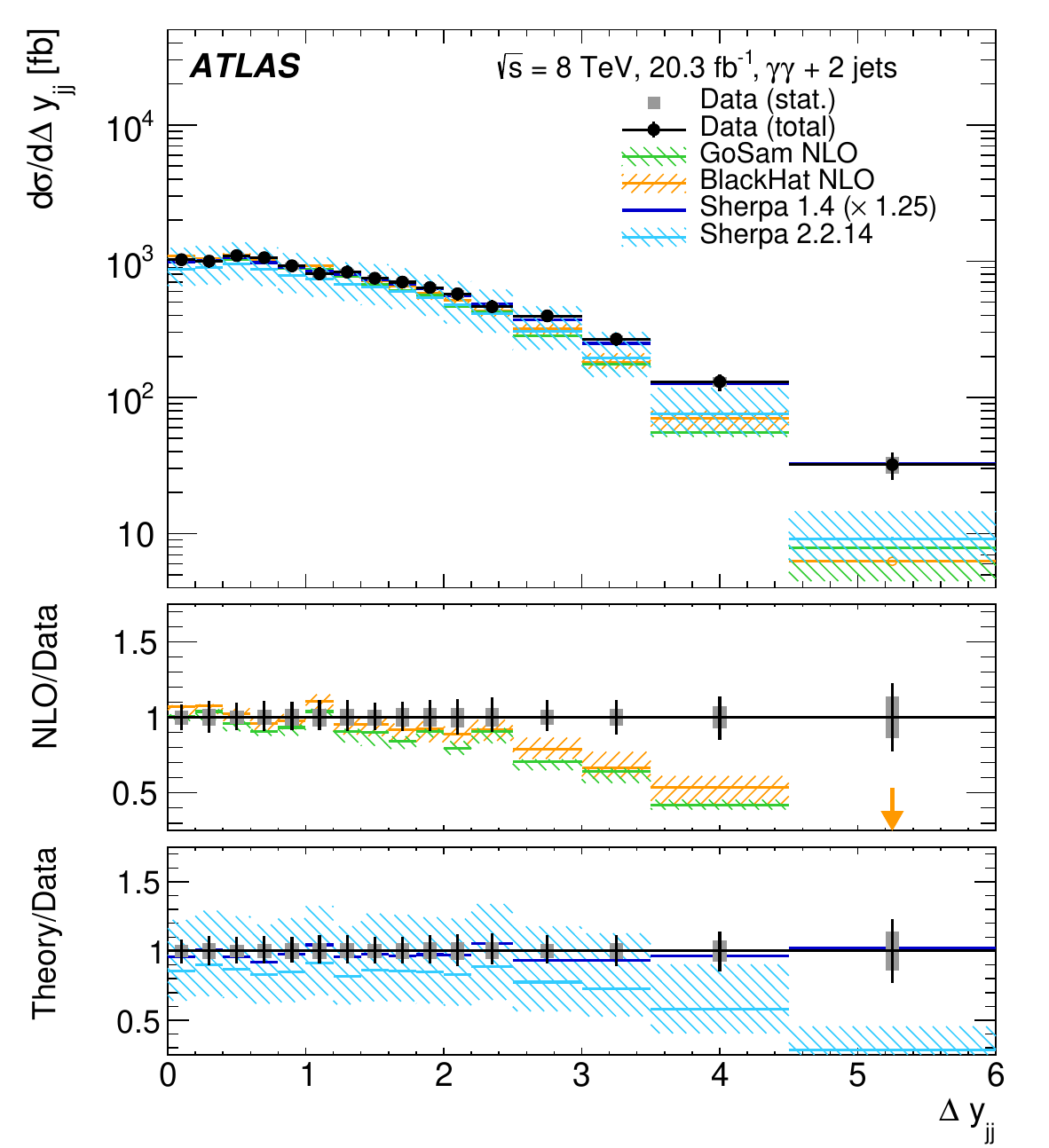}\label{fig:ggjj_2jet_excl_dyjj}}
\subfloat[]{\includegraphics[width= 0.49\textwidth]{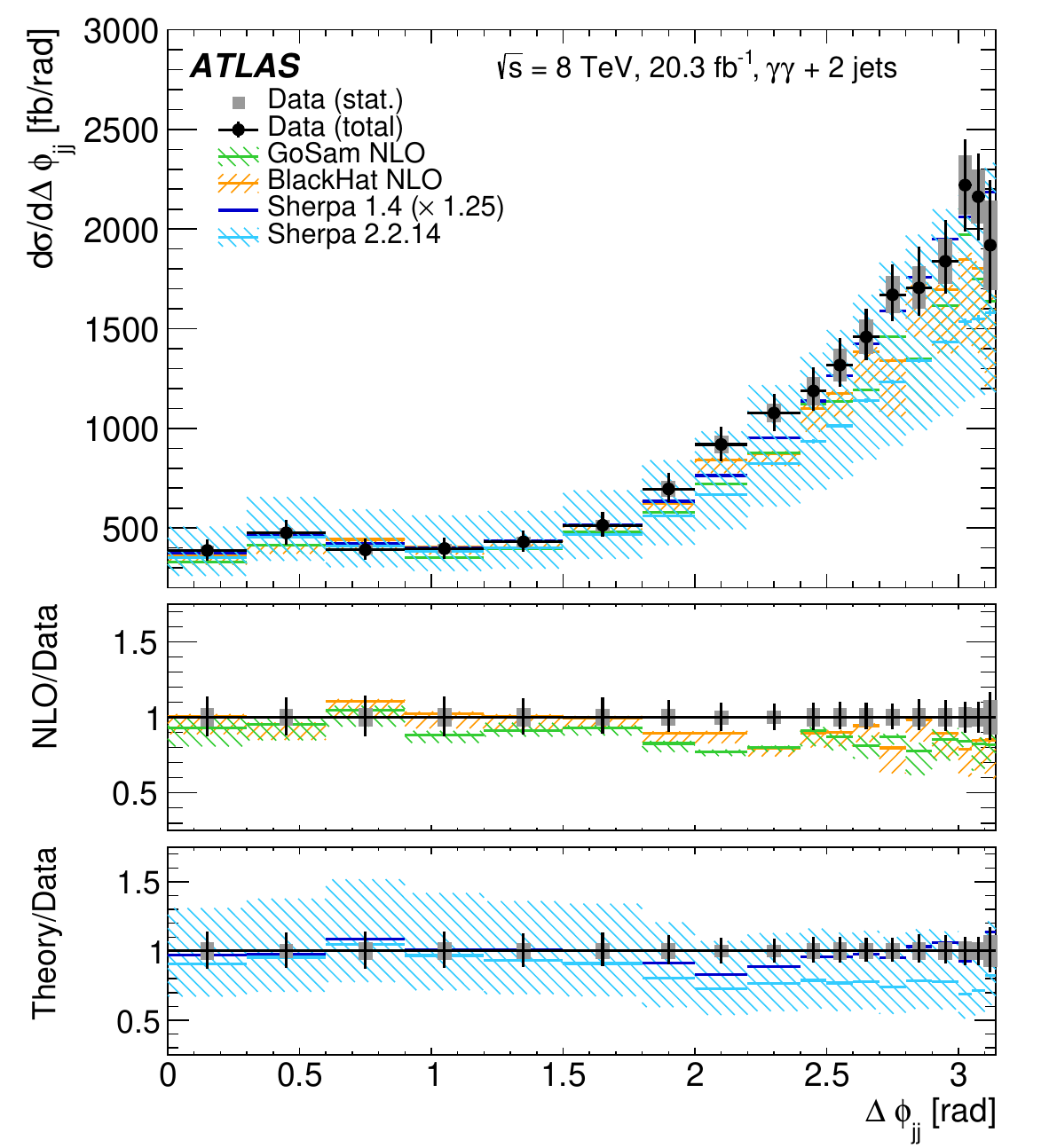}\label{fig:ggjj_2jet_excl_dphijj}}
\caption{Differential cross sections in the $\Nj = 2$ category as a function of (a) \mjj, (b) \pTjj, (c) \dyjj\ and (d) \dphijj,
compared with theoretical predictions from \gosam\ NLO, \blackhat\ NLO and \sherpa\ 2.2.14,
and to LO \sherpa\ 1.4.1 predictions normalized to data.
Ratios of theory to data are shown in the lower panel.
Arrows indicate data points outside the displayed ratio range.
Bins where the \gosam\ or \blackhat\ NLO scale uncertainty exceeds the central prediction are indicated by open markers without uncertainty bands.
The normalization factor for \sherpa\ 1.4.1 is quoted in the plot legend.
}
\label{fig:ggjj_2jet_excl_dijet}
\end{figure}

For the $\Nj=2$ category, the comparison of the measured data to \gosam\ and \blackhat\ predictions for the dijet kinematic observables is shown in Figure~\ref{fig:ggjj_2jet_excl_dijet}. \gosam\ and \blackhat\ give a similar description of the \mjj\ distribution: both lie slightly below the data at low masses, and the deficit grows rapidly with \mjj, from about 30--40\% at 300--500~\GeV\ to a factor of several in the highest-mass bins, where both calculations also suffer from large statistical and scale uncertainties. The separation of jets in rapidity \dyjj, is well described up to 2.5, where both \gosam\ and \blackhat\ start to deviate. \gosam\ and \blackhat\ describe the \dphijj\ distribution well below 1.8, above that, they are both below data but mostly within uncertainties. \gosam\ and \blackhat\ offer a similar description of \pTjj, both underestimate data below 90~\GeV\ and describe it well above that, but with data uncertainty rising from 10\% up to 200\% at high-\pTjj.
{\textsc{Sherpa~2.2.14}} provides a good description of the dijet observables, with the exception of \mjj. The \mjj\ distribution is described better than by the fixed-order calculations and agrees with the data within uncertainties up to about 250~\GeV, but the prediction falls increasingly below the data at higher masses, by about 50\% at 600--800~\GeV\ and about 70\% above 1~\TeV. The \pTjj\ distribution is well described across the full range. For \dyjj, good agreement is found up to $\dyjj \approx 3$, beyond which the prediction falls well below data in a region also poorly described by the fixed-order calculations. The \dphijj\ observable is very well described throughout.
\sherpa\ 1.4.1, once normalized to the data, is the only prediction that follows the \mjj\ shape over the full range, within about 20\%; it also describes the low-\pTjj\ region and \dyjj\ well within data uncertainties, and offers the best LO shape description for \dphijj.

\begin{figure}
\centering
\subfloat[]{\includegraphics[width= 0.47\textwidth]{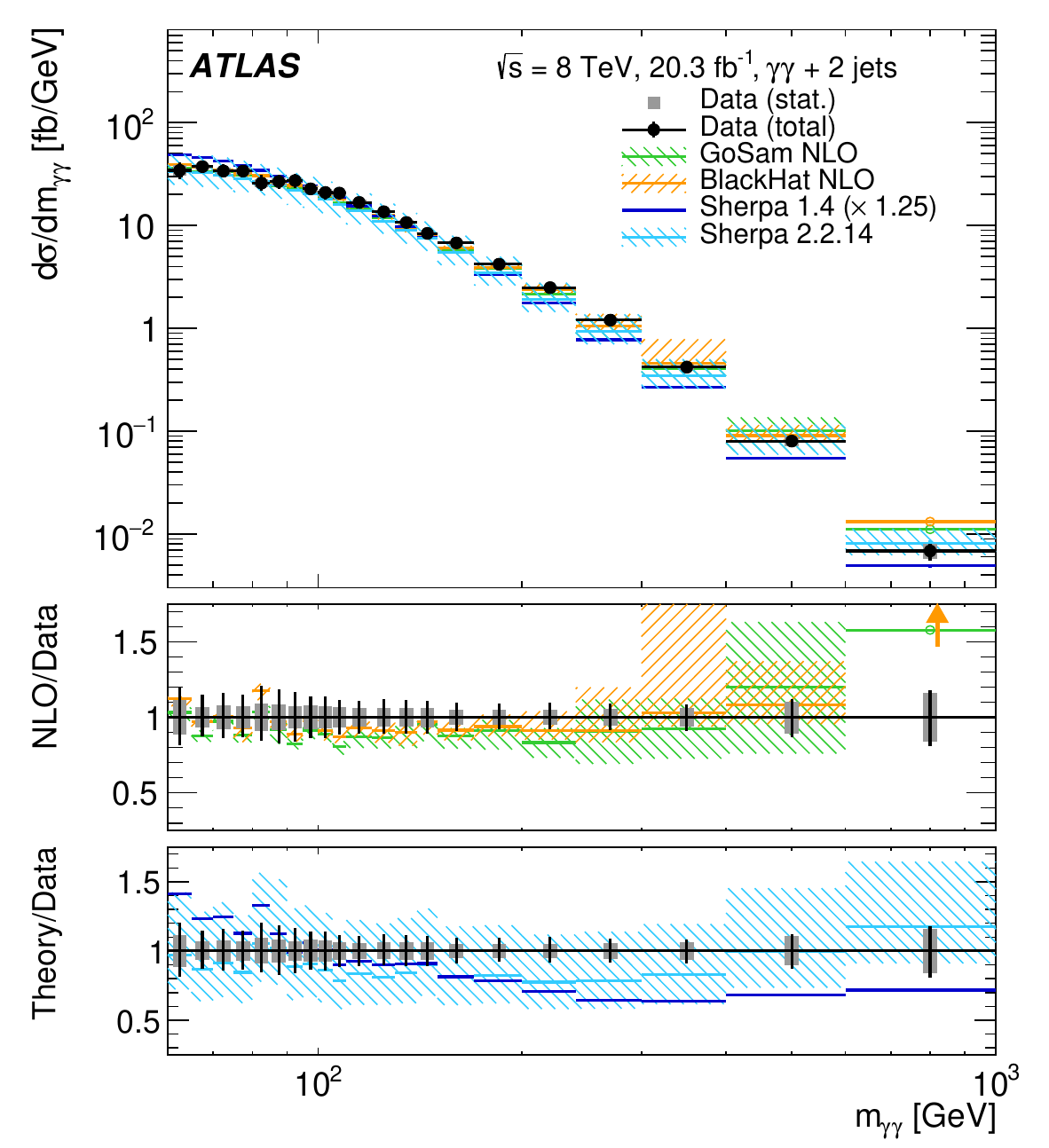}\label{fig:ggjj_2jet_excl_mgg}}
\subfloat[]{\includegraphics[width= 0.47\textwidth]{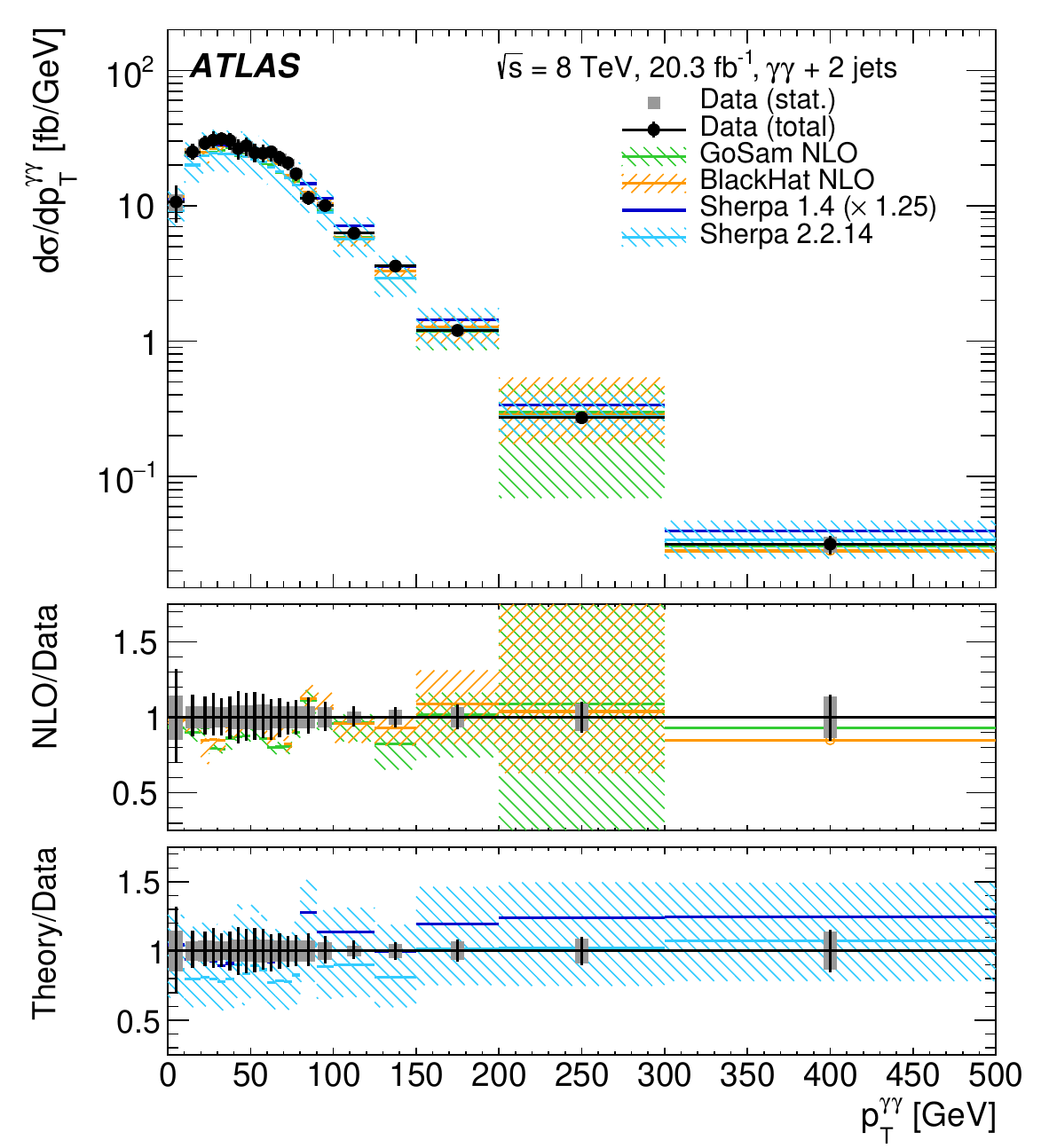}\label{fig:ggjj_2jet_excl_ptgg}} \\
\subfloat[]{\includegraphics[width= 0.47\textwidth]{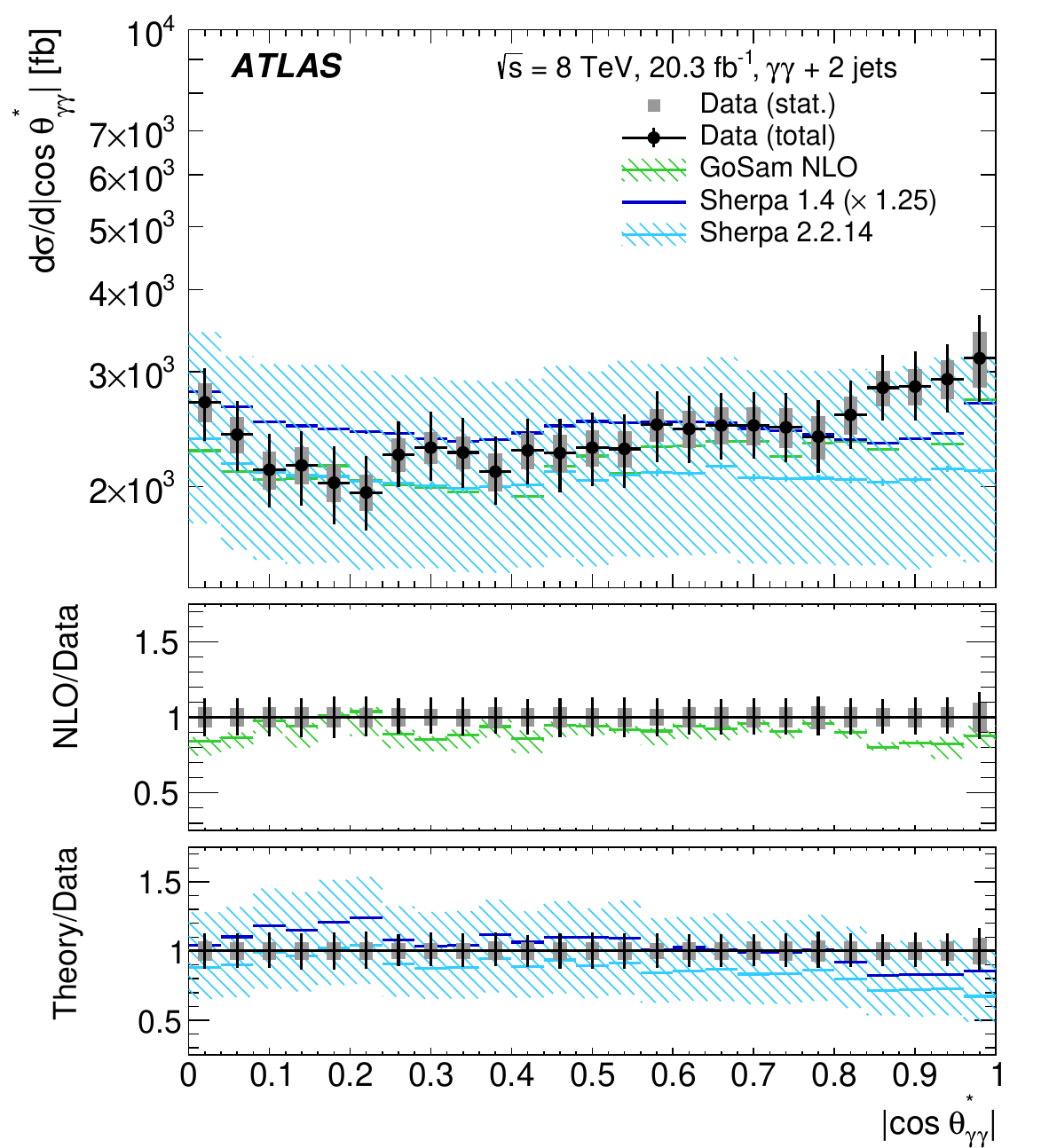}\label{fig:ggjj_2jet_excl_costhgg}}
\subfloat[]{\includegraphics[width= 0.47\textwidth]{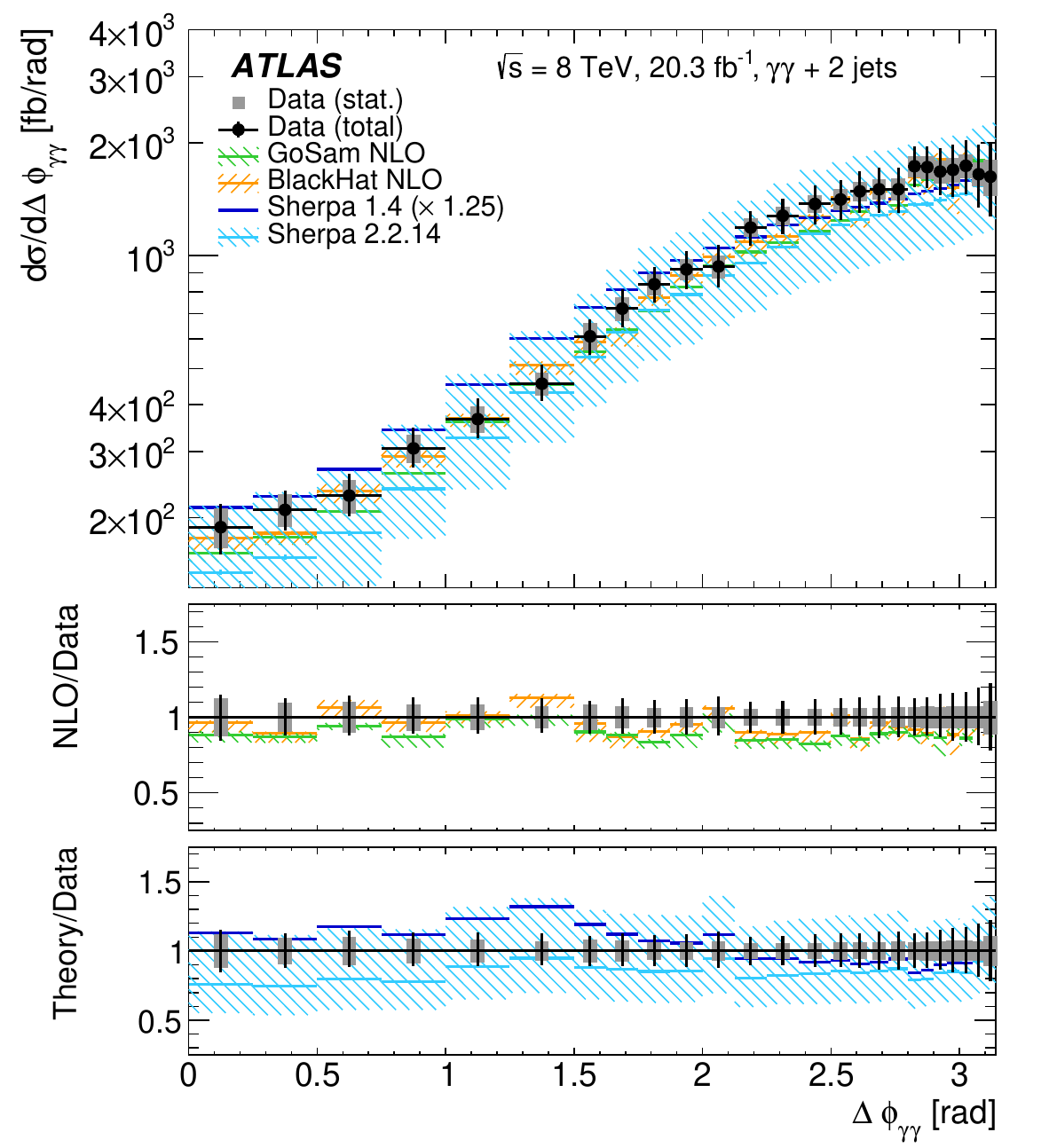}\label{fig:ggjj_2jet_excl_dphigg}}
\caption{
Differential cross sections as a function of (a) \mgg, (b) \pTgg, (c) \costhgg\ and (d) \dphigg\ for the $\Nj = 2$ category,
compared with \gosam\ NLO and \sherpa\ 2.2.14 predictions,
and to LO \sherpa\ 1.4.1 predictions normalized to data.
Ratios of theory to data are shown in the lower panel.
Arrows indicate data points outside the displayed ratio range.
Bins where the \gosam\ or \blackhat\ NLO scale uncertainty exceeds the central prediction are indicated by open markers without uncertainty bands.
The normalization factor for \sherpa\ 1.4.1 is quoted in the plot legend.
}
\label{fig:ggjj_2jet_excl_diphoton}
\end{figure}

The four diphoton observables in the $\Nj=2$ category are well described by both fixed order calculations, as seen in Figure~\ref{fig:ggjj_2jet_excl_diphoton}.
{\textsc{Sherpa~2.2.14}} provides a good description of all diphoton observables in the $\Nj=2$ category, broadly comparable to the fixed-order predictions. The \mgg\ shape is well reproduced from threshold to high masses. For \pTgg, the prediction agrees well with data for $\pTgg < 150~\GeV$ and remains consistent at higher values within the larger uncertainties. The \costhgg\ distribution is described well for $\costhgg < 0.8$, with a modest undershoot at higher values that is also present in \gosam. For \dphigg, a reasonable description is found across the full range, with a slight undershoot at intermediate separations.
\sherpa\ 1.4.1 fails to describe the shape of \mgg, \pTgg\ and \costhgg, and provides an adequate description of \dphigg.

\begin{figure}
\centering
\includegraphics[width= 0.49\textwidth]{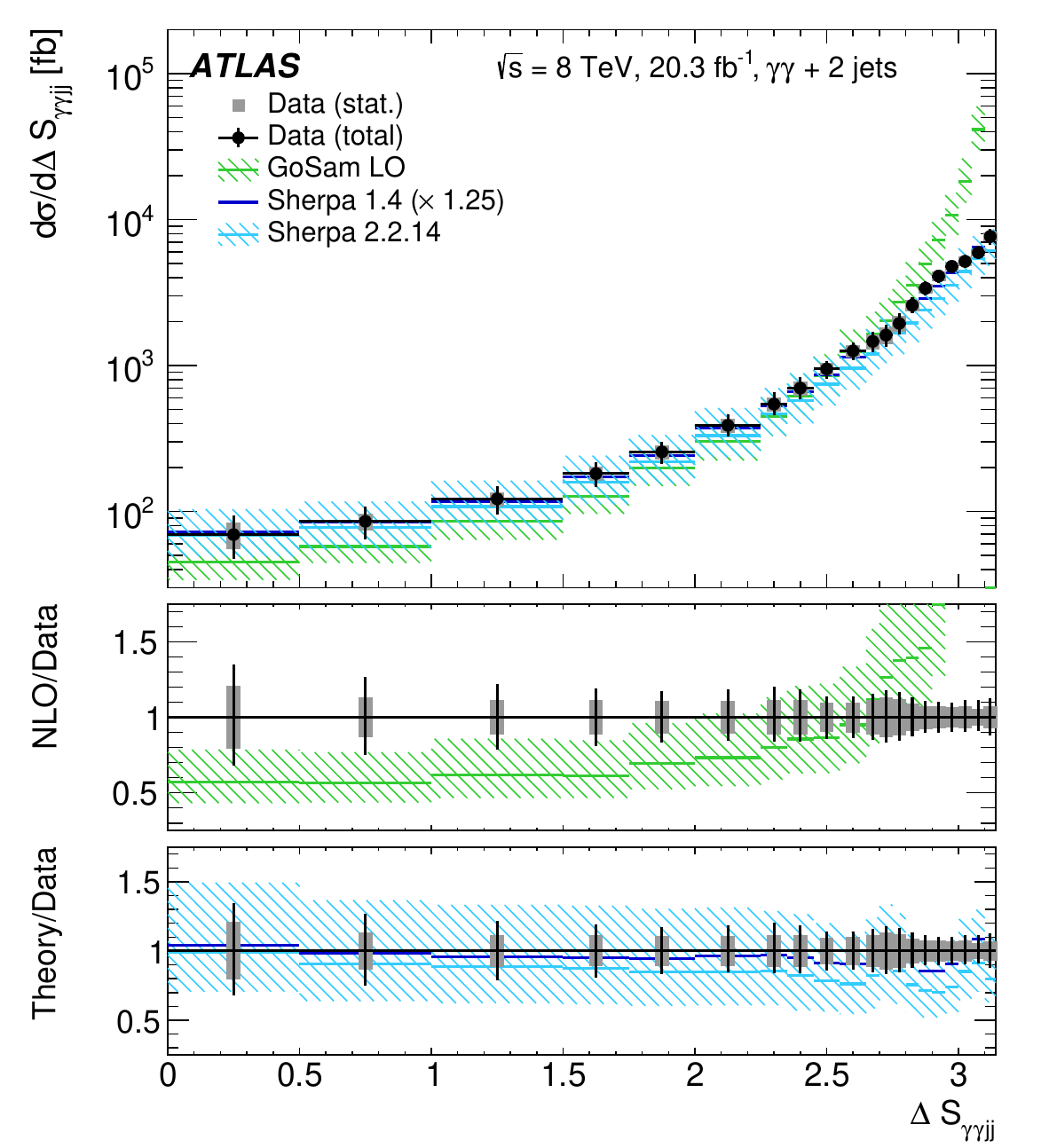}
\caption{
Differential cross section as a function of the \dSggjj\ observable for the $\Nj = 2$ category
compared with theoretical predictions from \gosam\ LO and \sherpa\ 2.2.14,
and to LO \sherpa\ 1.4.1 predictions normalized to data.
Ratios of theory to data are shown in the lower panel.
Arrows indicate data points outside the displayed ratio range.
Bins where the \gosam\ scale uncertainty exceeds the central prediction are indicated by open markers without uncertainty bands.
The normalization factor for \sherpa\ 1.4.1 is quoted in the plot legend.
}
\label{fig:ggjj_2jet_excl_dS}
\end{figure}

The \dSggjj\ ($\equiv |\Delta\phi_{\gamma\gamma,\text{jj}}|$) distribution is shown compared with a \gosam\ prediction, a \sherpa\ 2.2.14 prediction, and \sherpa\ 1.4.1 in Figure~\ref{fig:ggjj_2jet_excl_dS}. As noted in Section~\ref{sec:theory}, the \gosam\ prediction for this observable is computed at leading-order accuracy.
{\textsc{Sherpa~2.2.14}} provides a good description of the \dSggjj\ distribution across the bulk of the measured range (ratio within 15\% of data for $\dSggjj > 0.5~\text{rad}$).
\sherpa\ 1.4.1 describes the shape of \dSggjj\ very well. \gosam\ gives a good description up to $\dSggjj \simeq$~2.7, where it starts to deviate and the last bin is negative. This behavior is caused by a $\delta$ function in the calculation, used to describe the maximum azimuthal separation of the diphoton and dijet systems set at $\pi$. For a smoother and more accurate prediction, NLO computation of $\gamma\gamma+3\text{j}$ would be necessary~\cite{gosam_atlas}.

The \dSggjj\ distribution for the inclusive $\Nj \geq 2$ and $\Nj \geq 3$ categories is shown in Figure~\ref{fig:ggjj_23jet_incl_dS}. The qualitative features seen in the exclusive $\Nj = 2$ category are preserved: \sherpa~2.2.14 provides a good description across the bulk of the distribution in both categories, and \sherpa\ 1.4.1 gives a good shape description throughout. The low-\dSggjj\ region, where the diphoton and dijet systems are nearly collinear, is sensitive to multiple parton interactions and underlying event; all predictions tend to underestimate data in this region, with the deviation most pronounced in the $\Nj \geq 3$ category, consistent with the expected increase in MPI sensitivity at higher jet multiplicities.

\begin{figure}
\centering
\subfloat[]{\includegraphics[width= 0.49\textwidth]{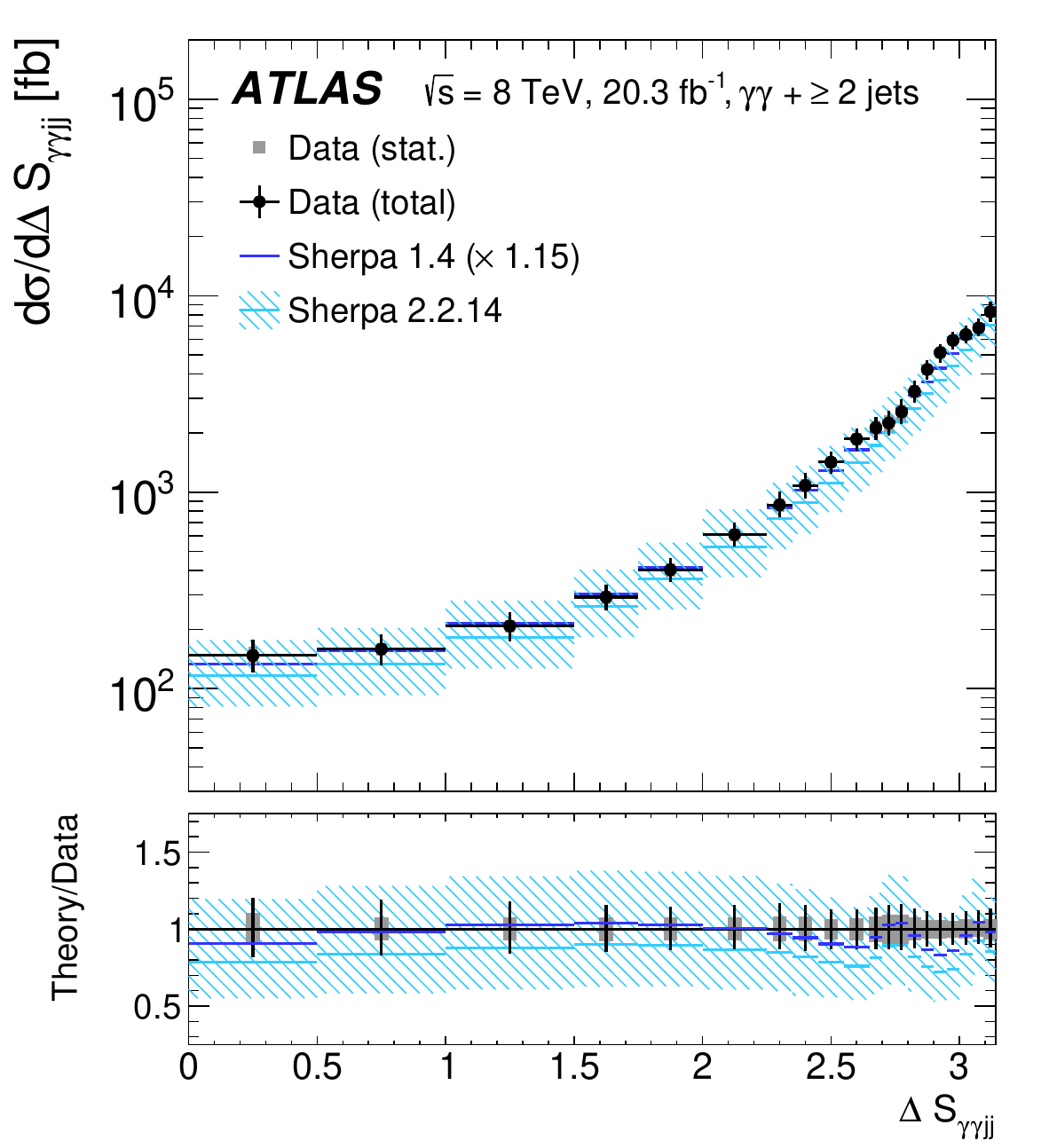}\label{fig:ggjj_2jet_incl_dSggjj}}
\subfloat[]{\includegraphics[width= 0.49\textwidth]{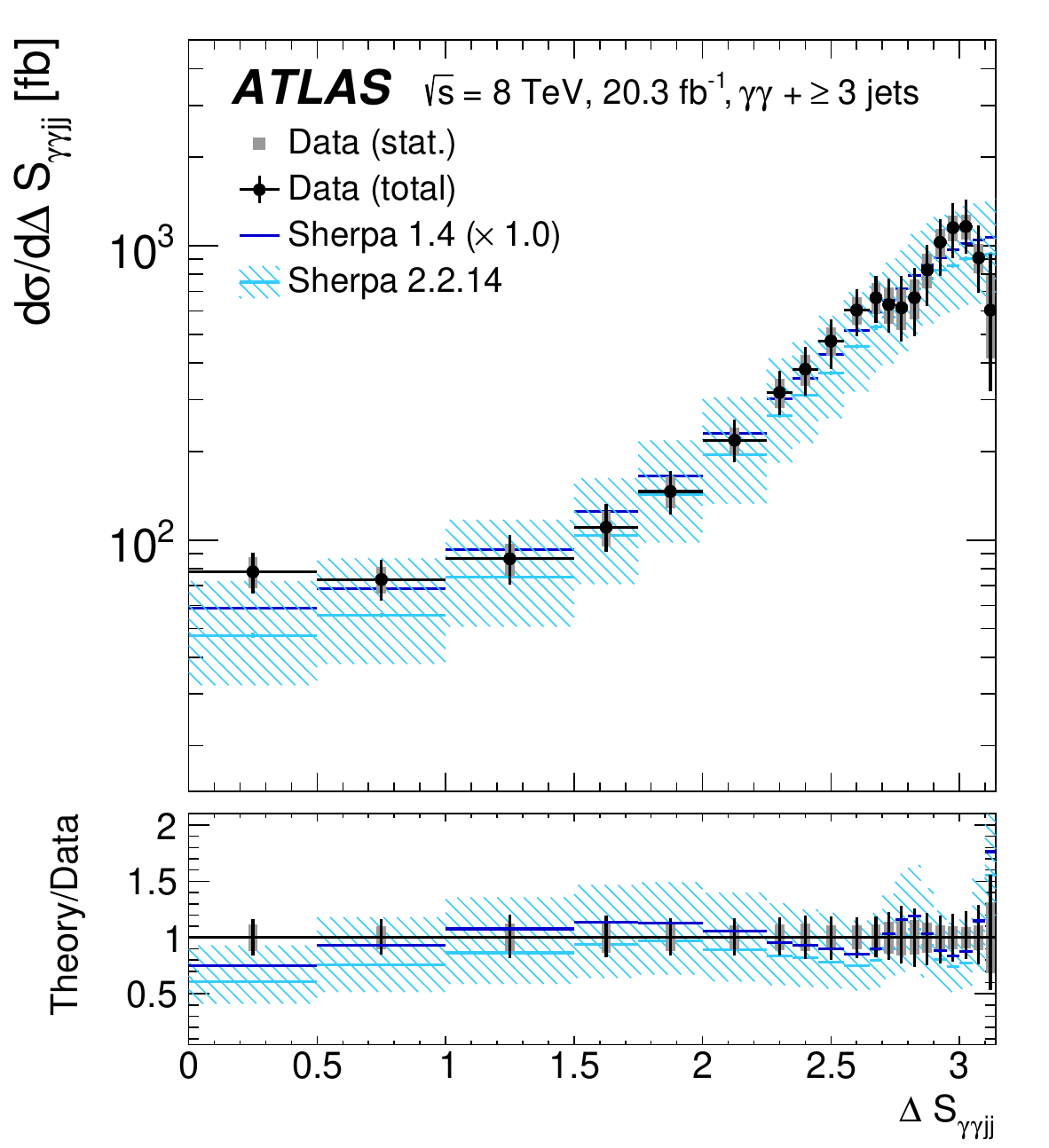}\label{fig:ggjj_3jet_incl_dS}}
\caption{Unfolded differential cross section as a function of the $\Delta S_{\gamma\gamma,\mathrm{jj}}$ observable for the (a) $\Nj \geq 2$ and (b) $\Nj \geq 3$ categories,
compared with {\textsc{Sherpa~2.2.14}} predictions and to LO \sherpa\ 1.4.1 predictions.
Ratios of theory to data are shown in the lower panel.
For the $\Nj \geq 2$ category, \sherpa\ 1.4.1 is normalized to data; the normalization factor is quoted in the plot legend.
For the $\Nj \geq 3$ category, \sherpa\ 1.4.1 is shown at absolute normalization since its predicted cross section matches the data.
Arrows indicate data points outside the displayed ratio range.}
\label{fig:ggjj_23jet_incl_dS}
\end{figure}

\begin{figure}
\centering
\subfloat[]{\includegraphics[width= 0.46\textwidth]{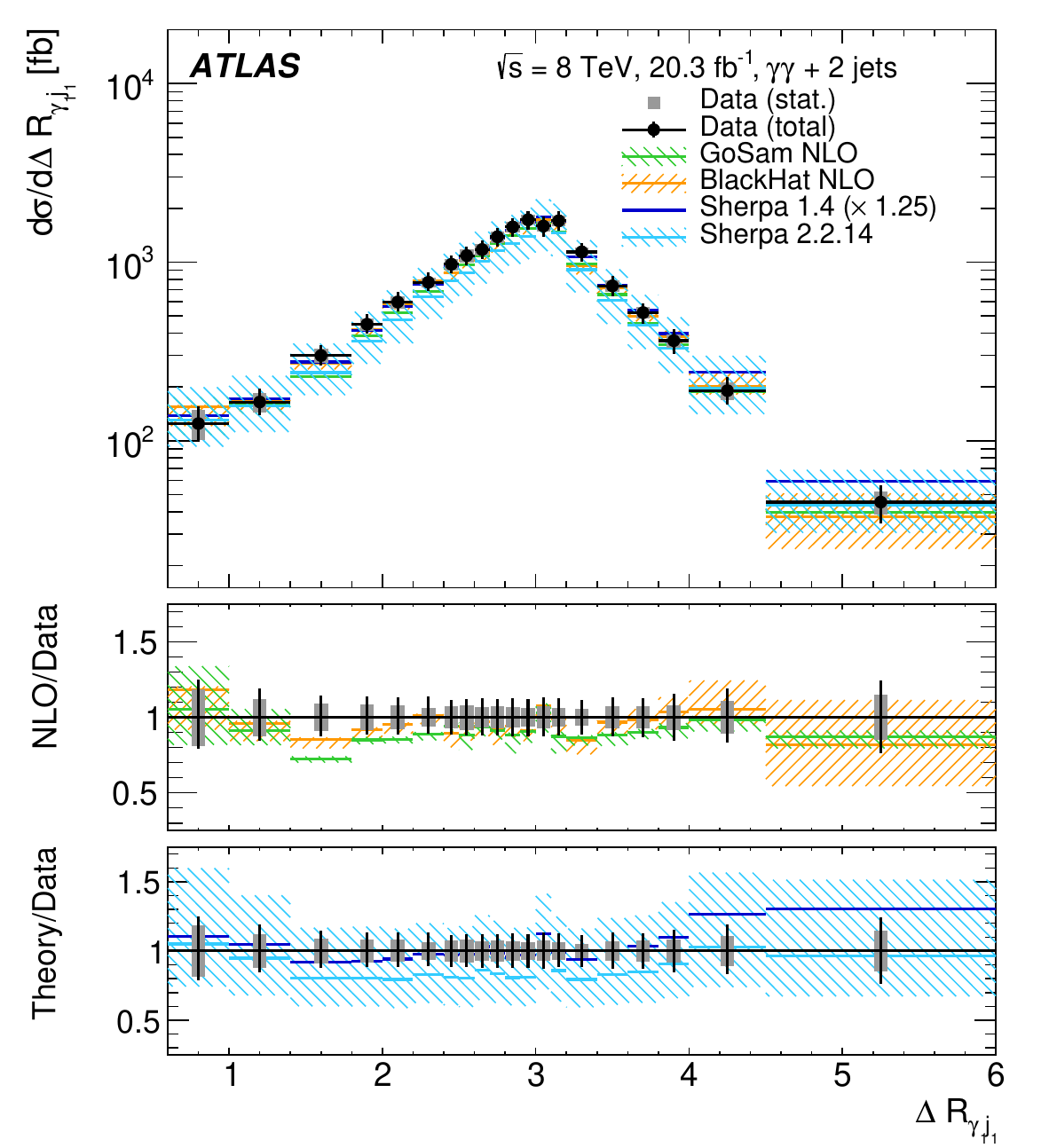}\label{fig:ggjj_2jet_excl_dRg1j1}}
\subfloat[]{\includegraphics[width= 0.46\textwidth]{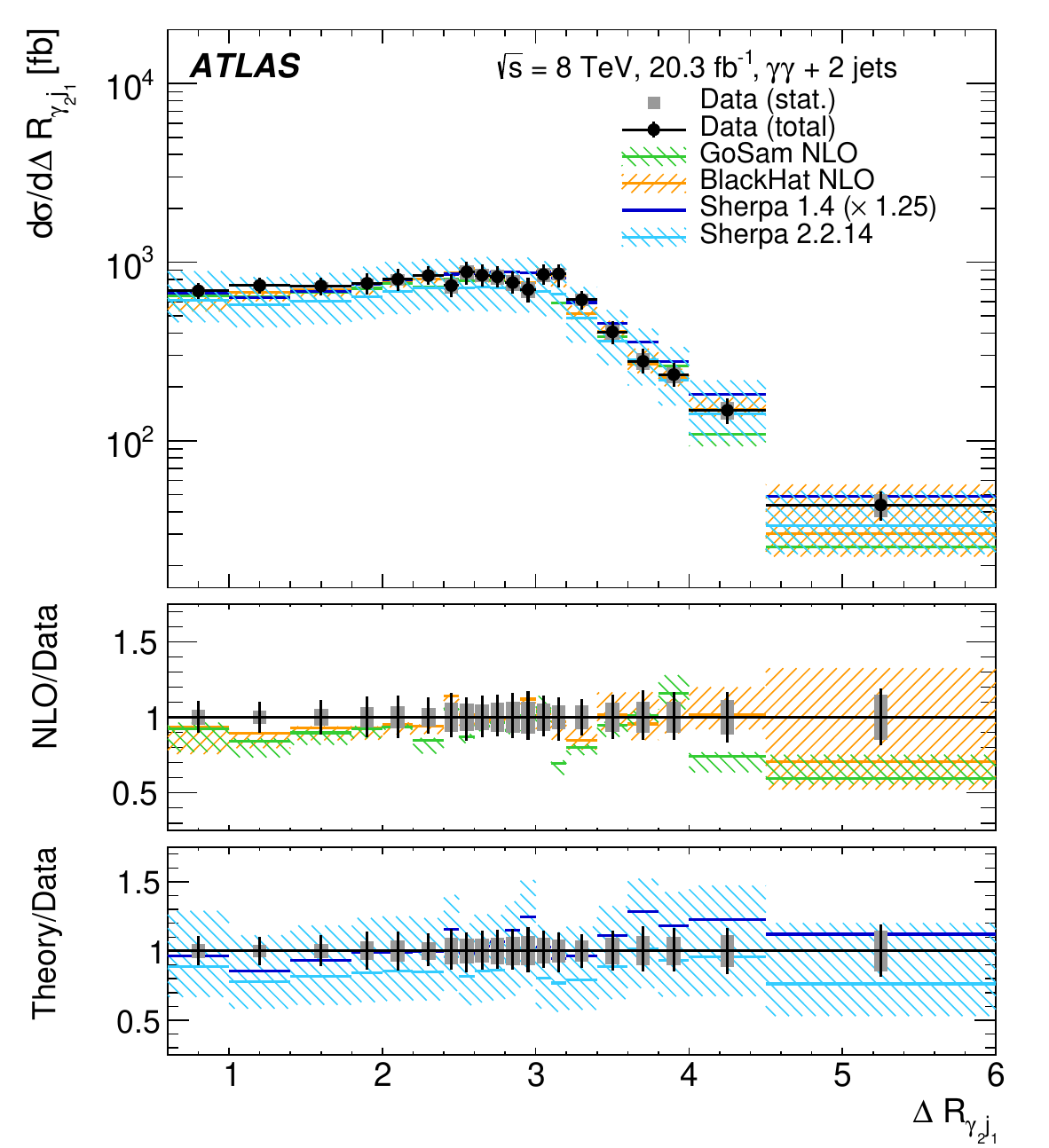}\label{fig:ggjj_2jet_excl_dRg2j1}} \\
\subfloat[]{\includegraphics[width= 0.46\textwidth]{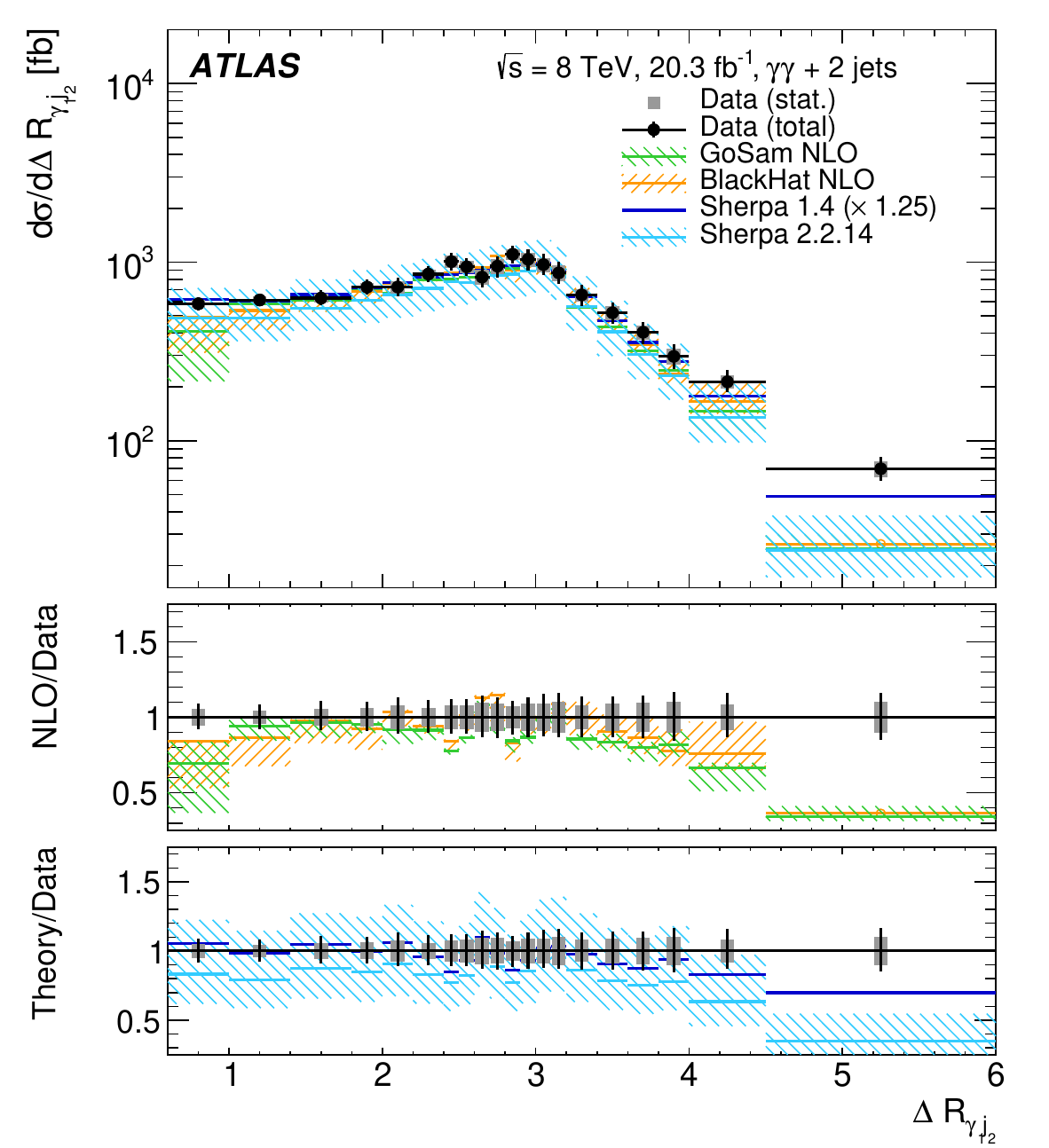}\label{fig:ggjj_2jet_excl_dRg1j2}}
\subfloat[]{\includegraphics[width= 0.46\textwidth]{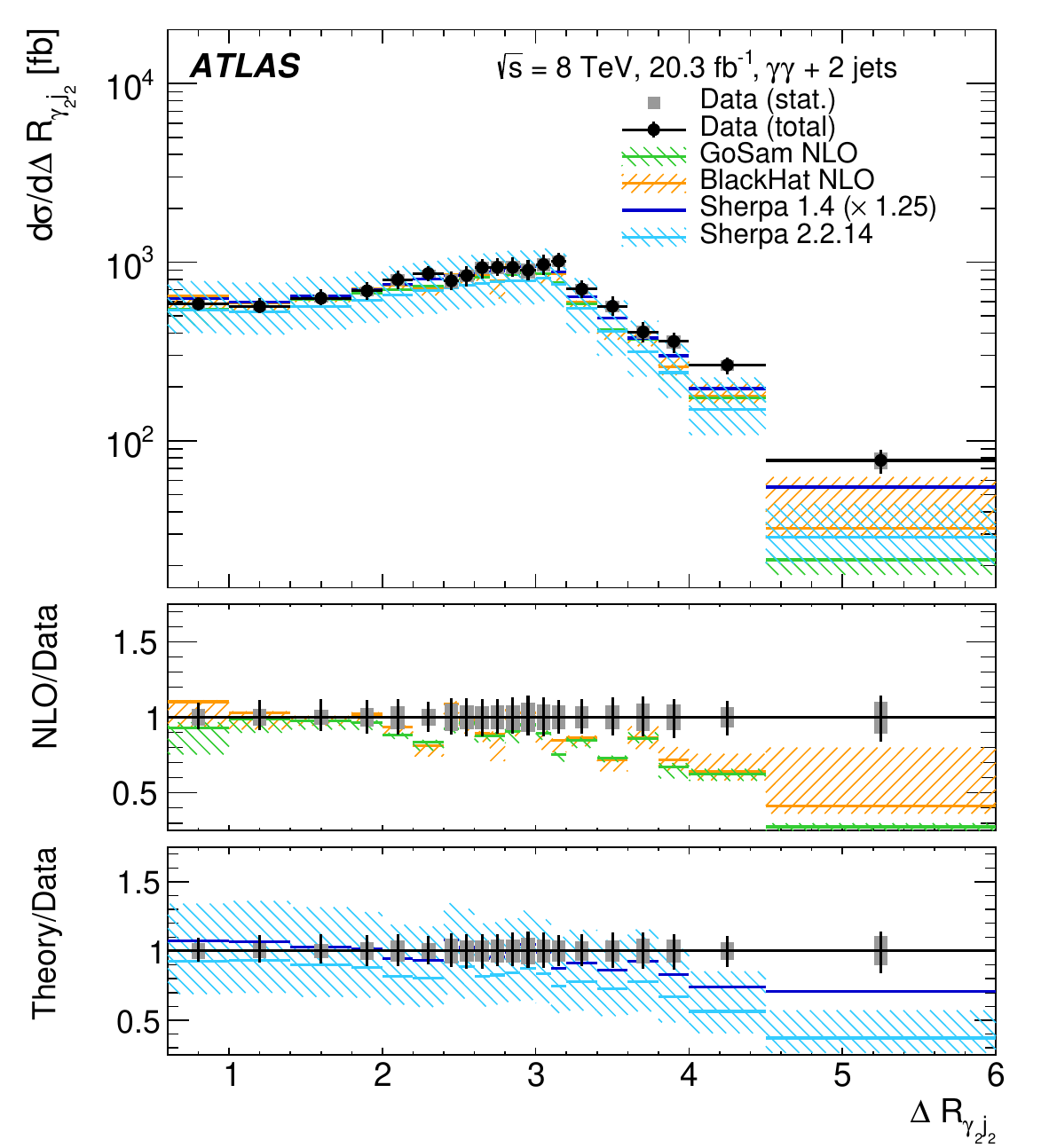}\label{fig:ggjj_2jet_excl_dRg2j2}}
\caption{Differential cross sections as a function of the $R$-separation for the leading jet and (a) the leading photon, (b) subleading photon and for the subleading jet and (c) the leading photon and (d) subleading photon in the $\Nj = 2$ category,
compared with theoretical predictions from \gosam, \blackhat\ and \sherpa\ 2.2.14,
and to LO \sherpa\ 1.4.1 predictions normalized to data.
Ratios of theory to data are shown in the lower panel.
Arrows indicate data points outside the displayed ratio range.
Bins where the \gosam\ or \blackhat\ NLO scale uncertainty exceeds the central prediction are indicated by open markers without uncertainty bands.
The normalization factor for \sherpa\ 1.4.1 is quoted in the plot legend.
}
\label{fig:ggjj_2jet_excl_dRgj}
\end{figure}

In Figure~\ref{fig:ggjj_2jet_excl_dRgj}, the photon--jet separation is very well described by both \gosam\ and \blackhat, except for the region of $\dRgtwojtwo>4$, where they both underestimate the data.
{\textsc{Sherpa~2.2.14}} provides a good description of all four photon--jet separation observables in the $\Nj=2$ category, remaining within approximately 15\% of data across the bulk of the distribution. In the highest separation bin, the prediction falls slightly below data, consistent with the behavior of the fixed-order calculations.
\sherpa\ 1.4.1 provides a good shape description of the \dRgj\ observables in the $\Nj =2$ category.

\begin{figure}
\centering
\subfloat[]{\includegraphics[width= 0.49\textwidth]{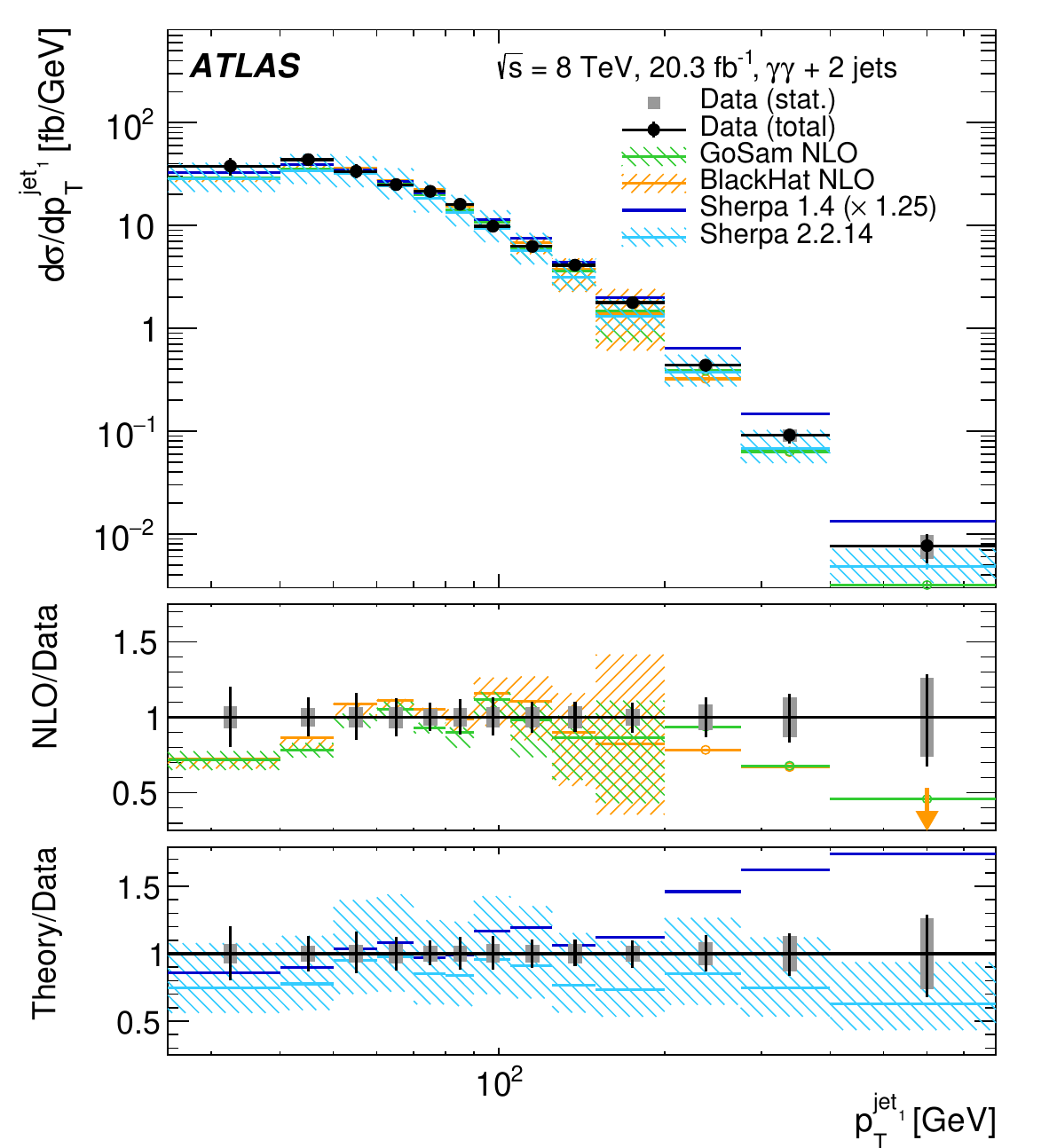}\label{fig:ggjj_2jet_excl_ptj1}}
\subfloat[]{\includegraphics[width= 0.49\textwidth]{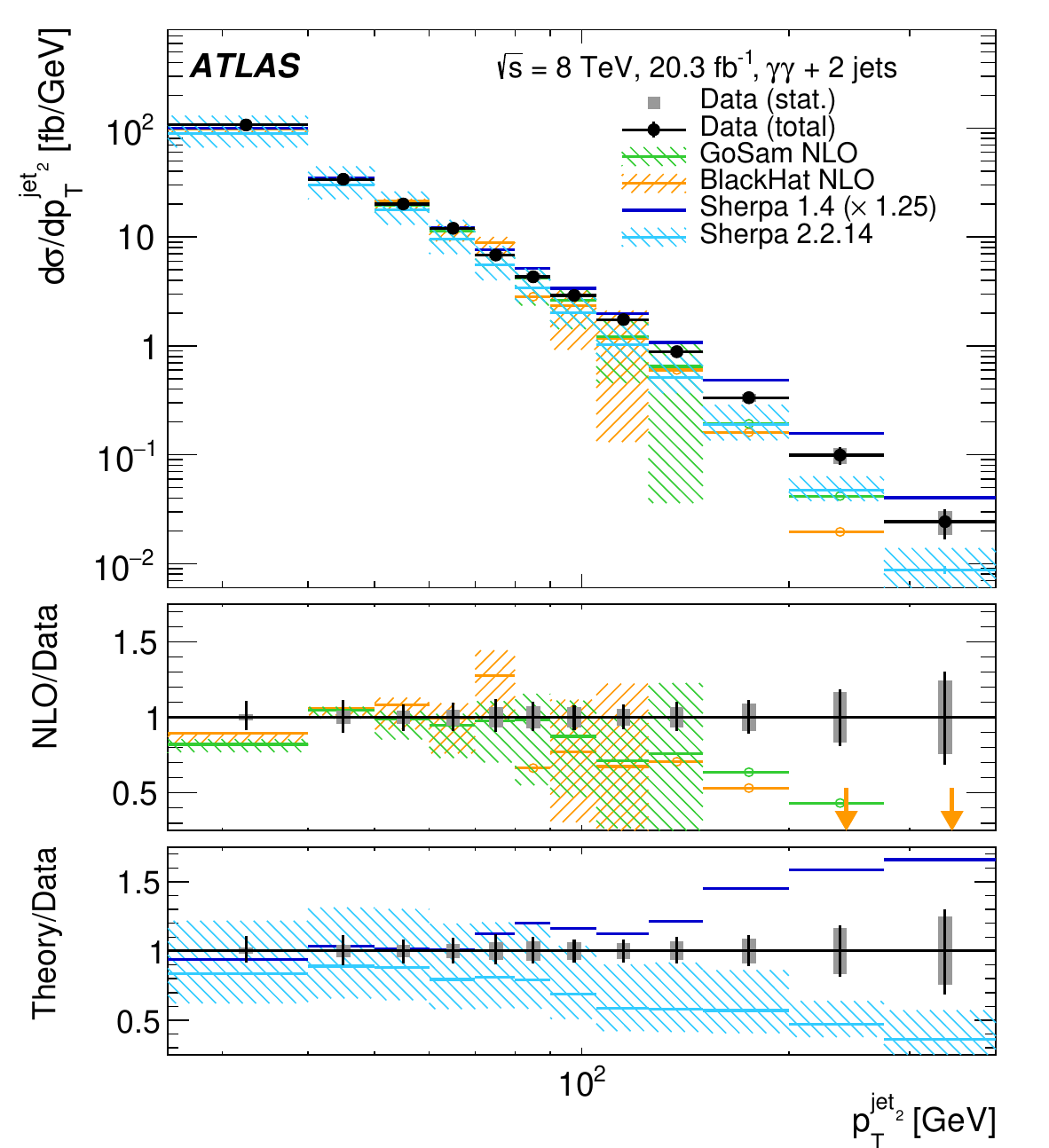}\label{fig:ggjj_2jet_excl_ptj2}}
\caption{
Differential cross sections as a function of the (a) leading jet transverse momentum $\pTjone$ and (b) subleading jet transverse momentum $\pTjtwo$ for the $\Nj = 2$ category,
compared with theoretical predictions from \gosam, \blackhat\ and \sherpa\ 2.2.14,
and to LO \sherpa\ 1.4.1 predictions normalized to data.
Ratios of theory to data are shown in the lower panel.
Arrows indicate data points outside the displayed ratio range.
Bins where the \gosam\ or \blackhat\ NLO scale uncertainty exceeds the central prediction are indicated by open markers without uncertainty bands.
The normalization factor for \sherpa\ 1.4.1 is quoted in the plot legend.
}
\label{fig:ggjj_2jet_excl_ptj12}
\end{figure}

Both fixed-order NLO calculations provide a good description of the leading and subleading \pTj, shown in Figure~\ref{fig:ggjj_2jet_excl_ptj1} and~\ref{fig:ggjj_2jet_excl_ptj2}, respectively.
For the jet transverse momenta, {\textsc{Sherpa~2.2.14}} describes the leading-jet \pT\ reasonably well for $\pTjone \lesssim 80~\GeV$, with an increasing undershoot at higher values. For the subleading-jet \pT, good agreement is found for $\pTjtwo \lesssim 60~\GeV$, after which the prediction falls significantly below data, consistent with the behavior of the fixed-order calculations.
\sherpa\ 1.4.1 fails to describe the shapes of both jet transverse momenta.

\begin{figure}
\centering
\subfloat[]{\includegraphics[width= 0.49\textwidth]{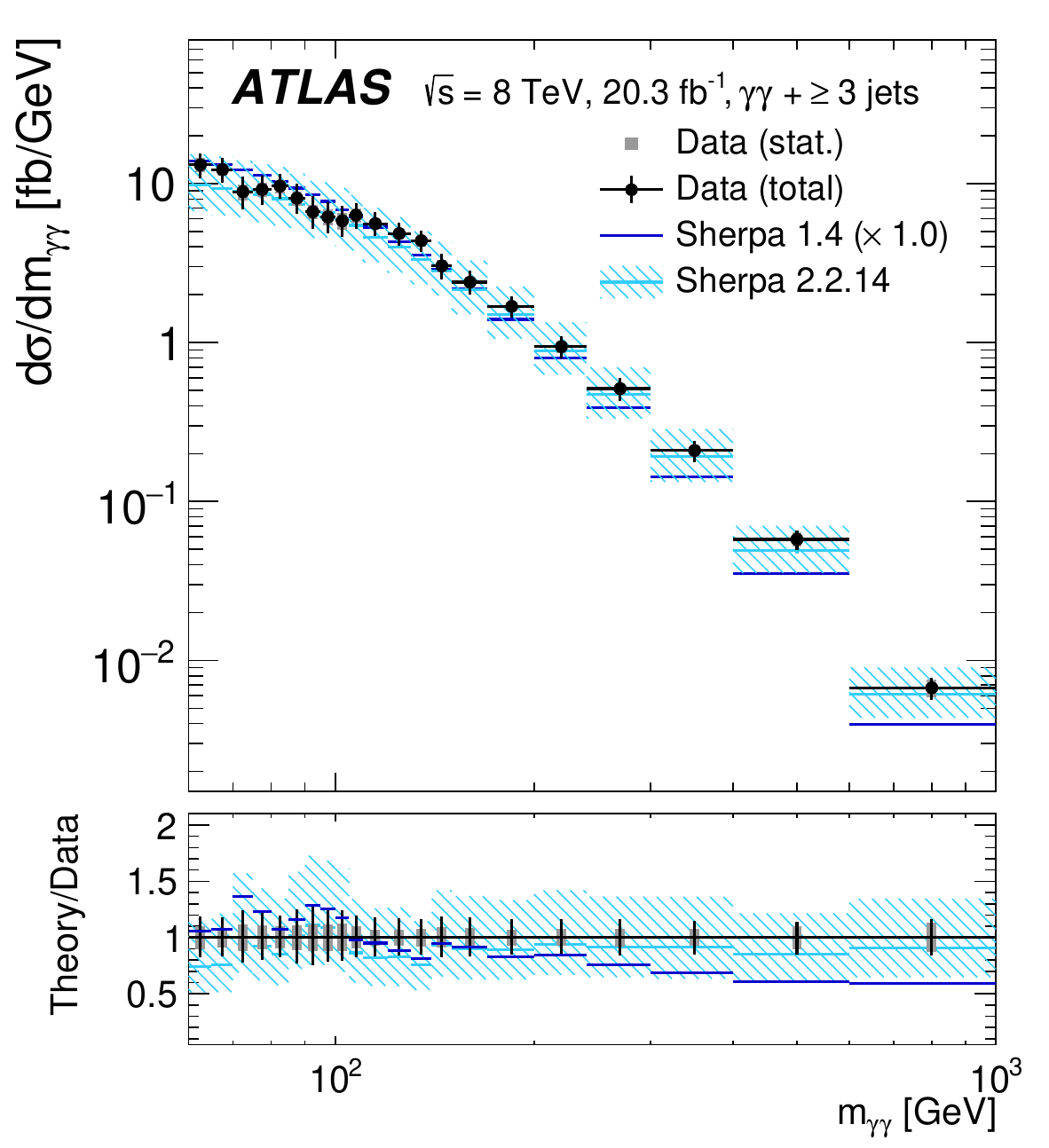}\label{fig:ggjj_3jet_incl_mgg}}
\subfloat[]{\includegraphics[width= 0.49\textwidth]{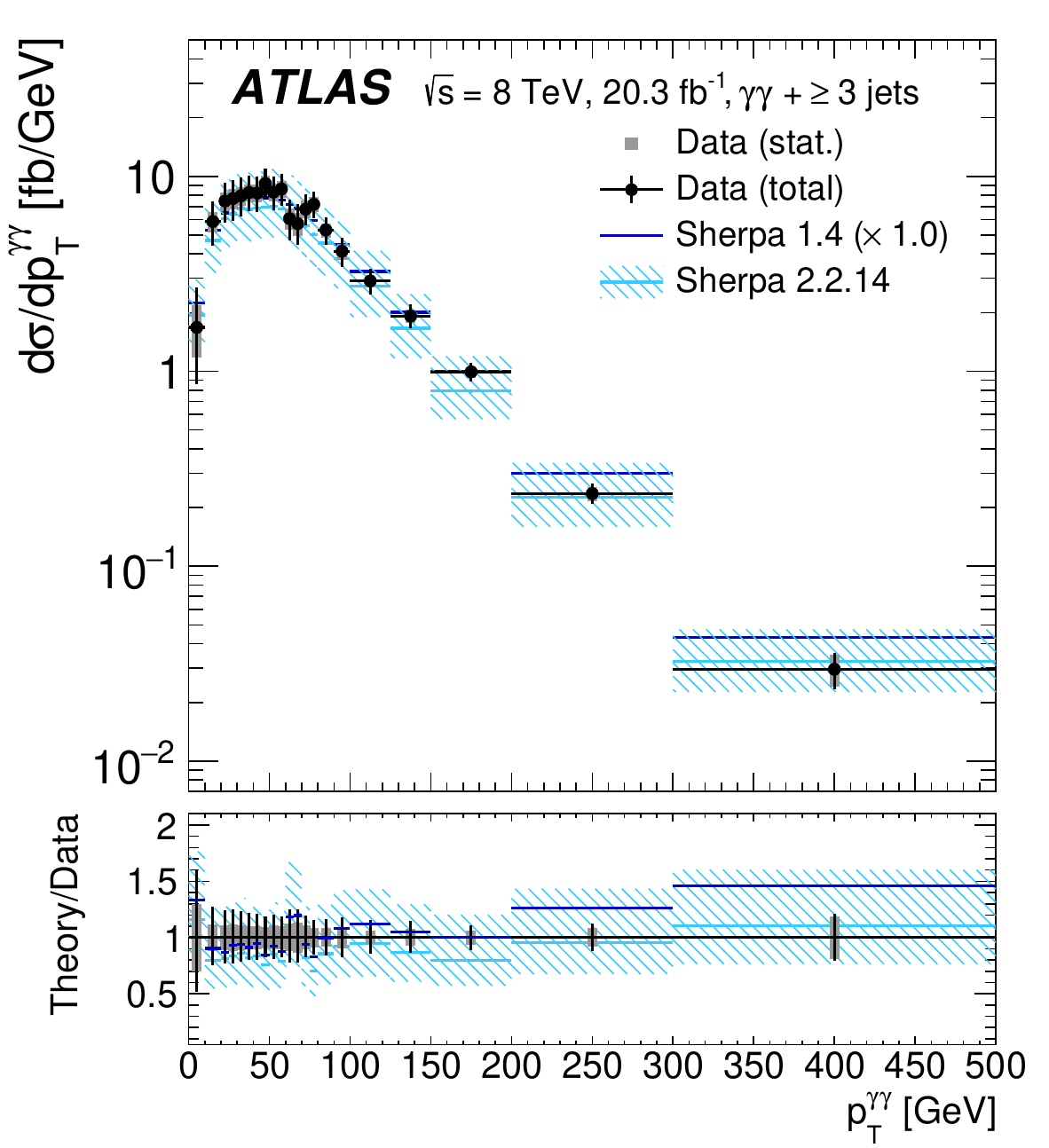}\label{fig:ggjj_3jet_incl_ptgg}} \\
\subfloat[]{\includegraphics[width= 0.49\textwidth]{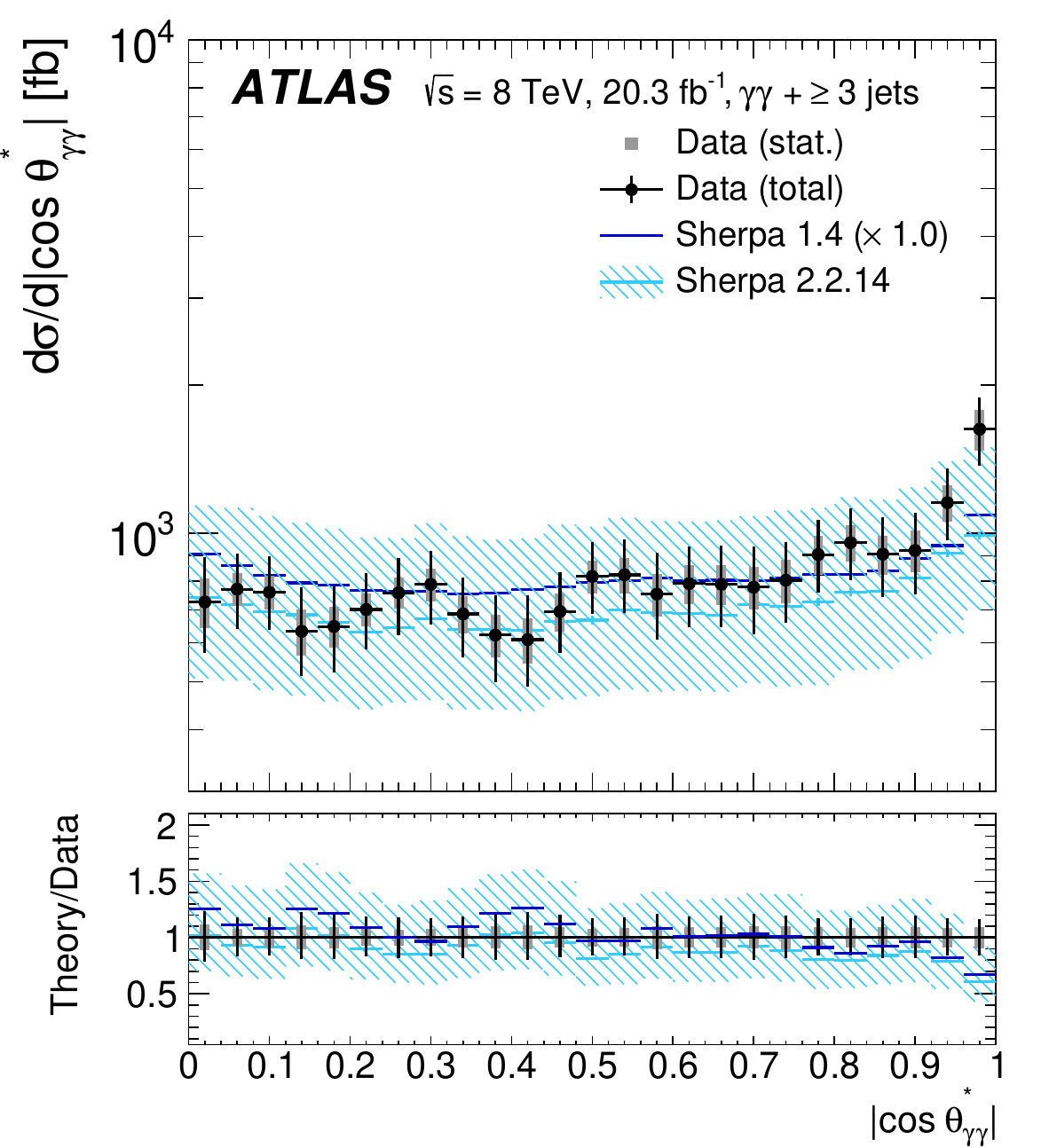}\label{fig:ggjj_3jet_incl_costhgg}}
\subfloat[]{\includegraphics[width= 0.49\textwidth]{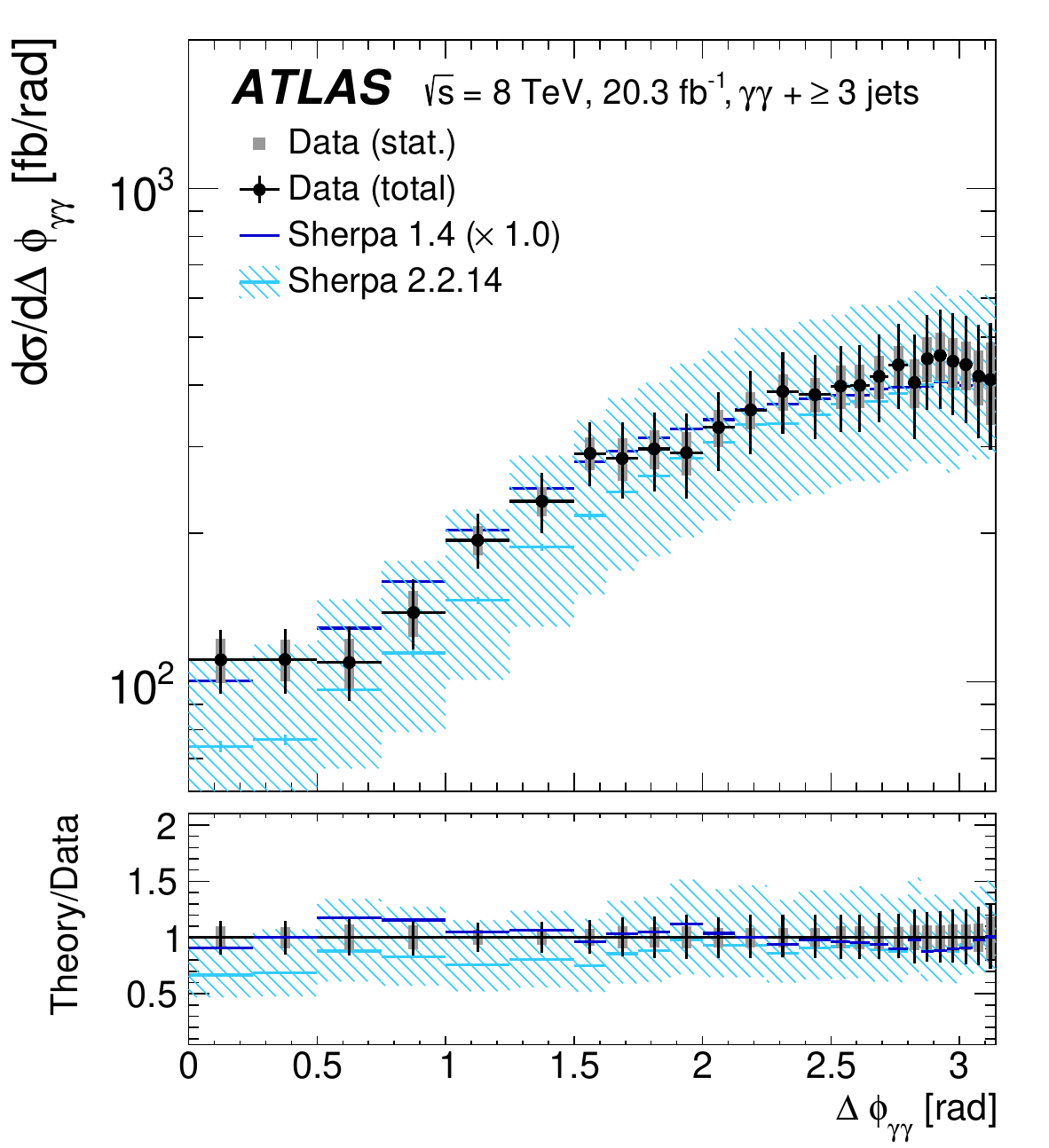}\label{fig:ggjj_3jet_incl_dphigg}}
\caption{
Differential cross sections as a function of (a) \mgg, (b) \pTgg, (c) \costhgg\ and (d) \dphigg\ for the $\Nj \geq 3$ category
compared with theoretical predictions from \sherpa\ 2.2.14,
and to LO \sherpa\ 1.4.1 at absolute normalization.
Ratios of theory to data are shown in the lower panel.
Arrows indicate data points outside the displayed ratio range.
}
\label{fig:ggjj_3jet_incl_diphoton}
\end{figure}

\begin{figure}
\begin{center}
\subfloat[]{\includegraphics[width= 0.49\textwidth]{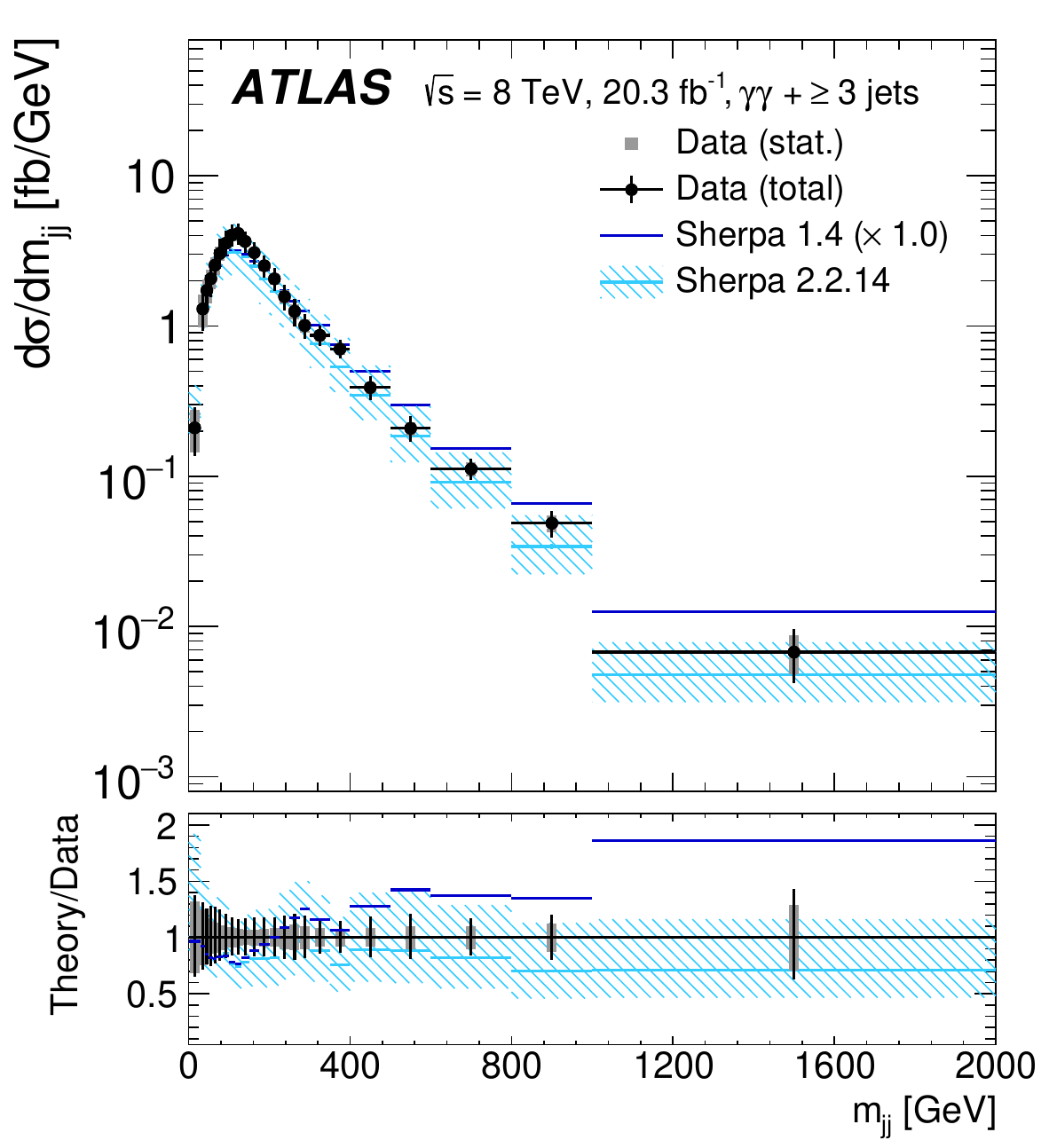}\label{fig:ggjj_3jet_incl_mjj}}
\subfloat[]{\includegraphics[width= 0.49\textwidth]{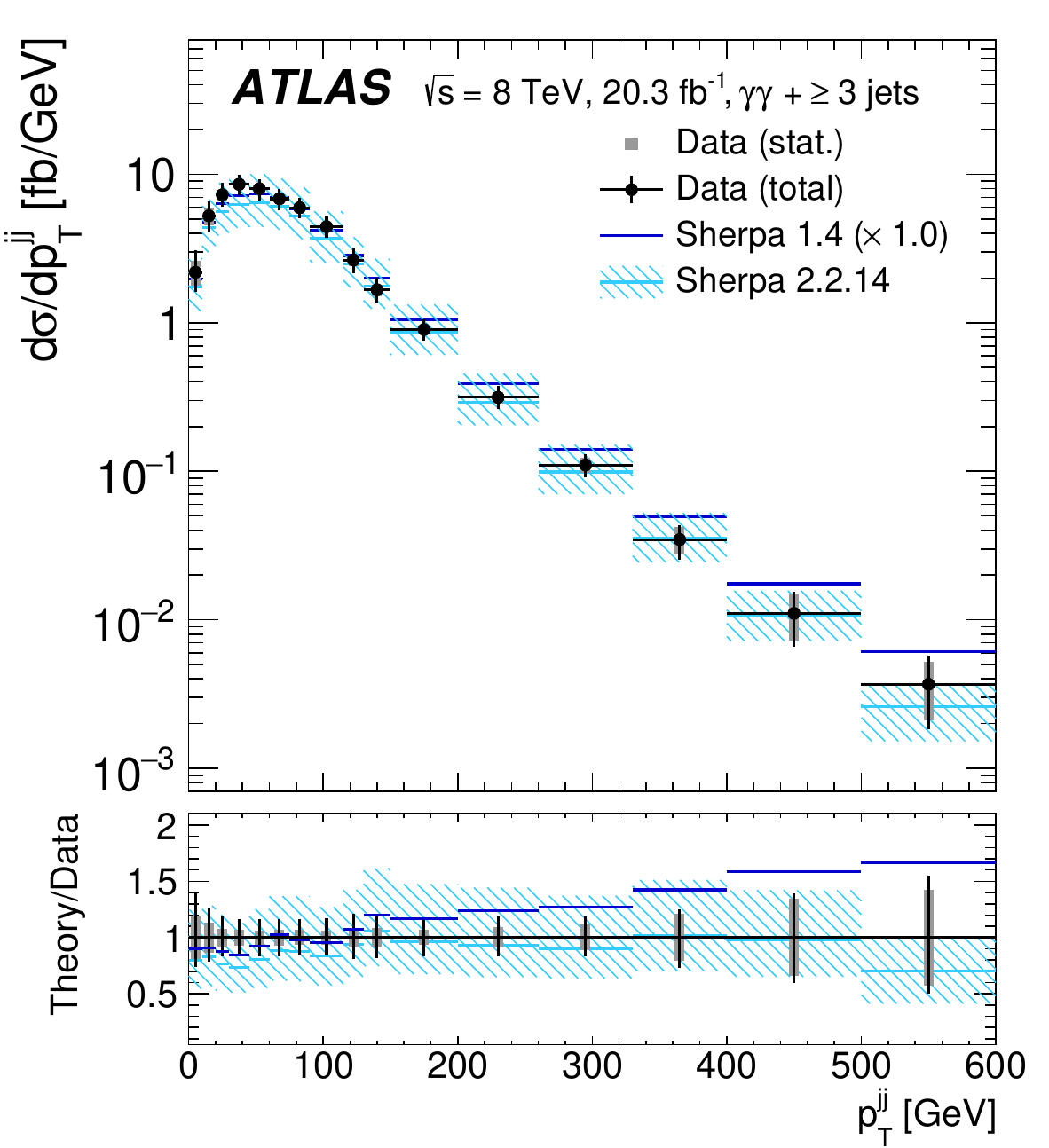}\label{fig:ggjj_3jet_incl_ptjj}} \\
\subfloat[]{\includegraphics[width= 0.49\textwidth]{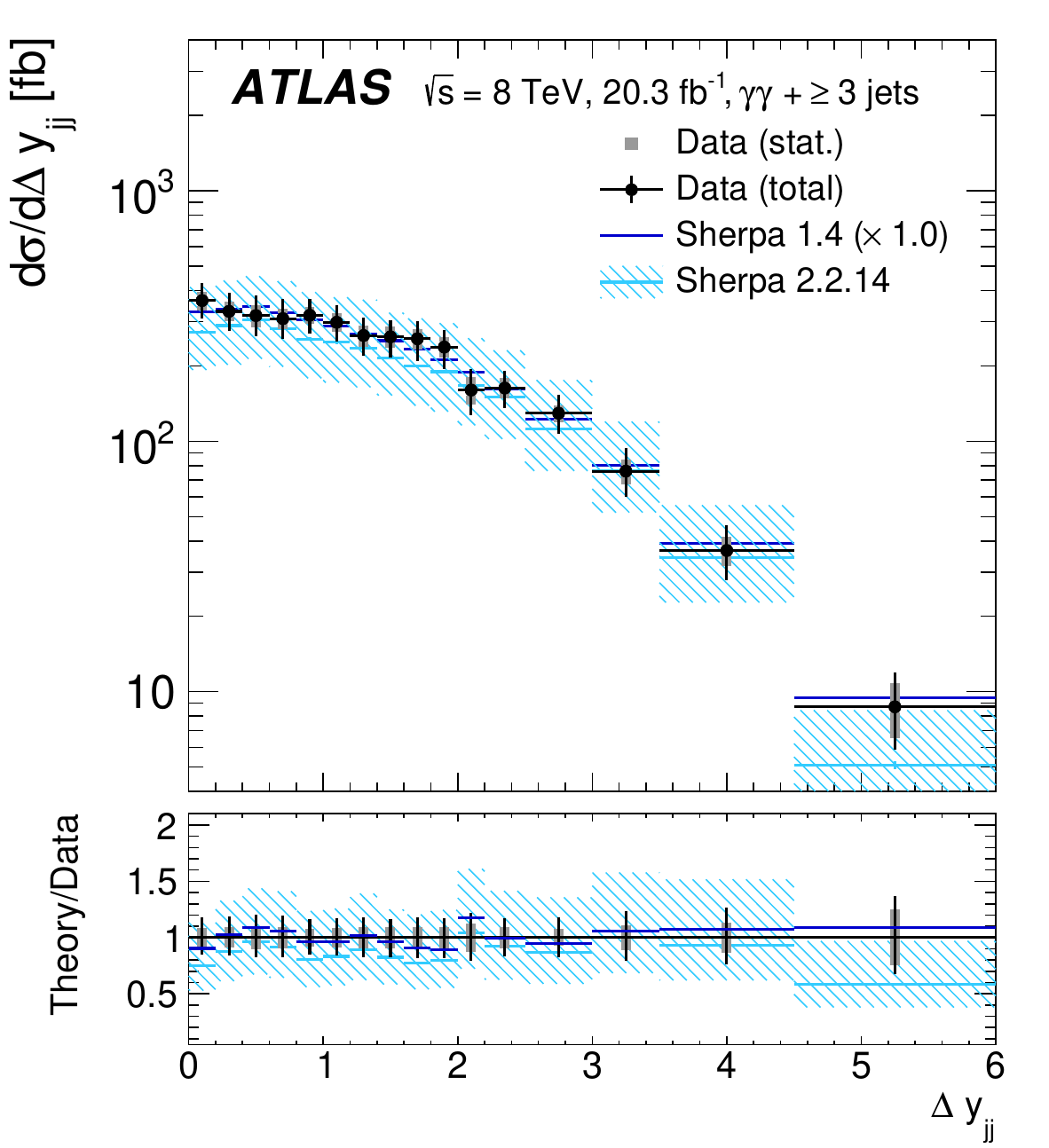}\label{fig:ggjj_3jet_incl_dyjj}}
\subfloat[]{\includegraphics[width= 0.49\textwidth]{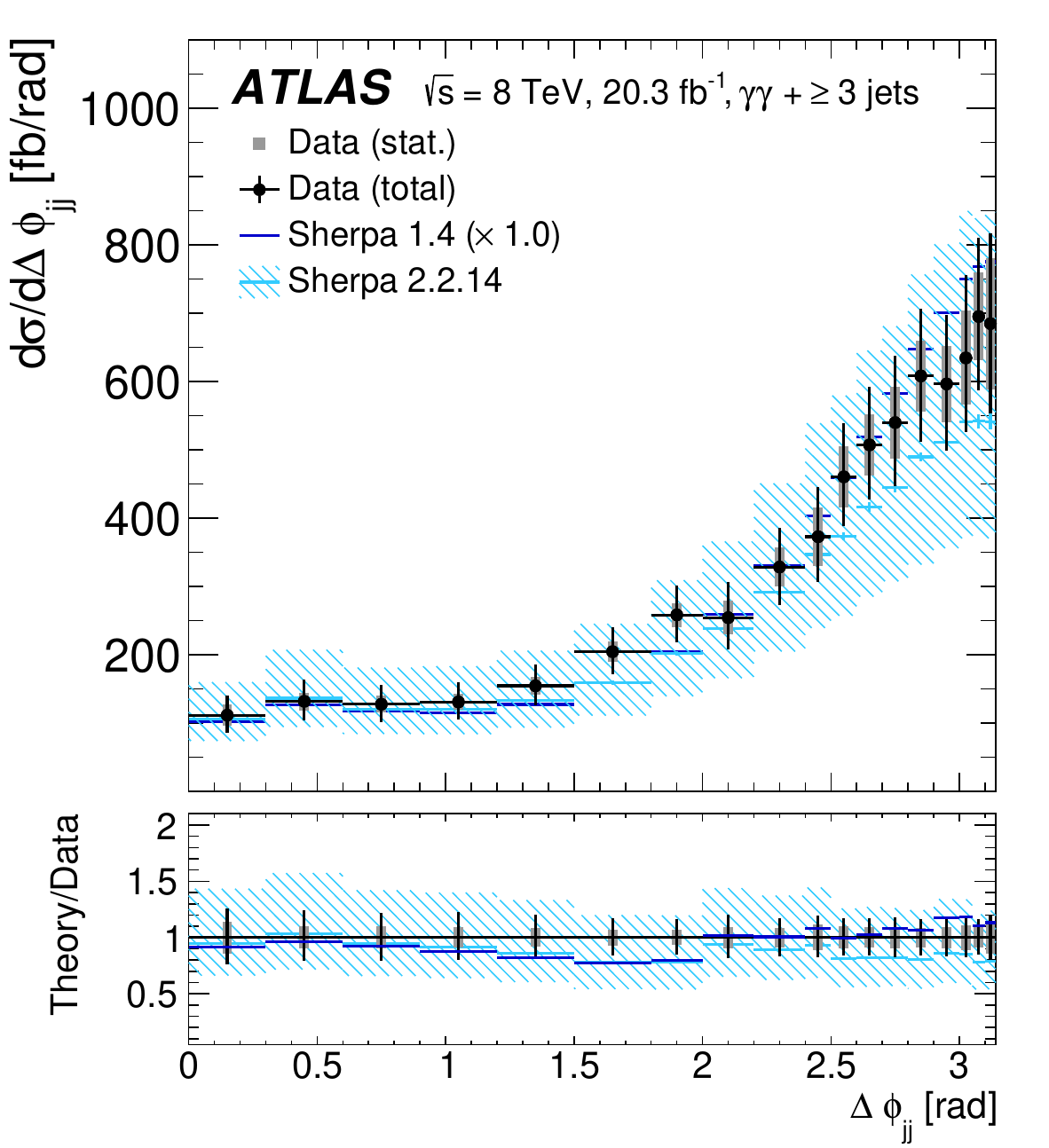}\label{fig:ggjj_3jet_incl_dphijj}}
\end{center}
\caption{
Differential cross sections as a function of (a) \mjj, (b) \pTjj, (c) \dyjj\ and (d) \dphijj\ for the $\Nj \geq 3$ category
compared with theoretical predictions from \sherpa\ 2.2.14,
and to LO \sherpa\ 1.4.1 at absolute normalization.
Ratios of theory to data are shown in the lower panel.
Arrows indicate data points outside the displayed ratio range.
}
\label{fig:ggjj_3jet_incl_dijet}
\end{figure}

\begin{figure}
\centering
\subfloat[]{\includegraphics[width= 0.49\textwidth]{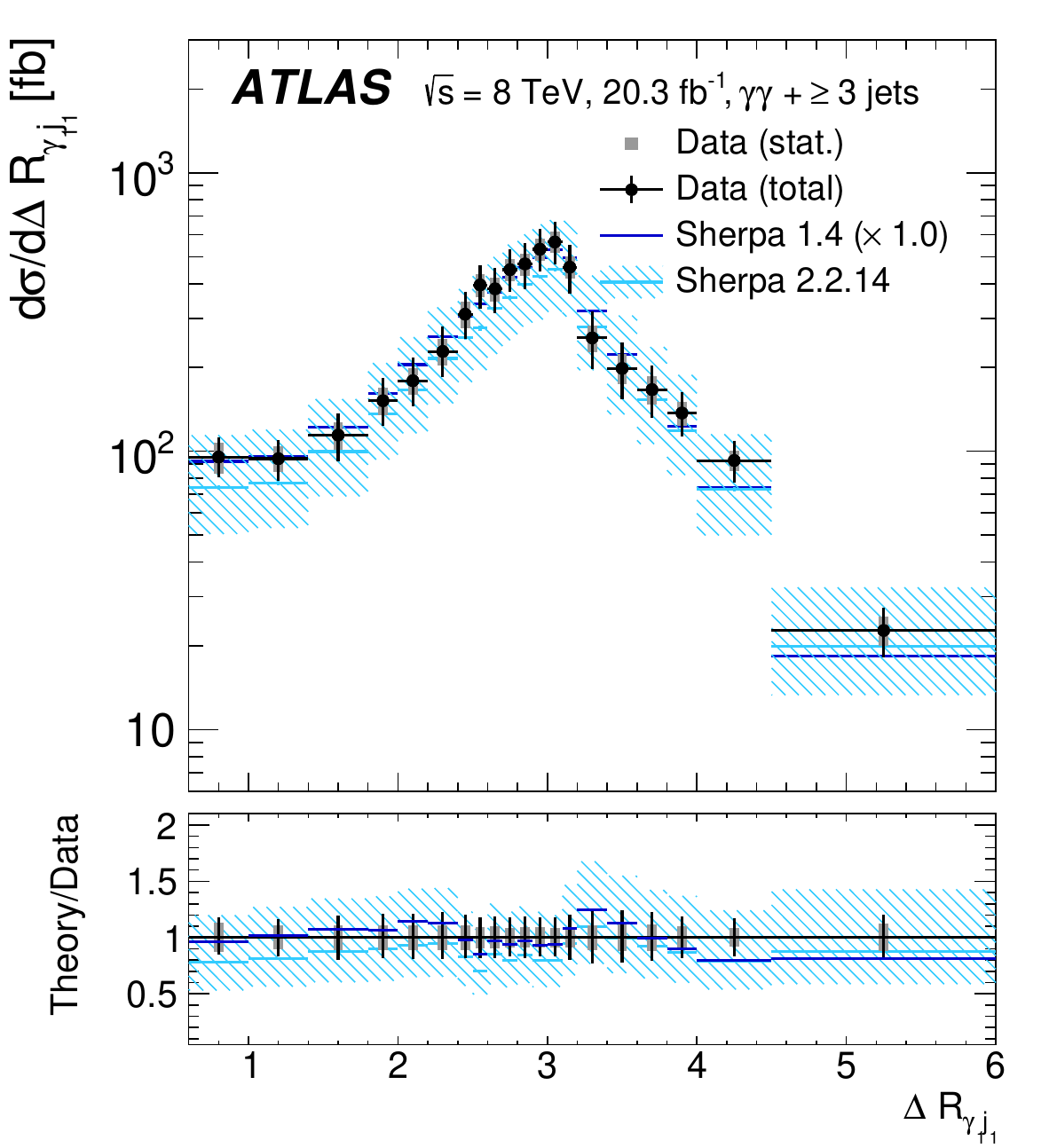}\label{fig:ggjj_3jet_incl_dRg1j1}}
\subfloat[]{\includegraphics[width= 0.49\textwidth]{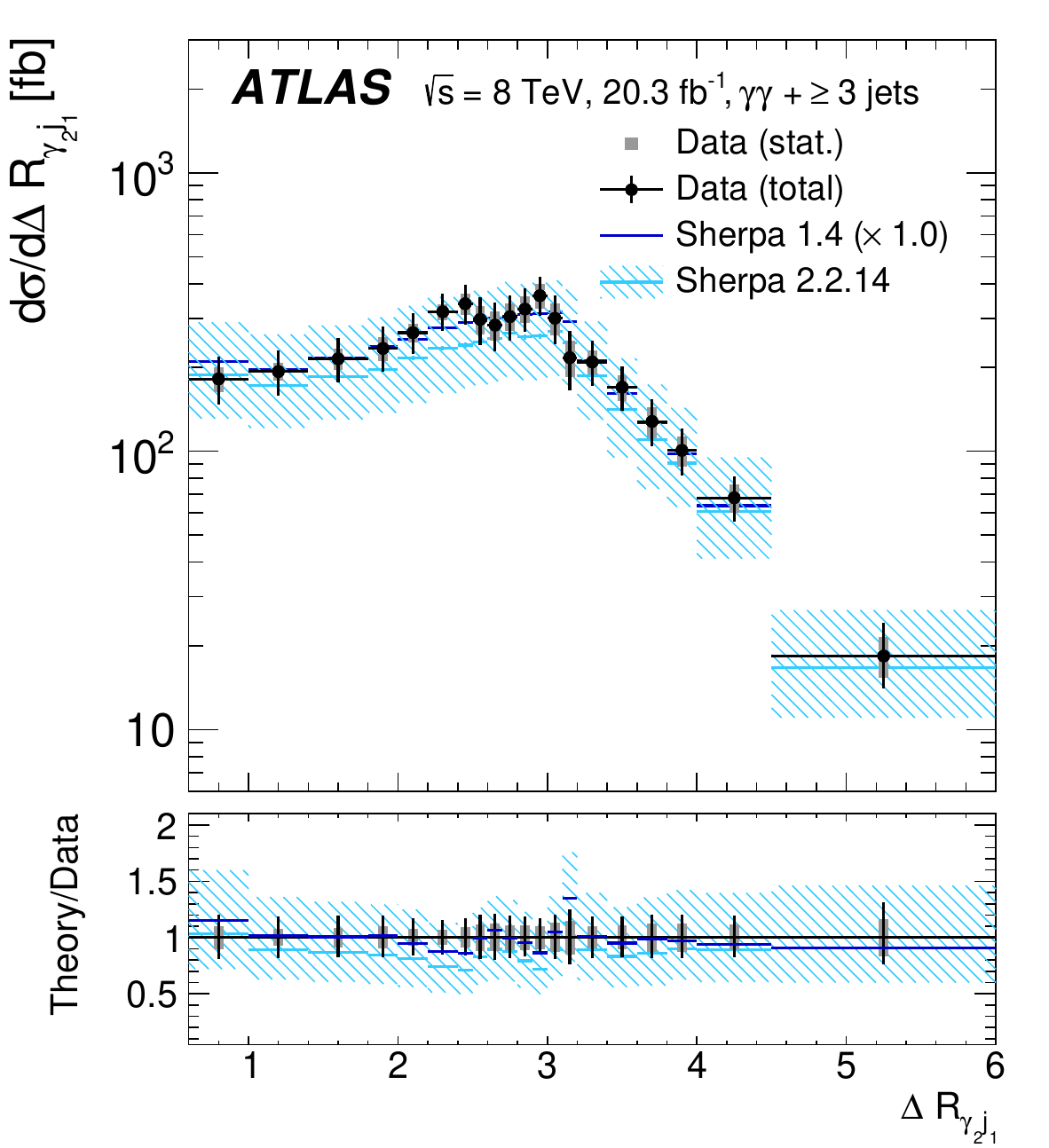}\label{fig:ggjj_3jet_incl_dRg2j1}} \\
\subfloat[]{\includegraphics[width= 0.49\textwidth]{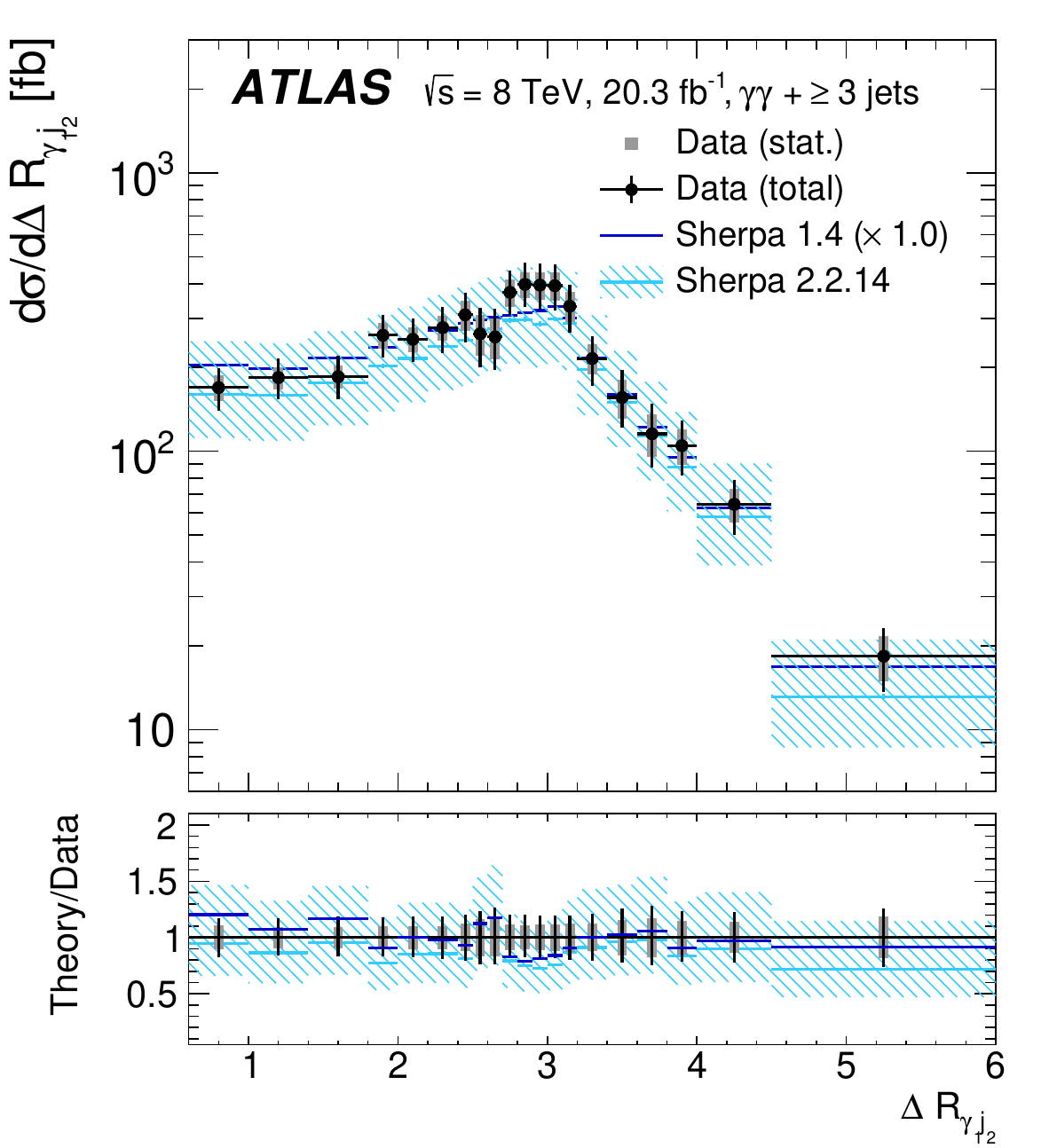}\label{fig:ggjj_3jet_incl_dRg1j2}}
\subfloat[]{\includegraphics[width= 0.49\textwidth]{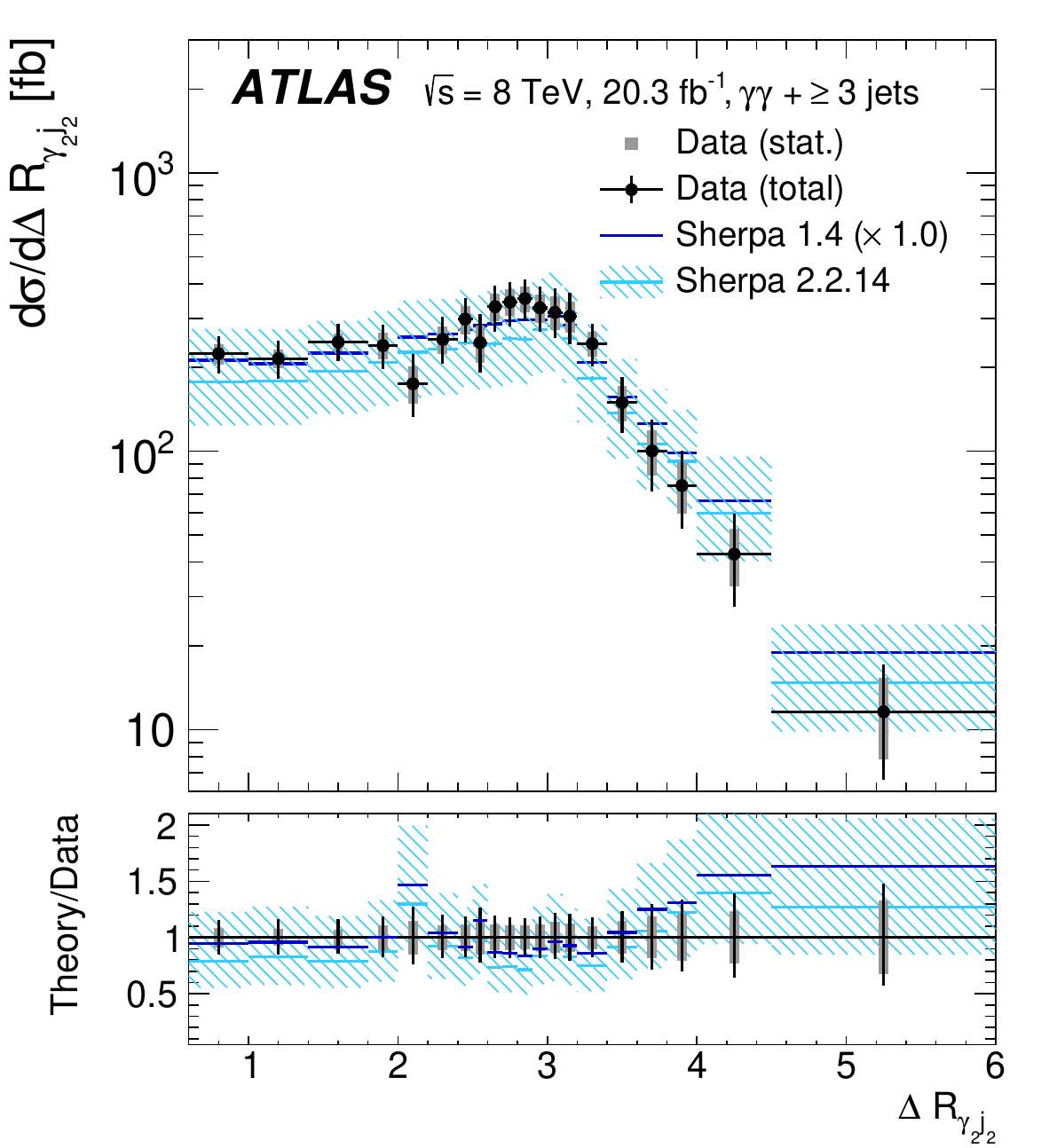}\label{fig:ggjj_3jet_incl_dRg2j2}}
\caption{
Differential cross sections as a function of the $R$-separation for the leading jet and (a) the leading photon, (b) subleading photon, subleading jet and the (c) leading photon (d) subleading photon for the $\Nj\geq 3$ category,
compared with theoretical predictions from \sherpa\ 2.2.14,
and to LO \sherpa\ 1.4.1 at absolute normalization.
Ratios of theory to data are shown in the lower panel.
Arrows indicate data points outside the displayed ratio range.
}
\label{fig:ggjj_3jet_incl_dRj}
\end{figure}

\begin{figure}
\centering
\subfloat[]{\includegraphics[width= 0.49\textwidth]{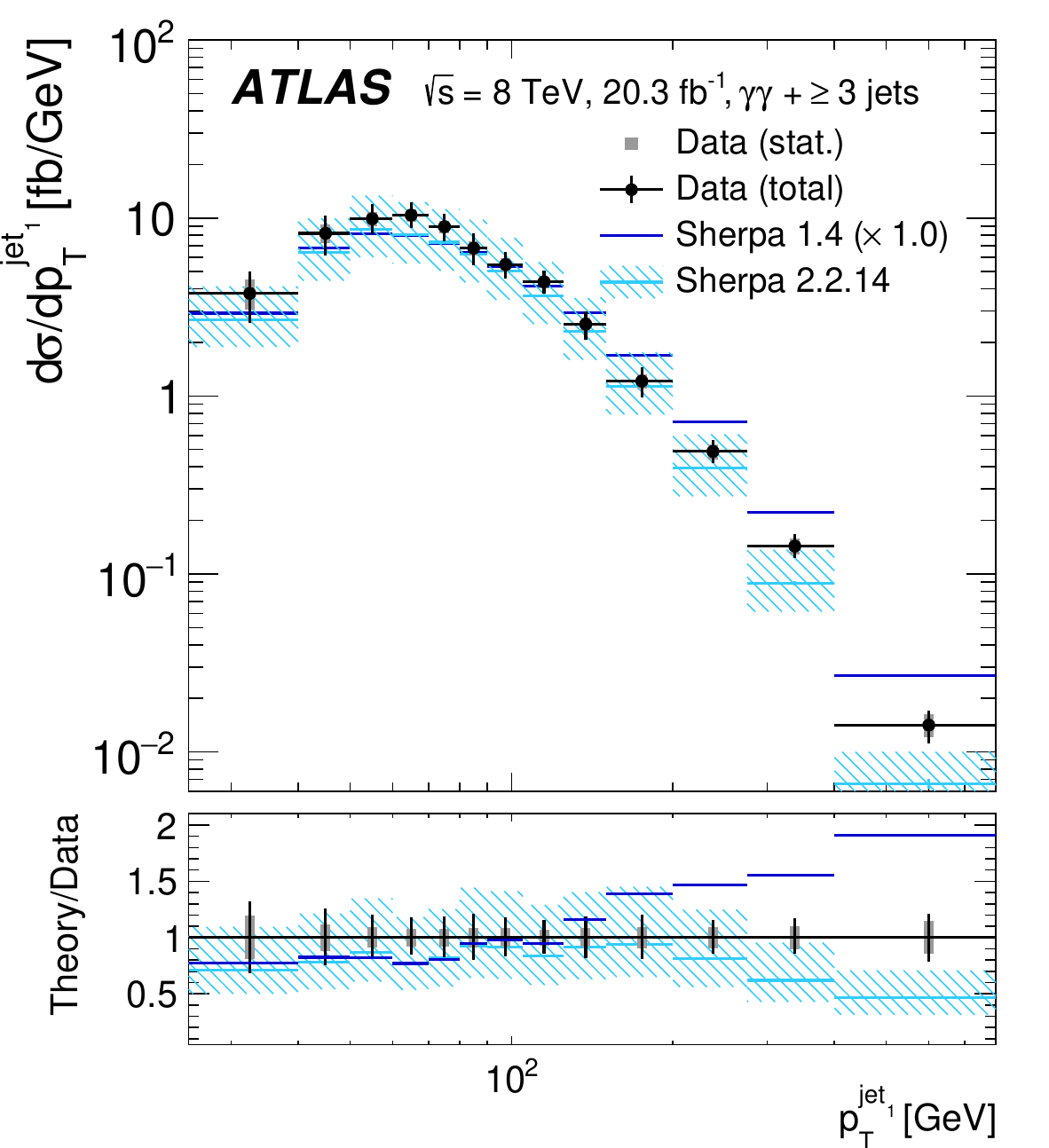}\label{fig:ggjj_3jet_incl_ptj1}}
\subfloat[]{\includegraphics[width= 0.49\textwidth]{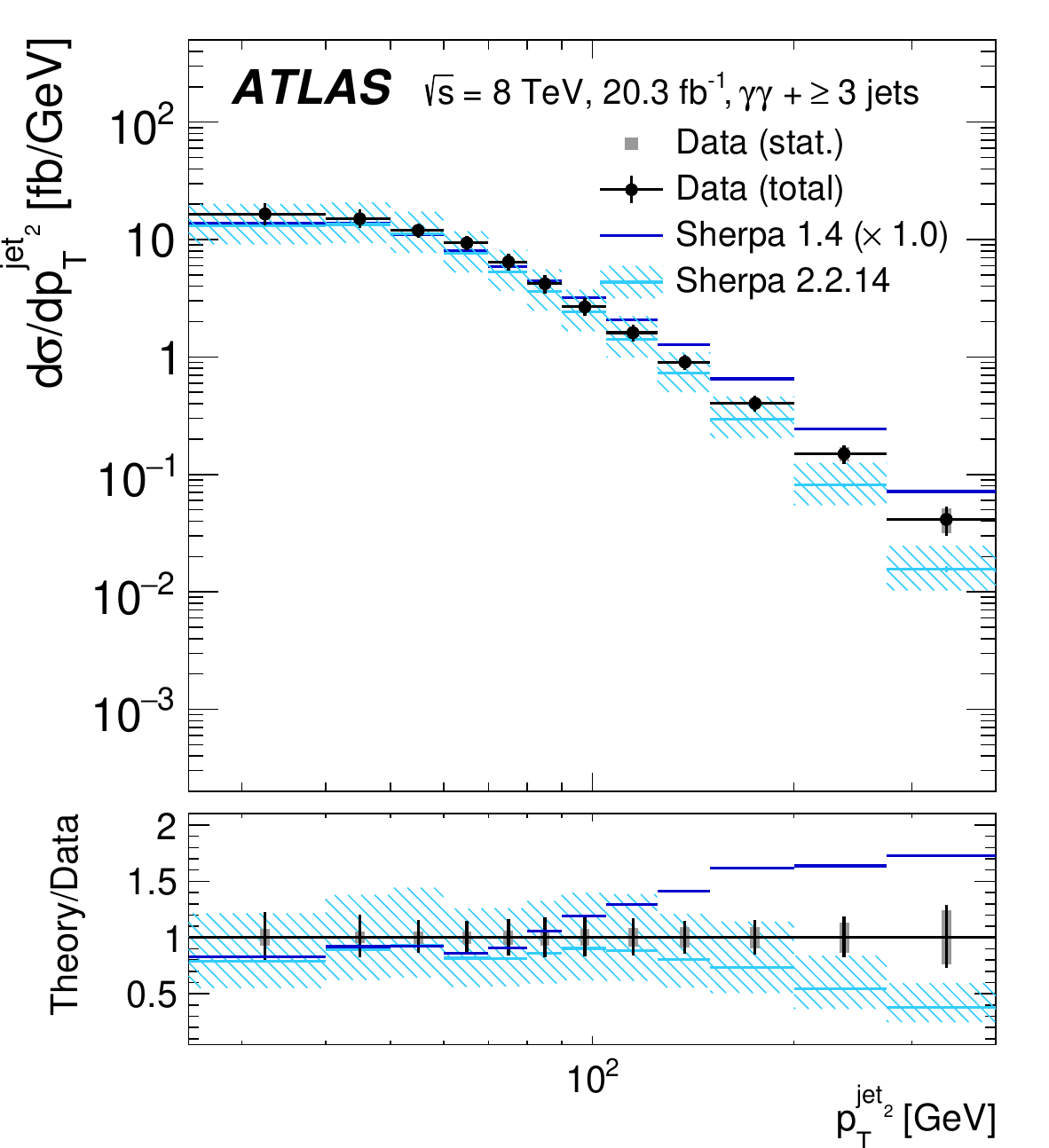}\label{fig:ggjj_3jet_incl_ptj2}} \\
\subfloat[]{\includegraphics[width= 0.49\textwidth]{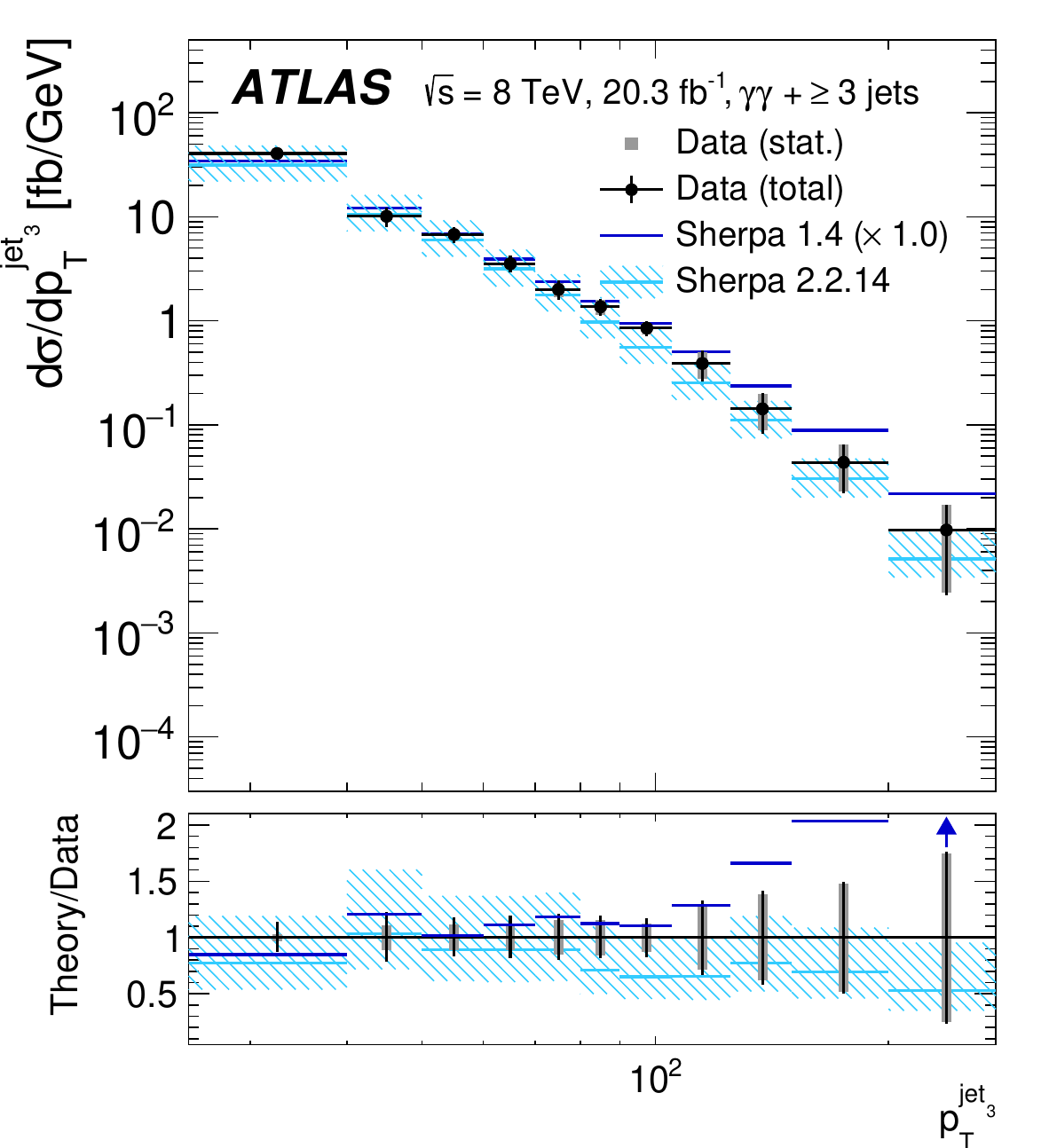}\label{fig:ggjj_3jet_incl_ptj3}}
\caption{
Differential cross sections as a function of the jet transverse momentum $\pTj$ for the (a) leading jet, (b) subleading jet and (c) third leading jet in the $\Nj \geq3$-jet category,
compared with theoretical predictions from \sherpa\ 2.2.14,
and to LO \sherpa\ 1.4.1 at absolute normalization.
Ratios of theory to data are shown in the lower panel.
Arrows indicate data points outside the displayed ratio range.
}
\label{fig:ggjj_3jet_incl_ptj123}
\end{figure}

Figures~\ref{fig:ggjj_3jet_incl_diphoton}--\ref{fig:ggjj_3jet_incl_ptj123} show a comparison of the measured distributions for the inclusive $\Nj\geq 3$ category to \sherpa\ 1.4.1, shown at absolute normalization since its integrated cross section matches the data in this category, and to the absolute {\textsc{Sherpa~2.2.14}} prediction. No fixed-order NLO predictions are available for this jet-multiplicity category.

{\textsc{Sherpa~2.2.14}} provides the best overall description among the predictions in the $\Nj\geq 3$ category (Figure~\ref{fig:ggjj_3jet_incl_diphoton}). The \mgg\ spectrum is well reproduced over the full measured range, and the \costhgg\ distribution is well described throughout. For \dphigg, good agreement is found for $\dphigg > 1~\text{rad}$, while a moderate undershoot at small separations is observed, consistent with the behavior seen in lower jet-multiplicity categories. The \pTgg\ distribution is reproduced at moderate values, with a progressive overestimate at high \pTgg\ similar to that observed in the 0-jet category.

The dijet observables for the $\Nj\geq 3$ category are shown in Figure~\ref{fig:ggjj_3jet_incl_dijet}. {\textsc{Sherpa~2.2.14}} provides a good description of the dijet system, consistent with the trends observed in the 2-jet category. The \mjj\ distribution is described within uncertainties over the full range, although the prediction lies 20--30\% below the data above about 100~\GeV. The \pTjj\ and \dphijj\ distributions are well reproduced across the measured range. For \dyjj, the agreement deteriorates at large rapidity separations, consistent with the behavior seen in the 2-jet category.

The photon--jet separation observables for the $\Nj\geq 3$ category are shown in Figure~\ref{fig:ggjj_3jet_incl_dRj}. {\textsc{Sherpa~2.2.14}} provides a good description of all four \dRgj\ observables, broadly consistent with the level of agreement observed in the 2-jet category, with a slight underestimate at the largest separation values.

The transverse momenta of the three leading jets are shown in Figure~\ref{fig:ggjj_3jet_incl_ptj123}. {\textsc{Sherpa~2.2.14}} provides a reasonable description of the leading and subleading jet \pT\ distributions, with an increasing undershoot at higher \pT\ values consistent with the behavior seen in the 2-jet category. The third-jet \pT\ distribution is also reasonably reproduced over the measured range.

Overall, \sherpa\ 1.4.1 describes the shapes well across most observables in the $\Nj\geq 3$ category.

The trends observed in the exclusive and $\Nj\geq 3$ categories persist for all inclusive categories: {\textsc{Sherpa~2.2.14}} provides the best overall description, while \sherpa\ 1.4.1 provides a reasonable shape description across most variables.


\section{Conclusion}
\label{sec:conclusion}

A measurement of the production cross section of isolated photon pairs in association with jets in $pp$ collisions at a center-of-mass energy $\sqrt{s} = 8~\TeV$ is presented. This measurement uses an integrated luminosity of 20.3~fb$^{-1}$ recorded in 2012 by the ATLAS detector at the LHC. Measurements in both inclusive and exclusive categories of the number of jets are presented. Cross sections are also measured differentially in up to sixteen kinematic and angular observables, depending on the jet category. The four main diphoton observables are presented: the diphoton invariant mass, \mgg, transverse momentum of the diphoton system \pTgg, and angular observables \costhgg\ and \dphigg. Transverse momenta of up to the first three leading jets are also measured. The photon--jet separations between the photons and the leading and subleading jets are presented, as are the dijet observables \mjj, \pTjj, \dyjj\ and \dphijj\ and the azimuthal separation between the diphoton and dijet systems \dSggjj. Relative statistical and systematic uncertainties related to jets become larger with increasing jet multiplicity, ranging from 3\% for $\Nj=0$ to 15\% for $\Nj \geq 3$, resulting in a total measurement uncertainty of 10\% to 16\% across jet categories.

Measured cross sections are compared with the {\textsc{Sherpa~1.4.1}} LO MC generator interfaced with a parton shower, fixed-order NLO predictions (\gosam\ and \blackhat), and NLO-matched predictions from {\textsc{Sherpa~2.2.14}}.
Integrated cross sections are generally underestimated by all generators in all categories (except \sherpa\ 1.4.1 in the $\geq 3$-jet inclusive category, where the predicted cross section matches the measurement), even if the NLO predictions are globally compatible with the measurements within uncertainties. Overall, the NLO predictions provide a reasonable description of the data in most regions of the differential distributions. At the same time, significant discrepancies are observed for increasing jet multiplicity and especially for angular variables, attributable to the absence of higher-order terms in the calculations. Fixed-order NLO predictions from \gosam\ and \blackhat\ describe the data with similar accuracy, since the predictions were generated with similar inputs. In the $\Nj = 2$ category, the \blackhat\ prediction is slightly closer to the measured integrated value, but the differences between the two predictions in the differential distributions are small and both provide a good description.
{\textsc{Sherpa~2.2.14}}, which provides NLO matrix elements for diphoton$+0,1$ jet production, shows overall compatibility with the data within the scale uncertainties, with some underestimation of the integrated cross sections in the 0-jet (9\%) and 2-jet (15\%) exclusive categories. Differentially, {\textsc{Sherpa~2.2.14}} provides the best overall description of the data across all jet categories and all observables, generally remaining within 15\% of the data over the bulk of each distribution. Common limitations include a progressive undershoot at large $|\cos\theta^*_{\gamma\gamma}|$ values across all categories, a slight deficit at the highest jet transverse momenta, an underestimate at large dijet rapidity separations $\Delta y_{\mathrm{jj}}$ in the two-jet category, and an underestimate at small \dphigg\ separations observed consistently across multiple jet-multiplicity categories. The same trends are observed in the inclusive jet categories. These results provide a stringent test of perturbative QCD predictions for diphoton production in association with jets and constitute a benchmark for future improvements to NLO-matched event generators.


\section*{Acknowledgements}

%
%

%
%

We thank CERN for the very successful operation of the LHC and its injectors, as well as the support staff at
CERN and at our institutions worldwide without whom ATLAS could not be operated efficiently.

The crucial computing support from all WLCG partners is acknowledged gratefully, in particular from CERN, the ATLAS Tier-1 facilities at TRIUMF/SFU (Canada), NDGF (Denmark, Norway, Sweden), CC-IN2P3 (France), KIT/GridKA (Germany), INFN-CNAF (Italy), NL-T1 (Netherlands), PIC (Spain), RAL (UK) and BNL (USA), the Tier-2 facilities worldwide and large non-WLCG resource providers. Major contributors of computing resources are listed in Ref.~\cite{ATL-SOFT-PUB-2026-001}.

We gratefully acknowledge the support of ANPCyT, Argentina; YerPhI, Armenia; ARC, Australia; BMWFW and FWF, Austria; ANAS, Azerbaijan; CNPq and FAPESP, Brazil; NSERC, NRC and CFI, Canada; CERN; CAS, MOST and NSFC, China; MSMT CR, MPO CR and VSC CR, Czech Republic; DNRF and DNSRC, Denmark; IN2P3-CNRS and CEA-DRF/IRFU, France; SRNSFG, Georgia; BMFTR, HGF and MPG, Germany; GSRT, Greece; RGC and Hong Kong SAR, China; ISF and Benoziyo Center, Israel; INFN, Italy; MEXT and JSPS, Japan; CNRST, Morocco; NWO, Netherlands; RCN, Norway; MNiSW and NCN, Poland; FCT, Portugal; MNE/IFA, Romania; MESTD, Serbia; MSSR, Slovakia; ARRS and MIZ\v{S}, Slovenia; MINECO, Spain; SRC and Wallenberg Foundation, Sweden; SERI, SNSF and Cantons of Bern and Geneva, Switzerland; MOST, Taipei; TAEK, Turkey; STFC, United Kingdom; DOE and NSF, United States of America.

Individual groups and members have received support from BCKDF and CANARIE, Canada; UNCE SCI/013, Czech Republic; ERDF, European Union; ANR, France; DFG and AvH Foundation, Germany; BSF-NSF, Israel; CERCA and AGAUR programs from Generalitat de Catalunya, Spain; The Royal Society and Leverhulme Trust, United Kingdom.

In addition, individual members wish to acknowledge support from Chile: Agencia Nacional de Investigaci\'on y Desarrollo (Fondecyt  3180653); European Union: European Research Council (ERC - 617185); Spain: Ministry of Science and Innovation (RYC2022-038164-I); Sweden: Swedish Research Council (SRC - 2017-05160), Knut and Alice Wallenberg Foundation (KAW 2013.0149, KAW 2015.0193, KAW 2017.0100).

%
%


\printbibliography

\clearpage
 
\begin{flushleft}
\hypersetup{urlcolor=black}
{\Large The ATLAS Collaboration}

\bigskip

\AtlasOrcid{M.~Aaboud}$^\textrm{\scriptsize 36c}$,
\AtlasOrcid[0000-0002-6665-4934]{G.~Aad}$^\textrm{\scriptsize 104}$,
\AtlasOrcid[0000-0002-5888-2734]{B.~Abbott}$^\textrm{\scriptsize 124}$,
\AtlasOrcid{B.~Abeloos}$^\textrm{\scriptsize 67}$,
\AtlasOrcid[0000-0002-2987-4006]{D.K.~Abhayasinghe}$^\textrm{\scriptsize 95}$,
\AtlasOrcid[0000-0002-8496-9294]{S.H.~Abidi}$^\textrm{\scriptsize 163}$,
\AtlasOrcid[0000-0002-8279-9324]{O.S.~AbouZeid}$^\textrm{\scriptsize 140}$,
\AtlasOrcid{N.L.~Abraham}$^\textrm{\scriptsize 154}$,
\AtlasOrcid[0000-0001-5329-6640]{H.~Abramowicz}$^\textrm{\scriptsize 159}$,
\AtlasOrcid[0000-0002-1599-2896]{H.~Abreu}$^\textrm{\scriptsize 158}$
% ...
\bigskip
\\

% ...
$^{36}$$^{(a)}$Facult\'e des Sciences Ain Chock, Universit\'e Hassan II de Casablanca;$^{(b)}$Facult\'e des Sciences Semlalia, Universit\'e Cadi Ayyad, LPHEA-Marrakech;$^{(c)}$LPMR, Facult\'e des Sciences, Universit\'e Mohamed Premier, Oujda;$^{(d)}$Facult\'e des sciences, Universit\'e Mohammed V, Rabat; Morocco.\\
% ...
$^{67}$IJCLab, Universit\'e Paris-Saclay, CNRS/IN2P3, 91405, Orsay; France.\\
% ...
$^{95}$Department of Physics, Royal Holloway University of London, Egham; United Kingdom.\\
% ...
$^{104}$CPPM, Aix-Marseille Universit\'e, CNRS/IN2P3, Marseille; France.\\
% ...
$^{124}$Homer L. Dodge Department of Physics and Astronomy, University of Oklahoma, Norman OK; United States of America.\\
% ...
$^{140}$Santa Cruz Institute for Particle Physics, University of California Santa Cruz, Santa Cruz CA; United States of America.\\
% ...
$^{154}$Department of Physics and Astronomy, University of Sussex, Brighton; United Kingdom.\\
% ...
$^{158}$Department of Physics, Technion, Israel Institute of Technology, Haifa; Israel.\\
$^{159}$Raymond and Beverly Sackler School of Physics and Astronomy, Tel Aviv University, Tel Aviv; Israel.\\
% ...
$^{163}$Department of Physics, University of Toronto, Toronto ON; Canada.\\
% ...
$^{a}$ Also Affiliated with an institute covered by a cooperation agreement with CERN.\\
$^{b}$ Also at Borough of Manhattan Community College, City University of New York, New York NY; United States of America.\\
$^{c}$ Also at Centre for High Performance Computing, CSIR Campus, Rosebank, Cape Town; South Africa.\\
$^{d}$ Also at CERN, Geneva; Switzerland.\\
% ...

\end{flushleft}


\clearpage

\end{document}